\pdfoutput=1 
\documentclass[12pt]{article}

\usepackage{inputenc}

\usepackage{amsmath,amssymb,bbm,color}
\usepackage{amsbsy}
\usepackage{euscript}
\usepackage{graphicx}
\usepackage{hyperref}
\usepackage{cite}
\usepackage{enumerate}

\newcommand{\afb}{{A_{\rm FB}}}
\newcommand{\vmax}{{v_{\max}}}
\newcommand{\cst}{{\triangle}}

\newcommand{\kksem}{{\tt  KKsem}}
\newcommand{\kkmc}{{\tt   KKMC}}
\newcommand{\kkfoam}{{\tt KKFoam}}
\newcommand{\foam}{{\tt FOAM}}
\newcommand{\dizet}{{\tt DIZET}}

\newcommand{\Mcal}{{\cal M}}

\newcommand{\veps}{{\varepsilon}}

\newcommand{\Peu}{\EuScript{P}}

\newcommand{\Meu}{\EuScript{M}}

\newcommand{\Dmf}{\mathfrak{D}}
\newcommand{\Bmf}{\mathfrak{B}}

\newcommand{\Rmf}{\mathfrak{R}}

\newcommand{\Mmf}{\mathfrak{M}}

\newcommand{\Mb}{\overline{M}}

\newcommand{\order}[1]{${\cal O}(#1)$}

\begin{document}

\begin{titlepage}

\begin{flushright}
\bf IFJPAN-IV-2017-11
\end{flushright}

\vspace{5mm}
\begin{center}
    {\Large\bf QED Interference in Charge Asymmetry\\
      Near the Z Resonance at Future
      Electron-Positron Colliders $^{\star}$ }
\end{center}

\vskip 5mm
\begin{center}
{\large S.\ Jadach$^a$, S.\ Yost$^{b,a}$}

\vskip 2mm
{\em $^a$Institute of Nuclear Physics, Polish Academy of Sciences,\\
  ul.\ Radzikowskiego 152, 31-342 Krak\'ow, Poland}\\
\vspace{1mm}
{\em $^b$The Citadel, Charleston, SC, USA\\ }
\end{center}

\vspace{2mm}
\begin{abstract}
\noindent
The measurement of the charge asymmetry 
$A_{\rm FB}(e^-e^+\to \mu^-\mu^+)$ will play an important role
at the high-luminosity circular electron-positron collider
FCCee considered for construction at CERN.  In particular,
near the $Z$ resonance, $\sqrt{s} \simeq M_Z \pm 3.5$ GeV,
$A_{\rm FB}$ will provide a very precise value of
the pure electromagnetic coupling constant $\alpha_{\rm QED}(M_Z)$,
which is vitally important for overall tests of the Standard Model.
For this purpose, $A_{\rm FB}$ will be measured at the FCCee with 
an experimental error better than
$\delta A_{\rm FB} \simeq 3\cdot 10^{-5},$ at least
a factor of 100 more precisely than at past LEP experiments!
The important question is whether the effect of interference between photon
emission in the initial and final state (IFI) can be removed from 
the $A_{\rm FB}$ data 
at the same precision level using perturbative QED calculations.  A first 
quantitative study of this problem is presented here, with the help of the
\kkmc\ program and a newly developed calculation based on soft photon 
resummation, matched with NLO and NNLO fixed-order calculations.
It is concluded that a factor of 10 improvement with respect to the LEP era 
is obtained.
We also present a clear indication that reducing the uncertainty
of charge asymmetry near the $Z$ peak due to IFI down to
$\delta A_{\rm FB} \simeq 3\cdot 10^{-5}$, 
{\em i.e.} the expected experimental precision at FCCee, is feasible.
\end{abstract}

\vspace{50mm}
\footnoterule
\noindent
{\footnotesize
$^{\star}$This work is partly supported by
 the Polish National Science Center grant 2016/23/B/ST2/03927,
 the Citadel Foundation 
 and the CERN FCC Design Study Programme.
}

\end{titlepage}

\newpage
\tableofcontents

\newpage
\section{Introduction}

At the future high-energy high-luminosity circular electron-positron collider
FCCee~\cite{Gomez-Ceballos:2013zzn,Mangano:2018mur,Benedikt:2018qee}
proposed for construction at CERN, 
the measurement of the muon charge asymmetry
$A^\mu_{\rm FB}=A_{\rm FB}(e^-e^+\to \mu^-\mu^+)$ 
will play an additional important role.
Near the $Z$ resonance, $\sqrt{s_\pm} \simeq M_Z \pm 3.5$ GeV, 
the measurement of $A^\mu_{\rm FB}$ may provide a very precise value of
the pure electromagnetic coupling constant $\alpha_{\rm QED}(M_Z)$,
which is vitally important for overall tests of the Standard Model (SM),
especially for the SM prediction of $\sin^2\theta^l_{\rm eff}$ and 
$M_W$ \cite{Blondel:2019qlh}, 
at a precision level at least a factor of 10 better than presently.
This kind of the measurement of $\alpha_{\rm QED}(M_Z)$ was proposed and 
analyzed in Ref.~\cite{Janot:2015gjr}.

In past measurements of the charge asymmetry $A^\mu_{\rm FB}$ (forward-backward
angular asymmetry) at electron-positron colliders LEP and SLC, 
the QED interference between photons emitted from the initial and final charged 
leptons had to be taken into account.
Let us abbreviate in the following $A^\mu_{\rm FB} \equiv \afb$.  In overall tests of the SM, 
the measurement of $\afb$ contributed mainly to knowledge of the $Z$ 
couplings and/or the effective electroweak mixing angle $\sin^2\theta_{\rm eff}^l$
\cite{ALEPH:2005ab}.

Thanks to the very high luminosity of the FCCee~\cite{Gomez-Ceballos:2013zzn}, 
the charge asymmetry 
$\afb (M_Z \pm 3.5 {\rm GeV})$ will possibly be measured with an error
$\delta \afb/ \afb \simeq 3\cdot 10^{-5}$ or even better 
\cite{Janot:2015gjr,Mangano:2018mur}.
This immediately poses the question of whether the effect of
QED initial-final interference can be removed from the data
at the same precision level.  
How big is the effect of IFI in $\afb$?
Far from the resonance, it is about $2-3\%$
and it is even bigger for a tight cutoff on the total energy of the 
emitted photons.  At the top of the $Z$ resonance, 
where $\afb$ was measured most precisely in the past LEP experiments, 
the IFI effect is suppressed by the ratio $\Gamma_Z/M_Z$ 
to the level of $\delta \afb \sim 0.1\%$,
due to the long time separation between the creation
and the decay stages of the $Z$ resonance, 
as elaborated in many LEP era works, see
Refs.~\cite{Bohm:1989pb,Jadach:1988zp,Bardin:1990fu,Bardin:1999gt,Kobel:2000aw,Jadach:2000ir}.
As we shall see in our analysis, at $\sqrt{s} \simeq M_Z \pm 3.5$ GeV,
the  same $\Gamma_Z/M_Z$ suppression of IFI in $A_{\rm FB}$
still works to some extent,
but the IFI effect is nevertheless at the $\delta A_{\rm FB} \sim 1\%$ level,
and growing for tight cutoffs, in spite of partial cancellations
in the difference between values at $\sqrt{s} \simeq M_Z \pm 3.5$,
as already noticed in Ref~\cite{Janot:2015gjr}.

This effect is huge with respect to the planned experimental precision at 
FCCee, and it would render measurement of the $A_{\rm FB} (M_Z \pm 3.5 {\rm GeV})$
completely useless unless the theoretical evaluation of IFI is equally precise!
Note that in the LEP data analysis, a cutoff on the total
photon energy was imposed by requiring a minimum value of the effective mass of
the muon pair, $M_{\mu^-\mu^+}$,  or a maximum acollinearity angle, 
which was translated into an upper limit on the total photon energy
$E_\gamma^{\rm tot}/s^{1/2}$ varying between $0.5$ to $0.998$
(see Table 2.1 in Ref.~\cite{ALEPH:2005ab}).
Due to the higher precision of the FCCee, a stronger cutoff probably will
be preferred
in order to minimize the background from hadronic and tau pair channels,
and for better control of the angular dependence of the muon detector 
efficiency\footnote{P. Janot, private communication.}.
Also, theoretical control over IFI may be better for a stronger cutoff
(in spite of its larger size) thanks to the power of soft photon resummation
and a better elimination of the four fermion contributions.
For this reason, we will use a photon energy cutoff 
$E_\gamma^{\rm tot}/s^{1/2} \leq 0.2$ or stronger.

How precise are the theoretical evaluations 
of the effect of IFI in $\afb$ presently available in perturbative QED?
In the pre-LEP era, ${\cal O}(\alpha^1)$ fixed-order calculations
were quoted to provide $\sim 0.3-0.5\%$ precision; 
see the review of Ref.~\cite{Bohm:1989pb}.
In the LEP1 phase near the Z resonance, thanks to  $\Gamma_Z/M_Z$ suppression, 
the IFI effect in $\afb$ at the $Z$ peak was not a burning issue.
For instance, in the work of Ref.~\cite{Bardin:1999gt} used in
the final data analysis of LEP1 of Ref.~\cite{ALEPH:2005ab},
the calculations of the IFI effect were done using
{\tt ZFITTER} \cite{Bardin:1999yd} 
and TOPAZ0 \cite{Montagna:1995ja,Montagna:1993py} programs, 
crosschecking them with the KORALZ Monte Carlo 
\cite{Jadach:1993yv,Jadach:1999tr}.

In all these calculations and programs, the treatment of IFI
was at the ${\cal O}(\alpha^1)$ fixed-order level,
without soft photon resummation.
Pioneering work on the resummation of soft photon effects 
near a narrow resonance, including resummation of $\ln(\Gamma_Z/M_Z)$, 
was already done earlier by the Frascati group;
see Refs.~\cite{Greco:1975rm,Greco:1975ke,Greco:1980mh},
but it was not exploited in the above studies, 
mainly because they did not include hard photon effects in a realistic way.

Significant progress on the IFI issue was made just before the end of the LEP 
era, with the advent of new method of the soft photon resummation
matched with fixed-order QED corrections up to \order{\alpha^2}
and electroweak (EW) corrections up to \order{\alpha^1},
the so-called coherent exclusive exponentiation~\cite{Jadach:2000ir,Jadach:1998jb} (CEEX)
and its implementation in the \kkmc\ program~\cite{Jadach:1999vf}.
The CEEX implementation in \kkmc\ has
included all the advances of soft photon resummation of the IFI contributions
of Refs.~\cite{Greco:1975rm,Greco:1975ke,Greco:1980mh}
relevant for narrow resonances.%
\footnote{In particular, resummation of $\ln(\Gamma_Z/M_Z)$ was included.}
The SM predictions of \kkmc\ for $\afb$ and other experimental 
observables were possible for arbitrary event selections (cuts),
because \kkmc\ is a regular Monte Carlo (MC) event generator.
Correct matching of the ${\cal O}(\alpha^1)$ IFI contributions
with other non-IFI corrections up to complete ${\cal O}(\alpha^2)$ QED
and  ${\cal O}(\alpha^1)$ electroweak
was implemented throughout the entire multiphoton phase space, 
including any number of soft and  hard photons.  

The CEEX/\kkmc\ calculation was instrumental in the analysis
of LEP2 data above the $Z$ peak and near the $WW$ threshold,
and helped to consolidate data analysis of $e^-e^+\to f\overline{f}$ processes
near the $Z$ peak.
The precision of the IFI calculations quoted at the end of the LEP era
was $\delta \afb \simeq 0.1\%$ at the $Z$ peak
and $\delta \afb \simeq 0.3\%$ far away from the $Z$ resonance;
see Refs.~\cite{Jadach:2000ir,Kobel:2000aw,Bardin:1999gt,Bardin:1999yd,Jadach:1999pp}.
These papers represent the state of the art in the perturbative QED 
calculation of the IFI contributions to $\afb$ until the present day.

The \kkmc\ precision tag on the IFI calculations, both near the $Z$ peak and 
away from it,
was more than sufficient for analyzing all LEP experimental data at the end of 
the LEP era.  However, this precision was quite clearly underestimated,
{\em i.e.}\ most likely it was far better.
However, it was difficult to better quantify the theoretical 
uncertainty of the IFI prediction of \kkmc, because there was 
no other calculation at a similar level of sophistication to compare with.
One of the main aims of this work will be to develop
a new alternative numerical calculation of the IFI contribution,
in order to compare with \kkmc\ and quantify
theoretical uncertainties of the IFI component in $\afb$
at the precision level higher than presently available.

Generally, one may be quite skeptical whether an improvement 
of the QED calculation of IFI
in $\afb$ from the LEP-era $\delta\afb \sim 10^{-3}$
down to $\delta\afb \sim 10^{-5}$, {\em i.e.}\ by a factor of 100,
is feasible at all!
However, there is an interesting precedence --
the prediction of perturbative QED for the $Z$ line shape
also progressed by a similar factor from the time before LEP started
until the end of the LEP era.
This was possible mainly due to soft photon resummation techniques.
The use of these techniques is again critical for
the present task of improving the QED calculation of IFI in $\afb$.
The aim of this paper is to check how far we can advance on the road
to the precision required for FCCee.

Let us stress, that present work is not a progress report on the development
of \kkmc, simply because \kkmc\ remains the same as in 1999.
It is, however, definitely a progress report on the understanding of the IFI
contribution, thanks to several newly developed analytic calculations 
implemented in the new computer code \kkfoam\ and a wealth of numerical results
for various kinds of matrix elements, phase space integrations, cutoffs, etc.
This work will have to be continued in the future,
including possible upgrades of the matrix element in \kkmc,
and/or development of new MC programs even more sophisticated than \kkmc.

Finally, in view of the growing interest in the higher order SM calculations
which would match the very high precision
of the FCCee experiment~\cite{Blondel:2018mad,Blondel:2019qlh},
it is important to note that the CEEX methodology of photon resummation 
and matching with fixed-order nonsoft QED and EW corrections also addresses
some important issues in the QED+EW perturbative calculations,
beyond what was typically done for the LEP data analysis, as quoted above.
The basic issue is that of the separation of QED and pure EW parts of the SM
in the perturbative expansion.
This is necessary, because QED corrections are larger, and 
their soft part has to be resummed to infinite order while the
nonsoft part must be included up to \order{\alpha^4},
while the perturbative series for more complicated EW corrections can be 
truncated earlier,
at \order{\alpha^2} or \order{\alpha^3} \cite{Blondel:2018mad}.

In the calculations for LEP data analysis (see~\cite{Bardin:1999gt} 
and other Refs.\ quoted above),
the issue of separating QED and EW parts was not a critical one,
because resummed higher-order QED was typically confined to the initial-state radiation (ISR) effective 
radiator function, and in the remaining \order{\alpha^1} calculations, the
 QED and EW parts enter additively, and thus are well separated 
(except of negligible IFI which was controlled up to \order{\alpha^1}.
Beyond \order{\alpha^1}, the QED and EW parts often enter multiplicatively,
for instance in 2-loop graphs with one loop involving photon exchange 
and another loop with massive bosons or fermions,
or one loop of pure EW origin with a hard photon emission insertion.
The CEEX technique provides for clear methodology of separating/factorizing
QED and EW parts at any order.

Omitting algebra which can be found in Refs.~\cite{Jadach:2000ir,Jadach:1998jb},
the main points of CEEX methodology can be summarized as follows:
\begin{itemize}
\item[(i)] In the first step of the {\em factorization} of the infrared (IR) factors
at the amplitude level,
for any group of multiloop graphs with one photon insertion,
the IR part is subtracted at the amplitude level using a well-known (1-loop) 
function defined in the classic Yennie-Frautschi-Suura 
work~\cite{Yennie:1961ad} times a finite contribution one order lower without 
a photon insertion.  (A similar procedure applies for multiloop corrections 
with two and more photon insertions.)
The remaining finite non-IR remnant will be used in the next step.
Similarly, for any group of real photon insertions into 
a given multiloop diagram with pure EW content, one subtracts, at the 
amplitude level, a well-known eikonal factor times a basic diagram with 
pure EW content.

A similar well-defined procedure applies for amplitudes with more 
real and virtual photon insertions.
The first step is finalized by constructing spin amplitudes for an arbitrary 
number of real photons distributed over the entire phase space in which 
non-IR remnants after IR subtractions are reinserted in a well-defined way, 
while IR virtual factors are exponentiated and explicit IR-divergent eikonal 
factors are ready for MC integration in the next step%
\footnote{Collinear contributions giving rise to nonsoft mass logarithms
are included order-by-order in the present version of CEEX.}.

\item[(ii)] The second step of {\em resummation}, that of squaring spin 
amplitudes, spin summation and phase space integration, is done numerically
in the Monte Carlo event generator. (There is no possibility of doing it 
analytically.) In the above CEEX scheme, the bulk of the IFI contribution is 
present in the resummed/exponentiated real+virtual form-factor 
and in the interferences emerging from squaring multiphoton spin amplitudes. 
Smaller contributions will remain hidden in the nonsoft finite remnants%
\footnote{For instance, spin amplitudes of the $\gamma-Z$ box are split into an
IR-divergent part, which is moved to an exponentiated form-factor, and 
the remaining IR-finite remnants are incorporated in the multiphoton spin 
amplitudes.}
The treatment of the nonfactorizable $\gamma\gamma$ and 
$\gamma Z$ \order{\alpha^1}
boxes in the above resummation scheme can be seen explicitly 
in Eqs.~\ref{eq:kkfoam0}, \ref{eq:boxing} for the matrix element in \kkfoam\
and in Eqs.~($21-24$) of Ref.~\cite{Jadach:1999vf} for the CEEX matrix element 
in \kkmc.
\end{itemize} 

A somewhat more detailed overview of the CEEX {\em factorization and 
resummation} in QED is given in Secs. C.2.7 and C.3 of the recent 
review~\cite{Blondel:2018mad}, while complete details can be found in 
Refs.~\cite{Jadach:2000ir,Jadach:1998jb,Jadach:1999vf}.
The above CEEX factorization and resummation of the universal QED corrections
will work equally well for extensions of the SM (BSM), 
provided the BSM predictions are formulated at the amplitude level.
So far, \kkmc\ for the $e^-e^+\to f\bar{f} +n\gamma$ process
is the only implementation of the CEEX scheme.
In Ref.~\cite{Jadach:2018jjo}, it is argued that 
in the context of the FCCee project, it is urgent to implement it
also for the Bhabha process.

The so-called deconvolution of QED effects from LEP experimental data which was
instrumental in the construction of 
pseudo-observables~\cite{Bardin:1999gt,ALEPH:2005ab}
can also be reorganized using CEEX technique.
In the LEP version, it was done using {\tt ZFITTER}, {\tt TOPAZ0} and 
{\tt KORALZ}, and it was proven to be acceptable within the LEP precision 
goals~\cite{Bardin:1999gt}, 
but the validity of this procedure is not automatically granted for 
FCCee precision.  In Sec. C.3 in Ref.~\cite{Blondel:2018mad}, 
a proposal is made for extending it to higher precision 
by exploiting the CEEX factorization and resummation scheme in the MC 
implementation.
An updated discussion on the same theme is also included in Sec.~5.7
of a more recent paper, Ref.~\cite{Jadach:2019bye}.
Validation of a more powerful scheme
of removing QED effects from experimental data at the precision 
level of the FCCee experiments will require a lot of numerical studies
of the type done in Ref.~\cite{Bardin:1999gt},
and most likely the development of the MC programs even more powerful
and versatile than \kkmc. 

The plan of the paper is the following:
Section~\ref{sec:physIFI} explains the origin and character of the IFI effect
in the angular distribution of the $e^-e^+\to \mu^- \mu^+$ process.

Section~\ref{sec:kkfoam} describes a new partly analytic, partly numerical 
calculation in the semisoft approximation and its software implementation, 
\kkfoam.  ``Semisoft'' means that the upper limit on the total photon energy 
is smaller than the total center-of-mass energy $s^{1/2}$, 
but in the presence of narrow resonance $Z$
it can be smaller or bigger than $\Gamma_Z/M_Z$.

Multiphoton spin amplitudes (spin amplitudes) are defined 
(Sec.~\ref{subsec:softME}),
in such a way that they reproduce the CEEX matrix element in the semisoft 
regime.  Squaring and summing spins is also done analytically, 
and finally, a phase space integration over photon angles and energies
is also performed analytically keeping total real photon energy fixed
(Secs.~\ref{subsec:IEX} and \ref{subsec:IEX2}).
Multiple sums over photons and phase space integrations are done 
(exactly) in a straightforward way, with a minimal use of Mellin-Fourier 
transforms%
\footnote{Mellin transforms are used merely
  as generating functionals for reorganizing combinatorics of the multiple
  sums over photons.}.

The IFI effect appears in the resulting muon angular distribution.
The final analytic result involves a fourfold convolution over radiator 
functions of the initial state radiation (ISR), final state radiation (FSR) 
and two functions due to initial-final state interference (IFI) 
(Sec.~\ref{subsec:IEX2}).
The remaining integration over the phase space is delegated to a numerical 
Monte Carlo method.  

Validity of the formula is formally extended to the 
full phase space, such that the ISR and FSR radiator functions can be upgraded 
with the known nonsoft QED contributions up to \order{\alpha^2} 
(Sec.~\ref{subsec:matching}).
Electroweak \order{\alpha^1} corrections are also included at the same
level as in \kkmc, that is using the \dizet\ library~\cite{Bardin:1989tq}.
The remaining 5-dimension integration over the fourfold convolution and muon
angle is slightly reorganized for numerical integration 
(Sec.~\ref{subsec:NumIntegr})
using the universal \foam\ MC tool~\cite{Jadach:1999sf,Jadach:2002kn} 
for integration and simulation.
The resulting MC generator \kkfoam\ is ready for use in
 Sec.~\ref{sec:Numeric}.

With all the above distributions and technicalities in place, 
a wealth of numerical results produced by \kkmc\ and \kkfoam\ is presented 
in Sec.~\ref{sec:Numeric}.
In particular, in Sec.~\ref{subsec:baseline} 
it will be checked in the calibration exercise
that for the matrix elements with resummation and without IFI, the
three programs \kkmc, \kkfoam and \kksem\ agree within $\sim 10^{-5}$ precision
for $\sigma_{\rm tot}$ and $\afb$.
Note that in the case where IFI is switched off, this kind of comparison
of \kkmc\ with the numerical tool {\tt KKsem} based on analytic
exponentiation was already done in Ref.~\cite{Jadach:2000ir}.
The new thing here is the inclusion of the IFI.

In Sec.~\ref{subsec:IFItoAFB}, the IFI effect in $\afb$ will be examined 
for three energies $\sqrt{s}=10,87.9,94.3GeV$ 
as a function of the cutoff on total photon energy,
comparing results of \kkmc\ and \kkfoam.
In Sec.~\ref{subsec:AFBdif}, we shall focus on the difference of $\afb$
between $\sqrt{s_+}=94.3GeV$ and $\sqrt{s_-}=87.9GeV$, which is directly
related to the measurement of $\alpha_{QED}(M_Z)$.
Secs.~\ref{subsec:AFBord} and \ref{subsec:ISRinAFB} will be devoted
to estimating higher-order QED uncertainties by means of comparing
results for $\afb$ for several variants of the QED matrix element in \kkmc.

Section~\ref{sec:Summary} summarizes the results,
focusing on the uncertainties in the QED calculation of the IFI effect 
in $\afb$.

Three appendices include details of the analytic phase space integration
in the semisoft approximation (Appendix \ref{appendixA}),
the kinematical mapping used in \kkfoam\ for the four-fold convolution 
(Appendix \ref{AppendixFoam}) and a recollection of some old 
\order{\alpha^1} analytic formulas without exponentiation
(Appendix \ref{sec:appendixC}) to be used in the numerical studies in 
Sec.~\ref{sec:Numeric}.


\section{Physics of IFI} 
\label{sec:physIFI}

\begin{figure}[!t]
  \centering
  \includegraphics[height=45mm]{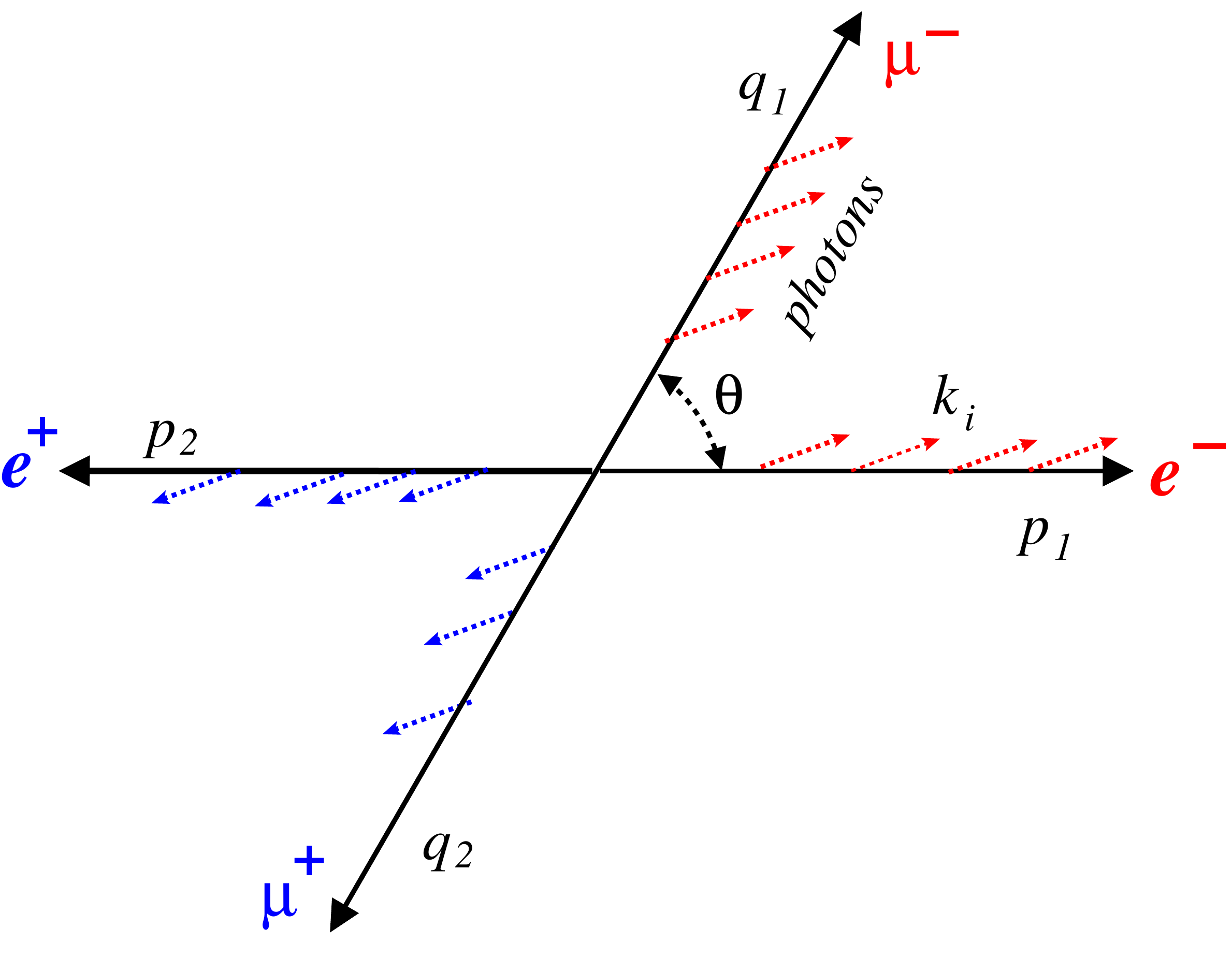}~~~ 
  \includegraphics[height=50mm]{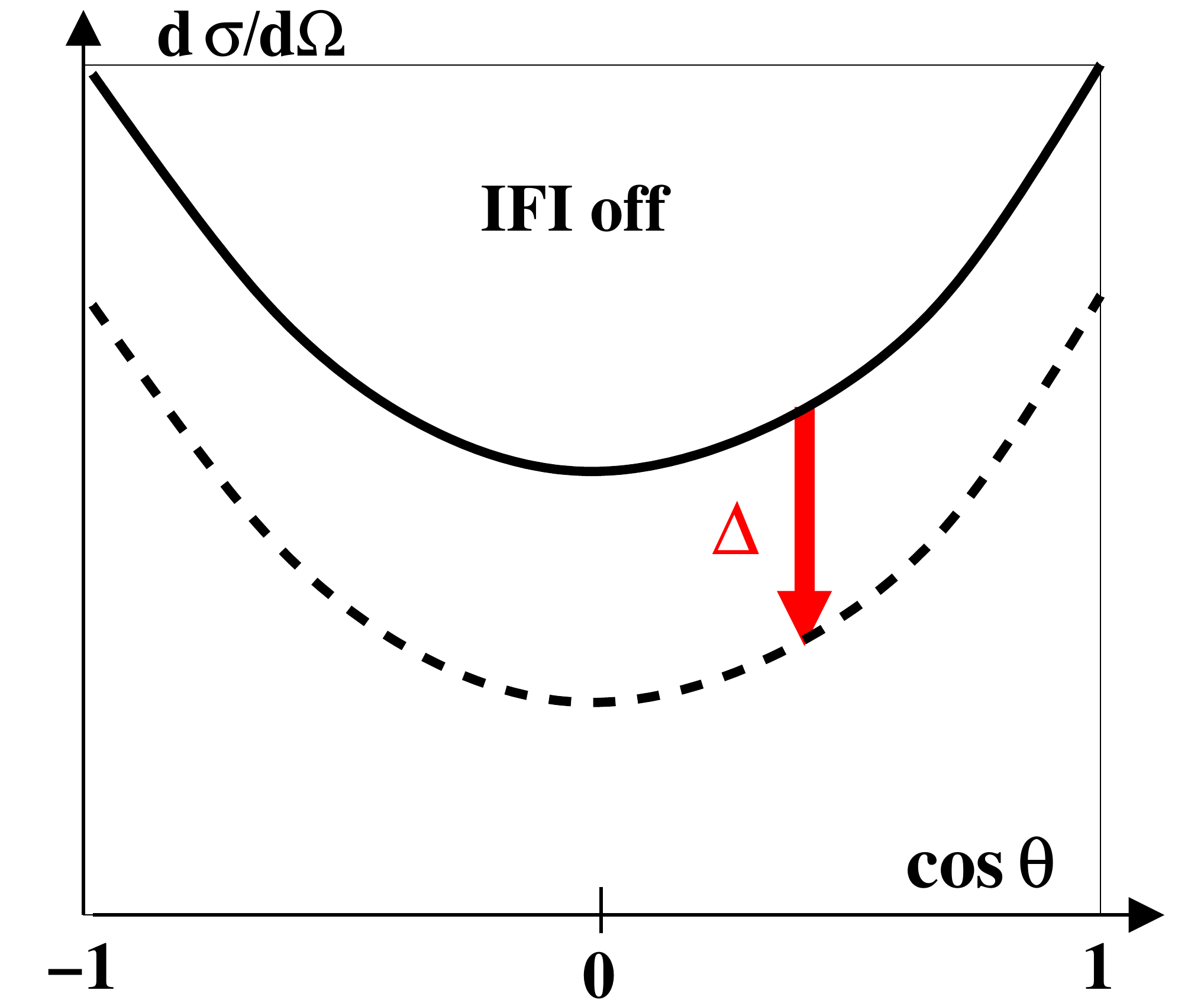}  
  \caption{\sf
  Multiple photon emission at a wide scattering angle.
  }
  \label{fig:thLarge}
\end{figure}

Any efficient evaluation of IFI in perturbative QED must
be based on a good understanding of the basic physics governing this phenomenon.
Let us consider the process $e^-e^+\to \mu^- \mu^+$ accompanied with any number
of real and virtual photons, illustrated schematically in
Fig.~\ref{fig:thLarge}.
In the case of final fermions emitted at wide angles, IFI can be neglected, 
and in the case of total photon energy (ISR+FSR) limited to $K< E=\sqrt{s}/2$,
the angular distribution is uniformly lowered by a $\theta$-independent
Sudakov form factor%
\footnote{The subscript ``virt'' appears because virtual corrections
 feature $ -\int_0^E $, while real emissions add $ +\int_0^K $,
 so that the uncompensated remnant $ -\int_K^E $ is of virtual origin.
}
\begin{equation}
\frac{d\sigma}{d\cos\theta}(K) \simeq
\frac{d\sigma_{\rm Born}}{d\cos\theta}
\exp
\Big[
  -\int_{K}^E \frac{dk^0}{k^0}
           \Big( 2\frac{\alpha}{\pi}\ln\frac{s}{m_e^2}
                +2\frac{\alpha}{\pi}\ln\frac{s}{m_\mu^2}
            \Big)_{\rm virt}
\Big]
=\frac{d\sigma_{\rm Born}}{d\cos\theta}\;
   e^{-\Delta(K/E)}.
\end{equation}
The above relation is illustrated schematically in Fig.~\ref{fig:thLarge}.

\begin{figure}[!t]
  \centering
  \includegraphics[height=45mm]{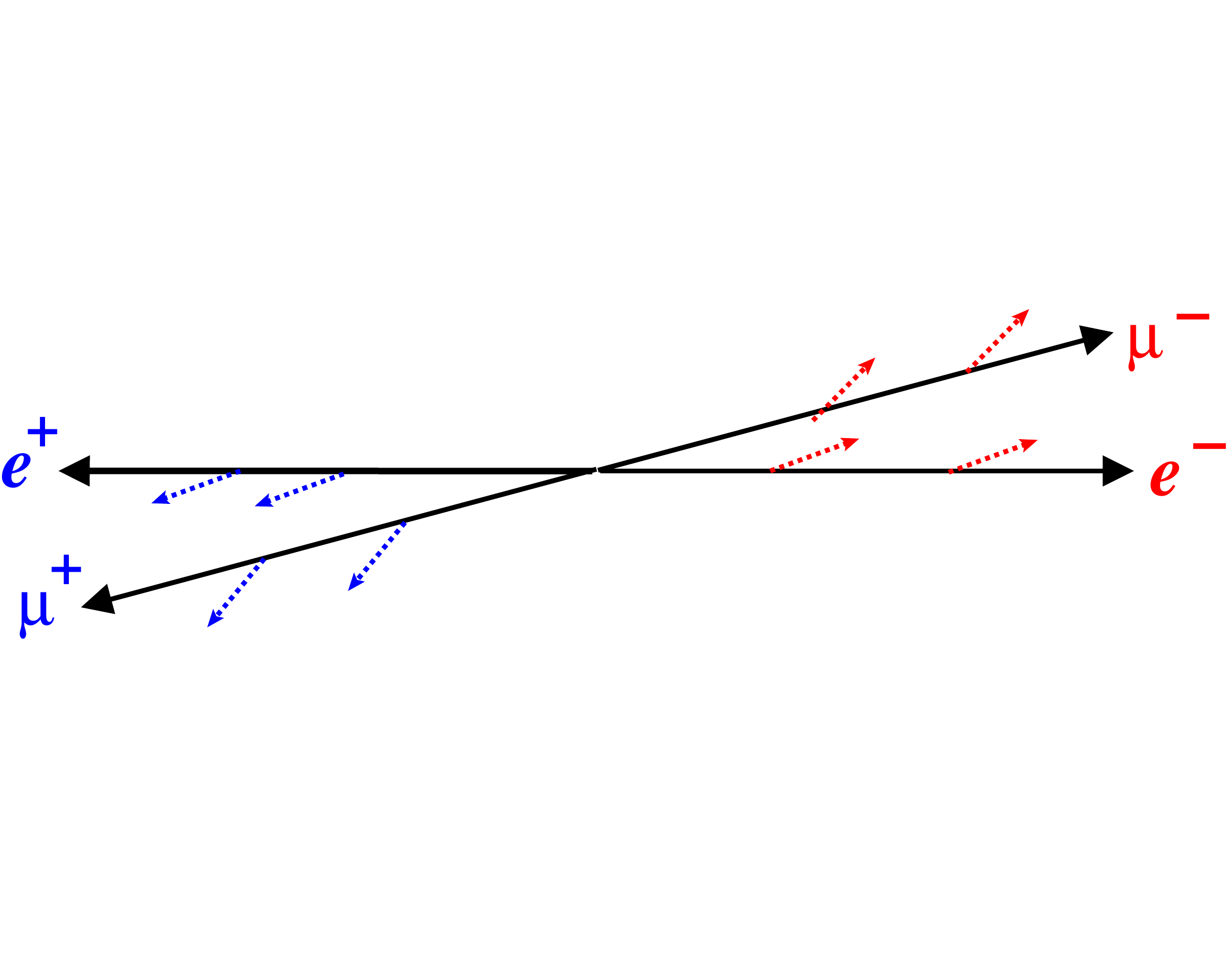}~~~ 
  \includegraphics[height=50mm]{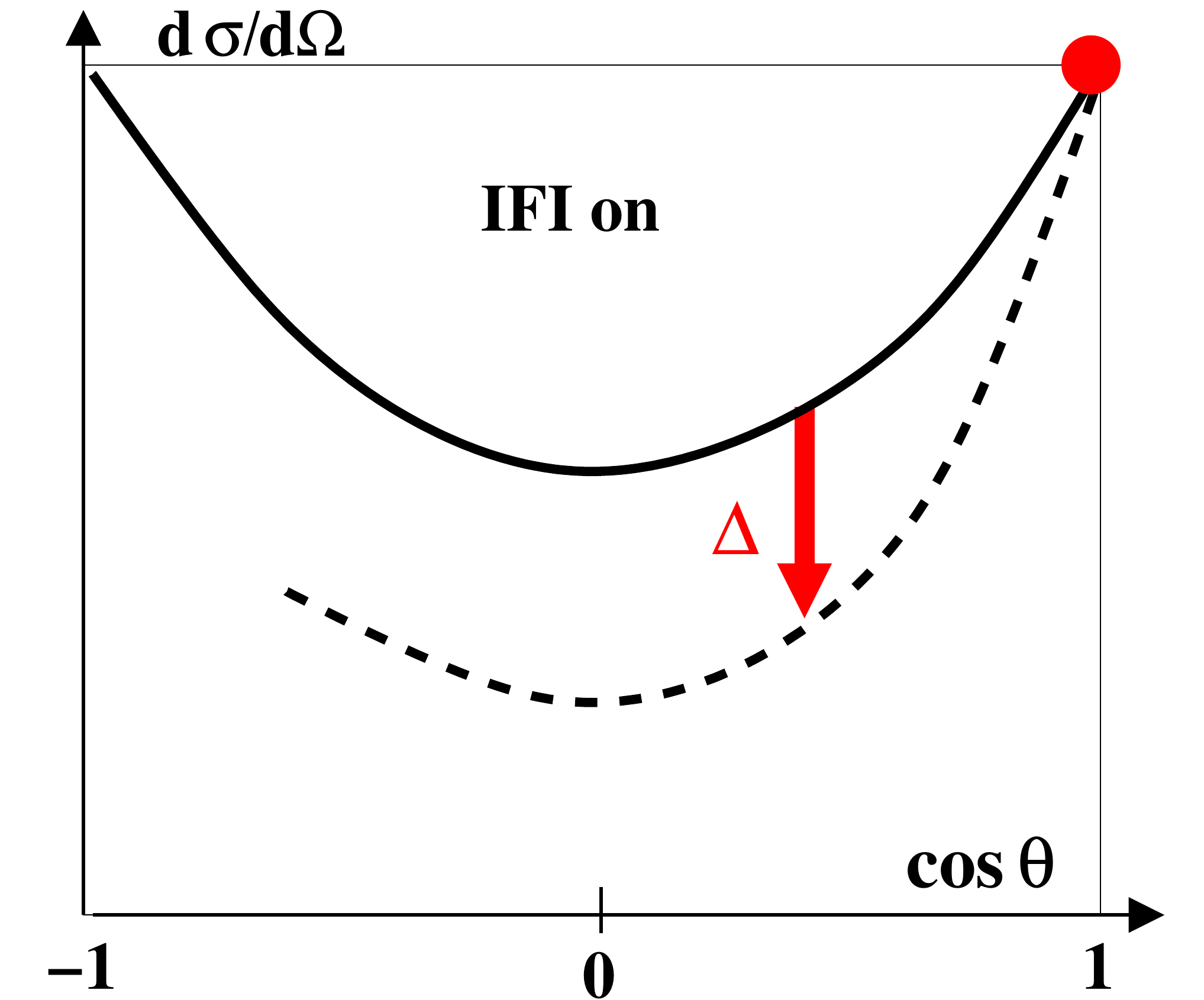}   
  \caption{\sf
  Multiple photon emission at a forward scattering angle.
  }
  \label{fig:thSmall}
\end{figure}

Photon emission is, however, suppressed in the small-angle limit $\theta \to 0$,
as illustrated schematically in Fig.~\ref{fig:thSmall}, simply because 
the outgoing muon inherits most of the electromagnetic field accompanying the
incoming electron of the same charge as the muon; hence there is no need for
the compensating action of bremsstrahlung.
In fact, bremsstrahlung dies out completely at $\theta=0$,
and it is the IFI contribution which kills both ISR and FSR.
The virtual form factor in the angular distribution 
at $t \to 0$, $s-|t|-|u|=0$, $|u|\to s$ becomes 
\begin{equation}
\Delta=
 \int_{K}^E \frac{dk^0}{k^0}
    \Big( 2\frac{\alpha}{\pi}\ln\frac{s}{m_e^2}
         +2\frac{\alpha}{\pi}\ln\frac{s}{m_\mu^2}
         -4\frac{\alpha}{\pi}\ln\frac{|t|}{|u|}
     \Big)
\to
 \int_{K}^E \frac{dk^0}{k^0}
    \Big( 2\frac{\alpha}{\pi}\ln\frac{|t|}{m_e^2}
         +2\frac{\alpha}{\pi}\ln\frac{|t|}{m_\mu^2}
     \Big)
\simeq 0.
\end{equation}

\begin{figure}[!t]
  \centering
  \includegraphics[height=45mm]{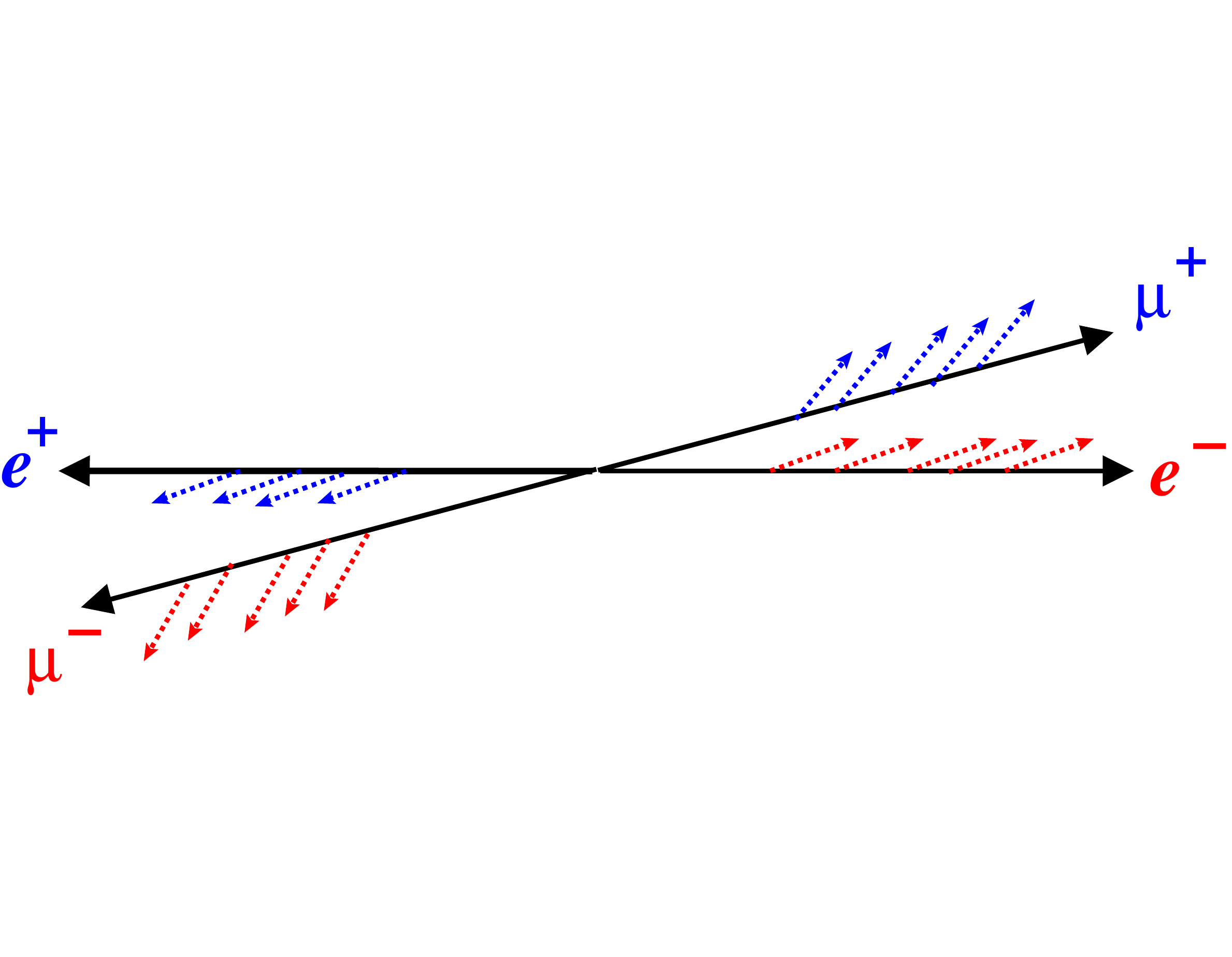}~~~ 
  \includegraphics[height=50mm]{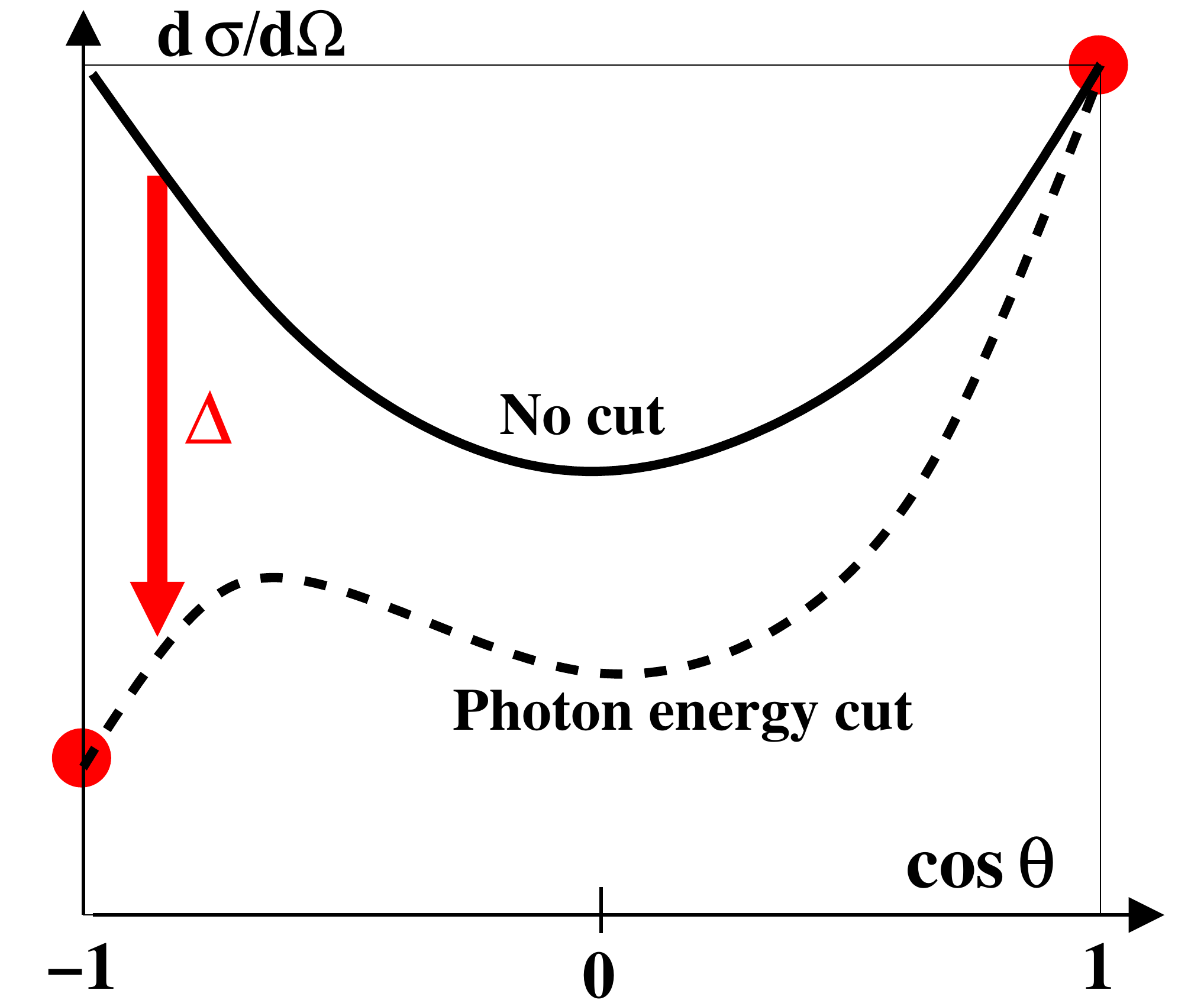} 
  \caption{\sf
  Multiple photon emission at a backward scattering angle.
  }
  \label{fig:thBackward}
\end{figure}

On the other hand, in backward scattering,
illustrated schematically in Fig.~\ref{fig:thBackward}, the situation 
is completely different.
The electromagnetic field accompanying $e^-$ has to be replaced by
that of $\mu^+$, hence the violent compensating action of the bremsstrahlung
is much stronger than for wide angles.
Here we have $u \to 0$ ($c\to -1$ side),$s-|t|-|u|=0$, $|t|\to s$. 
Thus, IFI enhances the total QED correction by a factor of 2:
\begin{equation}
\Delta=
 \int_{K}^E \frac{dk^0}{k^0}
    \Big( 2\frac{\alpha}{\pi}\ln\frac{s}{m_e^2}
         +2\frac{\alpha}{\pi}\ln\frac{s}{m_\mu^2}
         -4\frac{\alpha}{\pi}\ln\frac{|t|}{|u|}
     \Big)
\to
 \int_{K}^E \frac{dk^0}{k^0}
    \Big( 4\frac{\alpha}{\pi}\ln\frac{s}{m_e^2}
         +4\frac{\alpha}{\pi}\ln\frac{s}{m_\mu^2}
     \Big),
\end{equation}
creating a dip in the muon angular distribution for backward scattering
(in the presence of a cutoff on the total photon energy, as previously).

\begin{figure}[!t]
  \centering
  \includegraphics[height=65mm, width=90mm]{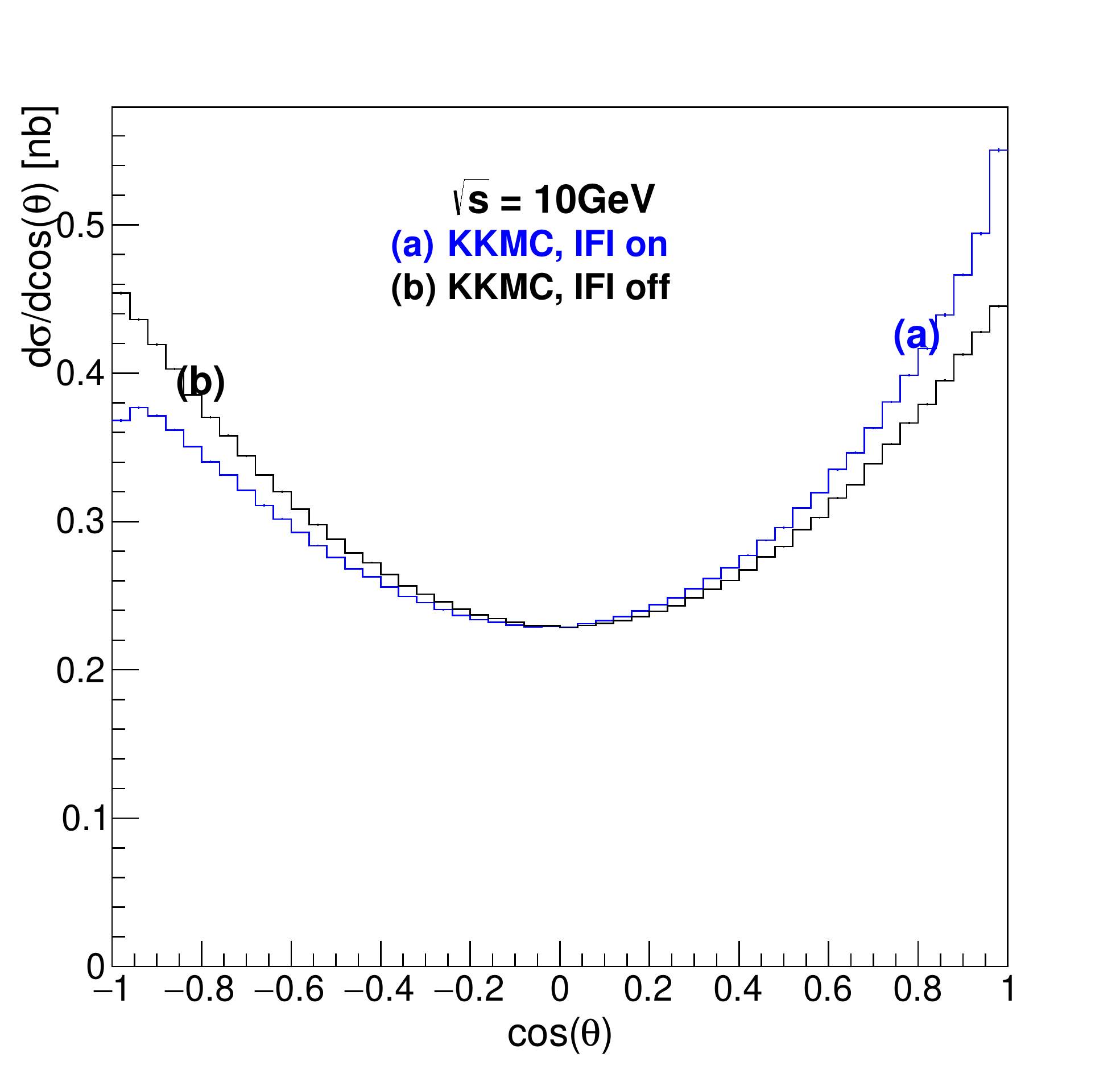}
  \caption{\sf
  Muon angular distribution for IFI switched on and off
  for total photon energy below 2\% of  $E_{beam}=\sqrt{s}/2=5$ GeV.
  The distribution is obtained from \kkmc.
  }
  \label{fig:cthKKMC}
\end{figure}
In reality, the distribution of $\cos\theta$ far from the resonance
appears as shown in Fig.~\ref{fig:cthKKMC}
for a relatively strong cutoff on total photon energy (2\% of the beam energy).

The presence of a narrow resonance significantly changes the pattern of 
QED cancellations.
Let us analyze briefly how the real and virtual corrections combine
at a resonance position $\sqrt{s}=M_Z$. 
\begin{itemize}                         
\item
For pure ISR, the virtual correction is
$\sim -\frac{2\alpha}{\pi}\ln\frac{s}{m_e^2} \ln\frac{E}{\lambda}$,
as without a resonance, while the real contribution
is cut by the resonance profile
$\sim +\frac{2\alpha}{\pi}\ln\frac{s}{m_e^2} \ln\frac{\Gamma_Z}{\lambda}$.
The resulting cross section $\sigma(K)$ is suppressed by
the remnant virtual factor
$\big[1-\frac{2\alpha}{\pi} \ln\frac{M_Z}{\Gamma_Z}\big]_{\rm virt}$ 
for any cut above the resonance width, $K>\Gamma_Z$.
\item
The effect of FSR is the same as in the case without a resonance, {\em i.e.}
\ $\sigma(K)$ is suppressed by the remnant virtual factor
$[1-\frac{2\alpha}{\pi}\ln\frac{s}{m_\mu^2} \ln\frac{E}{K}]_{\rm virt}$.
\item
The case of IFI is most complicated. The virtual correction
$\sim -\frac{4\alpha}{\pi}\ln\frac{t}{u} \ln\frac{\Gamma_Z}{\lambda}$ 
is cut by the resonance (contrary to the ISR case).
The real correction 
$\sim +\frac{4\alpha}{\pi}\ln\frac{t}{u} \ln\frac{\Gamma_Z}{\lambda}$ 
is also cut by the resonance (similar to the ISR case).
The resulting $d\sigma(K)/d\Omega$ is strongly power-suppressed 
by the $\Gamma_Z/M_Z$ factor for any cut above the resonance width, 
$K>\Gamma_Z$!  For an energy cut below the resonance width,  $K<\Gamma_Z$,
IFI starts to rise logarithmically,
{\em i.e.}\ the suppression factor is
$\sim  1-\frac{2\alpha}{\pi}\ln\frac{t}{u} \ln\frac{\Gamma_Z}{K}$.
\end{itemize}
Away from the resonance, IFI gradually changes
to the previous nonresonant case,
and in the entire neighborhood of the resonance,
a QED calculation including photon resummation
at the amplitude level (CEEX) is mandatory.

\begin{figure}[!t]
  \centering
  \includegraphics[height=75mm, width=90mm]{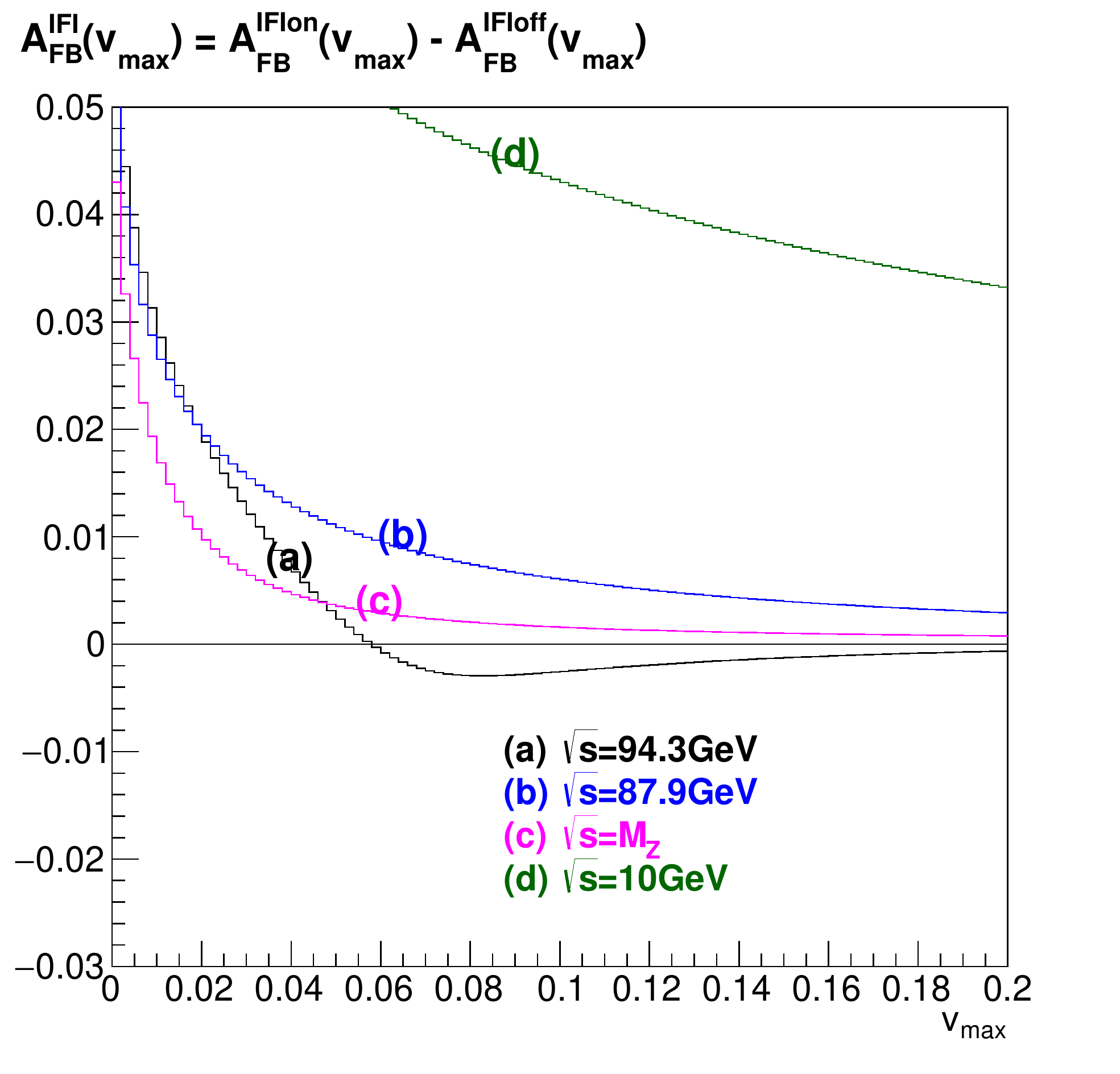}
  \caption{\sf
  The dependence of IFI contributions to charge asymmetry on
  total photon energy cutoff $v_{\max}$ for various beam energies.
  The distribution is obtained from \kkmc.
  }
  \label{fig:FigDifIFIb}
\end{figure}
The above mechanism is clearly illustrated in Fig.~\ref{fig:FigDifIFIb},
where the IFI contribution to $\afb$ is shown as a function of
$v_{\max}$, the cutoff on the total photon energy in units of the beam energy.%
\footnote{More precisely, $v=1-M^2_{\mu\mu}/s$. 
Here, we have temporarily used the ALEPH definition of $v$.}
As we see, far from the $Z$ resonance and for a loose
photon energy cutoff, $\afb \simeq 2\%$ and grows for stronger cutoffs.
In the middle of the resonance, it is strongly suppressed, $\afb < 0.1\%$,
and starts to grow below $v_{\max} \simeq \Gamma_Z/M_Z \simeq 0.02$.
Remarkably, at the other two energies $\sqrt{s} \simeq M_Z \pm 3.5$ GeV,
$\Gamma_Z/M_Z$ suppression is still quite strong, more than factor 1/5.

On the methodology side, although we are interested mainly in
the IFI effect off the $Z$ peak, at $\sqrt{s} \simeq M_Z\pm 3.5$ GeV,
it is worth also keeping an eye on $\sqrt{s}=M_Z$ and energies far away 
from the resonance.  This is simply because any technical problem or 
mistreatment of physics which may cause a small effect at 
$\sqrt{s} \simeq M_Z\pm 3.5$ GeV could be magnified there, making it
easier to trace it back and eliminate.
This is why we shall often compare our principal results 
with the results at  $\sqrt{s}=M_Z$ and  $\sqrt{s}=10$ GeV.

\section{New calculation in the semisoft approximation and \kkfoam\ Monte Carlo}
\label{sec:kkfoam}

As outlined in the Introduction, the aim of this section is to
describe in a detail all ingredients in the newly developed 
\kkfoam\ Monte Carlo which will be served in the next sections for 
validation of the \kkmc\ predictions for $\afb(s)$ at a precision at least a 
factor of 10 better than in the past.
\kkfoam\ is not a true event generator because photon momenta 
are partly integrated out analytically. 
Nevertheless, it provides weighted MC events with explicit muon four-momenta;
hence angular distributions of muons with an arbitrary cutoff on the total
photon energy can be calculated.

In Sec.~\ref{subsec:softME}, the multiphoton 
matrix element (spin amplitudes) will be defined.
In Secs.~\ref{subsec:IEX} and \ref{subsec:IEX2}
the above matrix element will be squared, and spin summation and
and phase space integration will be done partly analytically.
The IFI effect appears in the resulting muon angular distribution.
The resulting formula involves a fourfold convolution over radiator functions
of the initial state radiation (ISR), final state radiation (FSR) and
two functions due to initial-final state interference (IFI).
(Further analytic integration is not possible.)

In addition, in Sec.~\ref{subsec:matching}, the phase space integration 
is extended to the full phase space, and matching with the 
known \order{\alpha^1} and \order{\alpha^1} results for the ISR and FSR 
radiator functions is performed.
The radiator functions are convoluted with the 
effective Born spin amplitudes in which EW corrections are included.

In Sec.~\ref{subsec:NumIntegr}, it is explained how the remaining 
5-dimensional integration over the four radiator functions (ISR, FSR and 
2$\times$IFI) and the azimuthal angle $\theta$ of the muon is performed 
numerically using a Monte Carlo method.
It is not trivial due to the presence of new types of singularities 
in the IFI radiator functions, different from 
the standard ones of the ISR and FSR radiator functions.
The newly developed computer program \kkfoam\ is new software tool,
methodologically completely independent of \kkmc,
although it exploits some building blocks of \kkmc,
for instance the \dizet\ library of the EW corrections~\cite{Bardin:1989tq}.

\subsection{Matrix element of multi-soft-photon emission 
 in the semisoft approximation}
\label{subsec:softME}

Let us consider the matrix element of the process
\begin{equation}
 e^-(p_1) +e^+(p_2)\to \mu^-(q_1) +\mu^+(q_2) +\gamma(k_1)+\dots+\gamma(k_n)
\end{equation}
near the $Z$ resonance in the soft photon limit.
The standard kinematic variables
$ s=(p_1+p_2)^2,\quad t=(q_1-p_1)^2,\quad u=(q_2-p_1)^2 $
will be used.
Around any narrow resonance, the notion of the soft photon limit 
has to be refined.  In the framework of the
standard {\em Yennie-Frautschi-Suura} (YFS)~\cite{Yennie:1961ad}
soft photon resummation, one starts with all photons being very soft, 
{\em i.e.}\ $k_i^0 \ll \Gamma_Z \ll \sqrt{s}/2 $.
Near the resonance, however, it is worth considering a wider soft photon range,
with $k_i^0 \ll \sqrt{s}/2$, but allowing photon energies comparable to
or even greater than the resonance width $\Gamma_Z$.
In the following, we shall refer to this regime as the 
{\em semisoft approximation}.
Following the notation of Ref.~\cite{Jadach:2000ir}, in the semisoft regime,
the matrix element of our process reads as follows:
\begin{equation}
\begin{split}
&
\Meu^{\mu_1,\mu_2,...,\mu_n}(p_i,q_j,k_l)=
 \sum_{V=\gamma,Z}
 \sum\limits_{\Peu}\;
 e^{\alpha B_4^V(s_I,t,m_\gamma)}
 \prod_{i\in I} j^{\mu_i}_I(k_i)\;
 \prod_{r\in F} j^{\mu_r}_F(k_r)\;
 \Mcal_V\big(s_I,t ),
\\&
 s_I= P_I^2,\quad P_I=p_1+p_2-\sum_{i\in I} k_j,\quad
\\&
 j_I^\mu(k)=eQ_e \bigg( \frac{p_1^\mu}{kp_1}-\frac{p_2^\mu}{kp_2}  \bigg), \quad
 j_F^\mu(k)=eQ_f \bigg( \frac{q_1^\mu}{kq_1}-\frac{q_2^\mu}{kq_2}  \bigg),
\\&
\alpha B_4^V(s,t,m_\gamma)= 
\alpha B_4(s,t,m_\gamma)+ \alpha \Delta B_4^V(s,t,\overline{M}_V^2).
\label{eq:Mpartit}
\end{split}
\end{equation}
The above formula involves a sum over the set of $2^n$ partitions 
$\{ \Peu \} = \{ I,F\}^n $,
\begin{equation}
 \{ \Peu \} =\{
 (I,I,I,...,I),
 (F,I,I,...,I),
 (I,F,I,...,I),
 (F,F,I,...,I), ...,
 (F,F,I,...,F)\},
\end{equation}
of photons among the initial and final state.
The meaning of the shorthand notation
$i\in I$ is that $\prod_{i\in I}$ includes all photons
with $\Peu_i=I$ and similarly $\Peu_r=F$ for $r\in F$.

The form factor  $B_4(p_i,q_i,m_\gamma)$ is the standard one appearing
in YFS resummation~\cite{Yennie:1961ad} for four charged particles
in the scattering process.
As stressed in Refs.~\cite{Greco:1975rm, Greco:1975ke,Greco:1980mh},
in the semisoft regime, an additional term in the form factor
\begin{equation}
\alpha \Delta B_4^Z(s,t,\overline{M}^2)=
-2Q_e Q_f \frac{\alpha}{\pi} \ln\Big( \frac{t}{u} \Big)
  \ln\Big( \frac{\overline{M}_Z^2-s}{ \overline{M}_Z^2 } \Big),\quad
\overline{M}^2= M_Z^2-iM_Z\Gamma_Z,\quad
\Delta B_4^\gamma \equiv 0,
\end{equation}
must be included, but only in the resonant component of the amplitude.
For $\gamma$ exchange, only the standard $\alpha B_4$ of the YFS scheme
is needed, and $\alpha B_4^V$ is not present.
Most important is that, in the semisoft approximation,
the energy argument of the resonance propagator in the
Born matrix element $\Mcal_V$ must be shifted
by the total energy lost to initial state photons $j\in I$,%
\footnote{In the strict YFS soft limit this energy shift may be neglected.
  In the semisoft regime it could also be neglected
  for the $\gamma$-exchange part.
  For the sake of a better treatment of the collinear (mass) singularities,
  it is best to keep it everywhere.}
because of its strong energy dependence.
The same additional dependence on $s_I$ 
also enters into the form factor $\alpha B_4^V$.
The summation over all partitions of $n$ photons between 
the initial and final state $\{I,F\}$ is mandatory
in order to obey Bose-Einstein symmetry and gauge invariance.
Fermion spinor indices are implicit in $\Mcal_V$.
The standard YFS virtual form factor $B_4$
is usually regularized with a photon mass $m_\gamma$. The mass regulator can
be removed once the real and virtual calculations are combined.

In the framework of coherent exclusive
exponentiation (CEEX) \cite{Jadach:2000ir,Jadach:1999vf}, the above 
matrix element 
represents a {\em zeroth-order} CEEX matrix element defined throughout 
the entire phase space, including hard photons. 
Higher orders are also defined in the CEEX scheme, and implemented
for a finite number of the hard photons; see Ref.~\cite{Jadach:2000ir}.

The same matrix element can be rewritten in a compact form using a generating
functional formulation (Mellin-Fourier transform):
\begin{equation}
\begin{split}
&\Meu^{\mu_1,\mu_2,...,\mu_n}(p_i,q_j,k_1,...,k_n)=
\\&~~~
=\sum_{V=\gamma,Z}
\int \frac{ d^4 Q d^4 x}{(2\pi)^4}
e^{ix \cdot (P-Q)}\;
e^{\alpha B_4^V(Q^2,t,m_\gamma)}
 \Big[ \prod_{i=1}^n J^{\mu_i}(x,k_i) \big]
 \Mcal_V\big(Q^2,t )
\\&
J^\mu(x,k) =  e^{-ik \cdot x}  j^\mu_I(k) + j^\mu_F(k).
\label{eq:Mfunct}
\end{split}
\end{equation}
The corresponding total cross section reads:
\begin{equation}
\begin{split}
\sigma(s)=
&\frac{1}{{\rm flux}(s)}
\sum_{n=0}^\infty \frac{1}{n!}
\int \frac{d^3 q_1}{ q_1^0} \frac{d^3 q_2}{ q_2^0}\;
\prod_{i=1}^n \int \frac{d^3 k_i}{ k_i^0}\;
\delta\big(P-q_1-q_2-\sum_{i=1}^n k_i \big)\;
\\&\times
      \Meu^{\mu_1,\mu_2,...,\mu_n}(p,q,k_1,...,k_n)
\big[ \Meu_{\mu_1,\mu_2,...,\mu_n}(p,q,k_1,...,k_n)\big]^*,
\end{split}
\end{equation}
where $P=p_1+p_2$.
Note that in the above expression, the standard Lorentz invariant phase space
integral extends over the entire phase space,
as it would in the Monte Carlo implementation,
{\em i.e.}, energy conservation naturally limits photon energies from above.
A cutoff on the total photon energy will be imposed later 
in our analytic calculations, in order to perform phase
space integration analytically.

In the semisoft approximation, the matrix element 
of Eq.~(\ref{eq:Mpartit}) is simple enough
that in the absence of experimental cuts
one can perform an analytic integration over photon angles and energies
and sum explicitly over photon multiplicities.
This is what we refer to as an {\em analytic exponentiation}.

The first important step will be reorganization
of the multiphoton distributions
keeping phase space integration untouched.
Later on phase space
integrations will be done in the semisoft regime
and finally contributions from hard photon phase space
will be reintroduced in the standard matching procedure at the level
of partly integrated distributions.

\subsection{Reorganization of multiphoton distributions}
\label{subsec:IEX}

One may perform analytic reorganization of multiphoton distributions,
a necessary first step in the {\em analytic exponentiation},
by means of a combinatorial reorganization of
the sum over photons without using the generating functional
(Mellin-Fourier transform) formulation of Eq.~(\ref{eq:Mfunct}).
This method was developed in Ref.~\cite{Jadach:2000ir},
albeit for the resonant component only.
Another alternative method of integration/summation over semisoft photons
would employ a coherent states technique.  
This method was used, for instance, in Refs.~\cite{Greco:1975rm,Greco:1975ke}.
Let us start from the generating functional form of Eq.~(\ref{eq:Mfunct}),
which was used in the original YFS paper~\cite{Yennie:1961ad},
although it was applied there for the simpler nonresonant case.
Of course, all three methods lead to an identical final result.

In the first step, let us introduce the usual eikonal factors
for photon emission from the initial state, final state, and
their interference:
\begin{equation}
 S_I(k) = -j_I(k) \cdot j_I(k),\quad
 S_F(k) = -j_F(k) \cdot j_F(k),\quad
 S_X(k) = -j_I(k) \cdot j_F(k),\quad
\end{equation}
and Fourier representations of the $\delta$-functions
of the phase space:\footnote{
The overall 4-momentum conservation $\delta^{(4)}$-function will be present implicitly
in the next steps.}
\begin{equation}
\begin{split}
&\sigma(s)=
\\&
=\frac{1}{{\rm flux}(s)}
\sum_{n=0}^\infty \frac{1}{n!}
\sum_{V,V'}
\int \frac{d^3 q_1}{ q_1^0} \frac{d^3 q_2}{ q_2^0}\;
\frac{ d^4 Q d^4 x}{(2\pi)^4}
\frac{ d^4 Q' d^4 x'}{(2\pi)^4}
e^{ix \cdot (P-Q) -ix' \cdot (P-Q')}
\frac{ d^4 y}{(2\pi)^4}
e^{ iy \cdot (P-q_1-q_2) }
\\&~\times 
\prod_{i=1}^n \int \frac{d^3 k_i}{ k_i^0}\;
\big[ e^{-ik_i \cdot (y+x-x')} S_I(k_i) 
     +e^{-ik_i \cdot (y+x)}    S_X(k_i)
     +e^{-ik_i \cdot (y-x')}   S_X(k_i)
     +e^{-ik_i \cdot y}        S_F(k_i)
 \big]
\\&~\times
\Mcal_V(Q,t) \Mcal^*_{V'}(Q',t)\;
e^{\alpha B_4^V(Q^2,t,m_\gamma) +\alpha B_4^{*V'}(Q'^2,t,m_\gamma) }\;
\end{split}
\end{equation}

In the above functional representation, the
summation over photon multiplicities (exponentiation) is trivial:%
\footnote{Both the virtual functions $B_4^V$
  and real emission integrals over $S$-factors are regularized
  temporarily using a small photon mass $m_\gamma$, which will cancel
  in the final result.}
\begin{equation}
\begin{split}
&\sigma(s)=
\frac{1}{{\rm flux}(s)}
\sum_{V,V'}
\int \frac{d^3 q_1}{ q_1^0} \frac{d^3 q_2}{ q_2^0}\;
\frac{ d^4 Q d^4 x}{(2\pi)^4}
\frac{ d^4 Q' d^4 x'}{(2\pi)^4}
e^{ix \cdot (P-Q) -ix' \cdot (P-Q')}
\frac{ d^4 y}{(2\pi)^4}
e^{ iy \cdot (P-q_1-q_2) }
\\&~\times 
\exp\Big\{ \int \frac{d^3 k}{ k^0}\;
\big[ e^{-ik \cdot (y+x-x')} S_I(k) 
     +e^{-ik \cdot (y+x)}    S_X(k)
     +e^{-ik \cdot (y-x')}   S_X(k)
     +e^{-ik \cdot y}       S_F(k)
 \big]
\Big\}
\\&~\times
\exp\big\{
  \alpha B_4^V(Q^2,s,t) +\alpha( B_4^{V'}(Q'^2,s,t))^*
\big\}\;
\Mcal_V(Q,t) \Mcal^*_{V'}(Q',t)\;
\end{split}
\label{eq:expon0}
\end{equation}
The integrations over $x,x'$ and $y$ can be reorganized in order
to achieve a clear factorization into ISR, FSR, and IFI parts,
as shown in Appendix~\ref{appendixA}.

\begin{figure}[!t]
  \centering
  \includegraphics[width=0.80\textwidth]{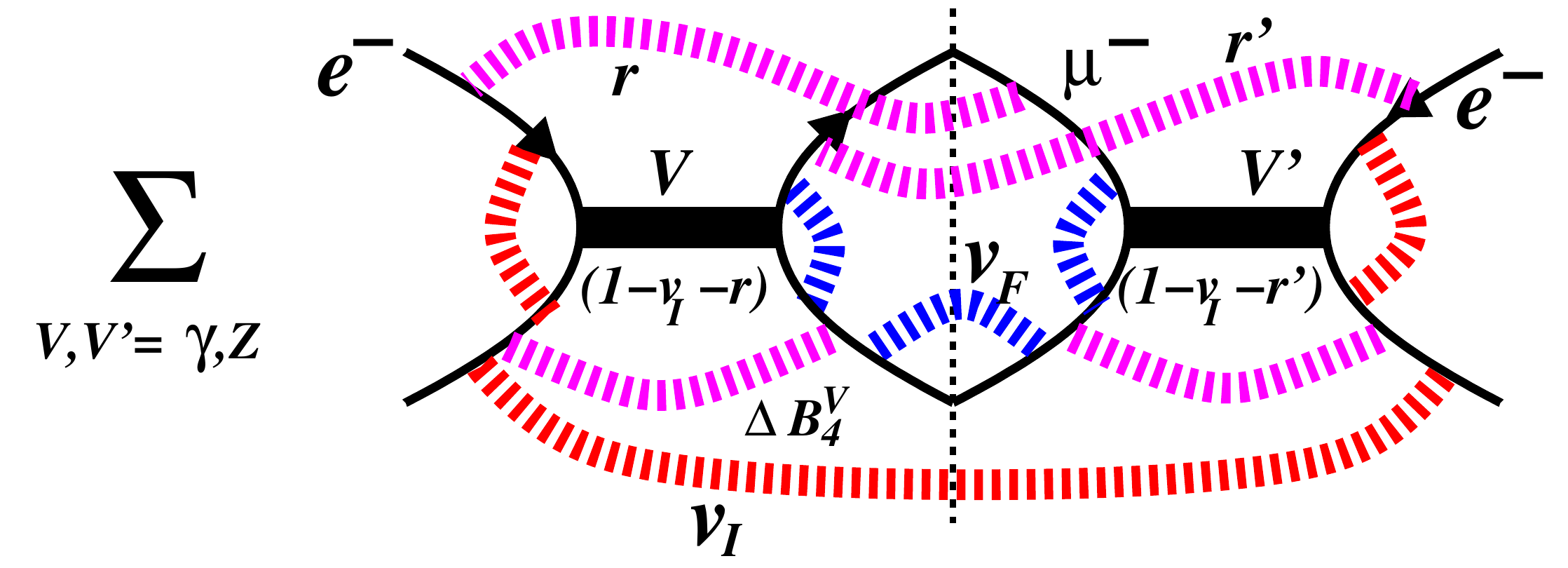}
  \caption{\sf
  Exponentiated multiple photon emission from initial and final fermions
  including ISR, FSR and IFI in the resonant process, 
  as in Eq.~(\ref{eq:factorized}).
  Dashed lines represent multiple real and/or virtual photons.
  }
  \label{fig:xKKsemIFI}
\end{figure}

A slightly reorganized form of Eq.~(\ref{eq:appA-final})
with $U^\mu$ representing the
total photon momentum of pure FSR emission,
$K^\mu$ representing the total momentum of pure ISR emission, 
and with $R^\mu$ and $R'^\mu$ aggregating IFI photons
present in $\Mcal_V$ and $\Mcal_{V'}^*$ correspondingly, 
reads as follows
\begin{equation}
\begin{split}
&\sigma(s)=
\frac{1}{{\rm flux}(s)}
\int \frac{d^3 q_1}{ 2q_1^0} \frac{d^3 q_2}{ 2q_2^0}\;
      d^4 K\; d^4 R\; d^4 R'\; d^4 U
\delta^4(P-q_1-q_2-K-R-R'-U)
\\&\times
\int \frac{d^4z }{ (2\pi)^4} \;
e^{ iz\cdot K +\int \frac{d^3 k}{ k^0}\; e^{-ik \cdot z}  S_I(k) }
\int \frac{ d^4u}{ (2\pi)^4} \;
e^{ iu\cdot R +\int \frac{d^3 k}{ k^0}\; e^{-ik \cdot u}  S_X(k) }
\\&\times
\int \frac{d^4u'}{ (2\pi)^4}\;
e^{ iu'\cdot R' +\int \frac{d^3 k}{ k^0}\; e^{-ik \cdot u'} S_X(k) }
\int \frac{d^4y}{ (2\pi)^4} \;
e^{  iy \cdot U  +\int \frac{d^3 k}{ k^0}\; e^{-ik \cdot y}  S_F(k) }
\\&\times
\sum_{V,V'=\gamma,Z} \Mcal_V(P-K-R)\; \Mcal^*_{V'}(P-K-R')
\\&\times
\exp\big\{
  2\alpha \Rmf B_4(s,t,m_\gamma) 
  +\alpha \Delta  B_4^V((P-K-R)^2) +(\alpha \Delta  B_4^{V'}((P-K-R')^2) )^*
\big\};
\end{split}
\label{eq:factorized}
\end{equation}
see also the illustration in Fig.~\ref{fig:xKKsemIFI}.

The role of the Mellin transform in the above
algebra was merely to provide compact bookkeeping of the complicated
sums in the multiphoton distributions,
without any modification of the underlying phase space integration.
At any step, we could go back to standard phase space without any cutoffs;
for instance Eq.~(\ref{eq:factorized}) can be rewritten as follows:
\begin{equation}
\begin{split}
&\sigma(s)=
\frac{1}{{\rm flux}(s)}
\int \frac{d^3 q_1}{ 2q_1^0} \frac{d^3 q_2}{ 2q_2^0}\;
      d^4 K\; d^4 R\; d^4 R'\; d^4 U
\delta^4(P-q_1-q_2-K-R-R'-U)
\\&\times
\sum_{n_1=0}^\infty \frac{1}{n_1!} 
\prod_{i_1=1}^{n_1} \frac{d^3 k_{i_1}}{ k^0_{i_1} } S_I(k_{i_1})
\delta_{K=\sum_{i_1=1}^{n_1} k_{i_1} }\;
\sum_{n_2=0}^\infty \frac{1}{n_2!} 
\prod_{i_2=1}^{n_2} \frac{d^3 k_{i_2}}{ k^0_{i_2} } S_X(k_{i_2})
\delta_{R=\sum_{i_2=1}^{n_1} k_{i_2} }\;
\\&\times
\sum_{n_3=0}^\infty \frac{1}{n_3!} 
\prod_{i_3=1}^{n_3} \frac{d^3 k_{i_3}}{ k^0_{i_3} } S_X(k_{i_3})
\delta_{R'=\sum_{i_3=1}^{n_3} k_{i_3} }\;
\sum_{n_4=0}^\infty \frac{1}{n_4!} 
\prod_{i_4=1}^{n_4} \frac{d^3 k_{i_4}}{ k^0_{i_4} } S_F(k_{i_4})
\delta_{U=\sum_{i_4=1}^{n_4} k_{i_4} }\;
\\&\times
\sum_{V,V'=\gamma,Z} \Mcal_V(P-K-R)\; \Mcal^*_{V'}(P-K-R')
\\&\times
\exp\big\{
  2\alpha \Rmf B_4(s,t,m_\gamma) 
  +\alpha \Delta  B_4^V((P-K-R)^2) +(\alpha \Delta  B_4^{V'}((P-K-R')^2) )^*
\big\}.
\end{split}
\label{eq:exclusive}
\end{equation}
This is a generalization of Eq.~(88) in Ref.~\cite{Jadach:2000ir}, which
was obtained there using pure combinatorics,
without any use of the Mellin-Fourier transform.
Both virtual and real photon integrals are IR-regularized using finite
photon mass $m_\gamma$.

Another advantage of the compact Eq.~(\ref{eq:factorized})
is that by means of adding and subtracting
\begin{equation}
\int_{k^0\leq E} \frac{d^3 k}{ k^0}\;
\big[ S_I(k) +2S_X(k) +S_F(k) \big],
\quad E=\frac{\sqrt{s}}{2},
\end{equation}
in the form-factor exponent, we obtain a manifestly IR-finite expression~\cite{Yennie:1961ad}:
\begin{equation}
\begin{split}
&\sigma(s)=
\frac{1}{{\rm flux}(s)}
\int \frac{d^3 q_1}{ 2q_1^0} \frac{d^3 q_2}{ 2q_2^0}\;
      d^4 K\; d^4 R\; d^4 R'\; d^4 U
\delta^4(P-q_1-q_2-K-R-R'-U)
\\&\times
\int \frac{d^4z }{ (2\pi)^4} \;
e^{ iz\cdot K +\int \frac{d^3 k}{ k^0}\; [e^{-ik \cdot z} -\theta(k^0<E)]  S_I(k) }
\int \frac{ d^4u}{ (2\pi)^4} \;
e^{ iu\cdot R +\int \frac{d^3 k}{ k^0}\; [e^{-ik \cdot u}-\theta(k^0<E) ]  S_X(k) }
\\&\times
\int \frac{d^4u'}{ (2\pi)^4}\;
e^{ iu'\cdot R' +\int \frac{d^3 k}{ k^0}\; [e^{-ik \cdot u'} -\theta(k^0<E)] S_X(k) }
\int \frac{d^4y}{ (2\pi)^4} \;
e^{  iy \cdot U  +\int \frac{d^3 k}{ k^0}\;[ e^{-ik \cdot y} -\theta(k^0<E) ] S_F(k) }
\\&\times
\sum_{V,V'=\gamma,Z}\;
\Mcal_V(P-K-R)\; \Mcal^*_{V'}(P-K-R'),
\\&\times
\exp\big\{
  Y(p_i,q_i)
  +\alpha \Delta  B_4^V((P-K-R)^2) +(\alpha \Delta  B_4^{V'}((P-K-R')^2) )^*
\big\},
\end{split}
\label{eq:factorized2}
\end{equation}
where $(e^{-ik \cdot y} -1)/k^0$ is IR-finite for $k^0\to 0$
and the classic YFS form factor
\begin{equation}
Y(p_i,q_i)=  2\alpha \Rmf B_4(s,t,m_\gamma)
      + \int_{k^0\leq E} \frac{d^3 k}{ k^0}\;
               \big[ S_I(k) +2S_X(k) +S_F(k) \big]
\end{equation}
is also finite in the $m_\gamma\to 0$ limit.

Reintroducing an IR cutoff $\veps$ on the real photon energies,
Eq.~\ref{eq:factorized2} can be also rewritten in
the following equivalent form with the standard phase space integration
and without any Mellin transforms~\cite{Jadach:1991ws}:
\begin{equation}
\begin{split}
&\sigma(s)=
\frac{1}{{\rm flux}(s)}
\int \frac{d^3 q_1}{ 2q_1^0} \frac{d^3 q_2}{ 2q_2^0}\;
      d^4 K\; d^4 R\; d^4 R'\; d^4 U
\delta^4(P-q_1-q_2-K-R-R'-U)
\\&\times
\sum_{n_1=0}^\infty \frac{1}{n_1!} 
\int\limits_{ k^0_{i_1}>\veps} \prod_{i_1=1}^{n_1} \frac{d^3 k_{i_1}}{ k^0_{i_1} } S_I(k_{i_1})
\delta_{K=\sum_{i_1=1}^{n_1} k_{i_1} }\;
\sum_{n_2=0}^\infty \frac{1}{n_2!} 
\int\limits_{ k^0_{i_2}>\veps} \prod_{i_2=1}^{n_2} \frac{d^3 k_{i_2}}{ k^0_{i_2} } S_X(k_{i_2})
\delta_{R=\sum_{i_2=1}^{n_1} k_{i_2} }\;
\\&\times
\sum_{n_3=0}^\infty \frac{1}{n_3!} 
\int\limits_{ k^0_{i_3}>\veps} \prod_{i_3=1}^{n_3} \frac{d^3 k_{i_3}}{ k^0_{i_3} } S_X(k_{i_3})
\delta_{R'=\sum_{i_3=1}^{n_3} k_{i_3} }\;
\sum_{n_4=0}^\infty \frac{1}{n_4!} 
\int\limits_{ k^0_{i_3}>\veps} \prod_{i_4=1}^{n_4} \frac{d^3 k_{i_4}}{ k^0_{i_4} } S_F(k_{i_4})
\delta_{U=\sum_{i_4=1}^{n_4} k_{i_4} }\;
\\&\times
\exp\Big( -\int_{\veps<k^0<E} \frac{d^3 k}{ k^0}  S(k) \Big)\;
\sum_{V,V'=\gamma,Z} \Mcal_V(P-K-R)\; \Mcal^*_{V'}(P-K-R')
\\&\times
\exp\big\{
  Y(p_i,q_i)
  +\alpha \Delta  B_4^V((P-K-R)^2) +(\alpha \Delta  B_4^{V'}((P-K-R')^2) )^*
\big\}.
\end{split}
\label{eq:exclusive2}
\end{equation}
The phase space integration in the above formula cannot be performed 
analytically.
(It is done numerically without any approximation in the \kkmc\ program.)
In the following, this phase space integration will be done analytically
in the semisoft approximation.

\subsection{Analytic integration over photon momenta}
\label{subsec:IEX2}
In the next step, we shall integrate over the photon angles
in Eq.~(\ref{eq:factorized2}),
introducing the cutoff 
$E_{\max} = v_{\max}\; \frac{1}{2} s^{1/2}$,
$ v_{\max} \ll 1 $,
on the total photon energy in order to simplify the phase space integral,
and staying within the semisoft approximation for the
multiphoton distributions,
as in Eq.~(\ref{eq:factorized2}) and Eq.~(\ref{eq:exclusive2}).

Let us show how it is done for the 
initial state part of this multiphoton integral.
In the semisoft photon limit, the integrand of Eq.~(\ref{eq:factorized2})
has no dependence on the spatial components of $K$ outside of the 
$e^{iz\cdot K}$ factor.  Typically, the Born matrix element and the resonant
form factor have a dependence on $K^0$ through
\begin{equation}
  (P-K-R)^2= P^2-2P\cdot(K+R)+(K+R)^2 \simeq s-2\sqrt{s} (K^0+R^0),
\end{equation}
but no dependence on the spatial components $\vec{K}$. Thus, the integral 
over $\vec{K}$ yields a factor $\delta^3(\vec{z})$ leads to%
\footnote{ See also Eq.~(\ref{eq:generic}) in the next subsection
  for versions without a Mellin transform.
}
\begin{equation}
\begin{split}
&
\int \frac{d^4 K\; d^4z }{ (2\pi)^4} \;
e^{ iz\cdot K +\int \frac{d^3 k}{ k^0}\;   S_I(k) [e^{-ik \cdot z} -\theta_{k^0<E}]   }
\\&
=\int  dK^0\;
\int \frac{dz^0 }{2\pi} \;
e^{ iz^0 K^0 +\int \frac{d k^0}{ k^0}\;  \gamma_I [e^{-ik^0 z^0} -\theta_{k^0<K^0}]   }
= \int\frac{d K^0}{K^0}\;  F(\gamma_I) \gamma_I  
  \left( \frac{K^0}{E} \right)^{\gamma_I},
\end{split}
\label{eq:init}
\end{equation}
where the integration over photon angles resulted in
\begin{equation}
\gamma_I=\gamma_I(s)=
 \int \frac{d^3 k}{ k^0}\;
      S_I(k)\; \delta( 2k^0/\sqrt{s}-1).
\end{equation}
The subtle point is that the elimination of
$\int d^3\vec{K} \delta^3(\vec{K}-\sum_{i=1}^n \vec{k}_i)  $
implies that we keep $ \vec{K}=\sum_{i=1}^n \vec{k}_i $
everywhere in the entire integrand.
Note that in \kkmc, the above ``recoil effect'' in the Born matrix element 
and phase space integral is taken into account correctly 
for hard photons as well.  The function 
\begin{equation}
F(\gamma)\equiv \frac{\exp(-\gamma C_E )}{\Gamma(1+\gamma)}
\end{equation}
is well known from YFS work (Eq.\ (2.44) in Ref.\ \cite{Yennie:1961ad}) and 
is due to the competition of soft real photons for the available fixed total energy.

Similarly, we are able to integrate over FSR and IFI photons:
\begin{equation}
\begin{split}
&
\int d^4U \frac{d^4y}{ (2\pi)^4} \;
e^{  iy \cdot U  +\int \frac{d^3 k}{ k^0}\; S_F(k) [e^{-ik \cdot y} -\theta_{k^0<E} ] }
=  \int \frac{dU^0}{U^0}\;\gamma_F 
 \left( \frac{K^0}{E} \right)^{\gamma_F} F(\gamma_F),
\\&
\int d^4 R \frac{ d^4u}{ (2\pi)^4} \;
e^{ iu\cdot R +\int \frac{d^3 k}{ k^0}\;   S_X(k) [e^{-ik \cdot u}-\theta_{k^0<E} ]  }
= \int \frac{dR^0}{R^0}\;\gamma_X  
   \left( \frac{K^0}{E} \right)^{\gamma_X} F(\gamma_X),
\\&
\int d^4 R'\frac{d^4u'}{ (2\pi)^4}\;
e^{ iu'\cdot R' +\int \frac{d^3 k}{ k^0}\; S_X(k) [e^{-ik \cdot u'}-\theta_{k^0<E} ] }
= \int \frac{dR'^0}{R'^0}\;\gamma_X 
  \left( \frac{K^0}{E} \right)^{\gamma_X} F(\gamma_X),
\end{split}
\end{equation}
where
\begin{equation}
\begin{split}
&
\gamma_F=\gamma_F(s)=
 \int \frac{d^3 k}{ k^0}\;
      S_F(k)\; \delta( 2k^0/\sqrt{s}-1),
\\&
\gamma_X=\gamma_X(\cos\theta)=
 \int \frac{d^3 k}{ k^0}\;
      S_X(k)\; \delta( 2k^0/\sqrt{s}-1),
\end{split}
\end{equation}
and $\theta$ is the angle the between the momenta $p_1$ of $e^-$ and 
$q_1$ of $\mu^-$.

Inserting all the above into Eq.~(\ref{eq:factorized2}),
we finally obtain a
{\em compact elegant formula}:
\begin{equation}
\begin{split}
&\sigma(s,v_{\max})=
\frac{3\sigma_0(s)}{8}
\sum_{V,V'}
\int_0^1 dv_I\;  dv_F\; dr\; dr'\;
\int \frac{d \cos\theta d\phi}{2}\; \theta(v_{\max}-v_I-r-r'-v_F)
\\&\times
 \rho(\gamma_I,v_I)\; \rho(\gamma_X,r)\;\rho(\gamma_X,r')\; \rho(\gamma_F,v_F)\;
 e^{Y(p_i,q_i)}\;
\\&\times
 \frac{1}{4}\sum_{\veps\tau}
 \Rmf\big\{e^{\alpha \Delta B_4^V(s(1-v_I-r))}   
           \Mmf^V_{\veps\tau}\big(v_I+r,c\big)\;
          [e^{\alpha \Delta B_4^{V'}(s(1-v_I-r'))}
           \Mmf^{V'}_{\veps\tau}\big(v_I+r',c\big)]^*
    \big\},
\label{eq:compact}
\end{split}
\end{equation}
where the Born spin amplitudes of Appendix~\ref{sec:appendixC} are used and we define
\begin{equation}
\begin{split}
&
 \rho(\gamma,v) = F(\gamma)   \gamma    v^{\gamma-1},\quad
  v_I=\frac{2K^0}{\sqrt{s}},\quad
  r=\ln\frac{2R^0}{\sqrt{s}},\quad
  r'=\ln\frac{2R'^0}{\sqrt{s}},\quad
  v_F=\frac{2U^0}{\sqrt{s}}.
\label{eq:variables1}
\end{split}
\end{equation}
The appearance of the real part $\Rmf[\Mcal_V \Mcal_{V'}^*]$ 
has resulted from symmetrization over $r$ and $r'$.
The overall structure of the above integral is illustrated 
in Fig.~\ref{fig:xKKsemIFI}.

Note that the YFS function  $\rho(\gamma,v)$ obeys the following nice 
convolution rule
(related to the fact that it represents a Markovian process):
\begin{equation}
\int dv_1 dv_2 \delta(v-v_1-v_2)\; \rho(\gamma_1,v_1) \rho(\gamma_2,v_2) 
 = \rho(\gamma_1+\gamma_2,v),
\label{eq:rho_convol}
\end{equation}
but this feature cannot be exploited to simplify 
the integral of Eq.~(\ref{eq:compact}),
because of the peculiar dependence of the matrix element on $r$ and $r'$.
Let us stress that  the double convolution over ISR photons, 
separately for the Born amplitude and its conjugate  
seen in Eq.~(\ref{eq:compact}),
is the landmark feature of the semisoft exponentiation
pioneered in Refs.~\cite{Greco:1975rm,Greco:1975ke} 
and implemented in \kkmc.

\subsection{Matching of analytic exponentiation with fixed orders}
\label{subsec:matching}

Any matching of analytic exponentiation with fixed-order calculations 
must address the inclusion of the hard photon phase space beyond the 
semisoft regime represented in Eq.~(\ref{eq:compact})
with the cutoff $v_{\max} \ll 1$ on the total photon energy.
The above matching will follow past examples 
in Refs.~\cite{Z-physics-at-lep-1:89,Bardin:1999gt,Jadach:2000ir}.
It will result in the formula valid for  $0<v_{\max} \leq 1$.

In order to match analytic exponentiation
with known analytic \order{\alpha^{1,2,3}} QED results for ISR and FSR
and compare with \kkmc\ over the entire phase space,
let us extrapolate the formula of Eq.~(\ref{eq:compact}) 
beyond the semisoft regime to the entire range of the variable
\begin{equation}
v= 1-M^2_{\mu\mu} / s, \quad v\in (0,1),
\end{equation}
replacing soft photon approximation
\[ v = v_I+v_F+r+r', \]
with a multiplicative ansatz guided by the collinear kinematics,
\[ 1-v = (1-v_I)(1-v_F)(1-r)(1-r'). \]
With this ansatz, Eq.~(\ref{eq:compact}) takes the form
\begin{equation}
\begin{split}
&\sigma^{(0)}(s,v_{\max})=
\frac{3\sigma_0(s)}{8}
\sum_{V,V'}
\int dv\; dv_I\;  dv_F\; dr\; dr'\;
\delta_{1-v=(1-v_I)(1-v_F)(1-r)(1-r')}\;
\theta_{v_{\max}>v}\;
\\&\times
\int \frac{d \cos\theta d\phi}{2}\;
 \rho(\gamma_I(s),v_I)\; \rho(\gamma_F(s(1-v_I)(1-v_F)),v_F)\; 
 \rho(\gamma_X(c),r)\;\rho(\gamma_X(c),r')\;
 e^{Y(p_i,q_i)}\;
\\&\times
 \frac{1}{4} \sum_{\veps\tau}
 \Rmf\big\{e^{\alpha \Delta B_4^V(  s(1-v_I)(1-r))} 
           \Mmf^V_{\veps\tau} \big(1-(1-v_I)(1-r), c\big)\;
\\&~~~~~~~~~~~~~
         [e^{\alpha \Delta B_4^{V'}(s(1-v_I)(1-r'))} 
           \Mmf^{V'}_{\veps\tau} \big(1-(1-v_I)(1-r'),c\big)]^*
    \big\},
\label{eq:kkfoam0}
\end{split}
\end{equation}
where $c=\cos\theta$ and he have
inserted also the Born spin amplitudes of Appendix~\ref{sec:appendixC},
From now on we may use $0<v_{\max}\leq 1$.

In the numerical comparison of the above ${\cal O}(\alpha^0)_{\rm exp}$ formula
with \kkmc, it is worth including numerically significant 
${\cal O}(\alpha^2)$ contributions from the trivial phase integration.
It was shown in Ref.~\cite{Jadach:2000ir} (see Eq.~(206) there)
that the following substitution does the job:
\begin{equation}
\begin{split}
&\rho(\gamma_I,v_I) \to 
 \rho_I^{(0)}(\gamma_I,v_I)=
  \rho(\gamma_I,v_I) 
  \exp\Big[ \frac{1}{4}\gamma_I
           +\frac{\alpha}{\pi}\Big( -\frac{1}{2} +\frac{\pi^2}{3} \Big)  \Big]
   \Big[ 1 -\frac{1}{4} \gamma_I\ln(1-v_I) \Big],
\\&
 \rho(\gamma_F,v_F) \to 
\rho_F^{(0)}(\gamma_F,v_F)=
  \rho(\gamma_F,v_F) 
\\&~~~~~~~~~~~~~~~~~~~
  \exp\Big[ \frac{1}{4}\gamma_I
           +\frac{\alpha}{\pi}\Big( -\frac{1}{2} +\frac{\pi^2}{3} \Big)
           -\frac{\gamma_F}{2}\ln(1-v_F)
      \Big]
  \Big[  1 -\frac{1}{4} \gamma_F\ln(1-v_F) \Big],
\end{split}
\label{eq:rho0}
\end{equation}
where $\gamma_F = \gamma_F(s(1-v_I)(1-v_F))$.
\footnote{We could also use  $\gamma_F = \gamma_F(s(1-v))$, but
we have checked that it leads to the same numerical results.}

In order to compare with  ${\cal O}(\alpha^2)$ \kkmc\ calculations
(including non-IR contributions of IFI up to ${\cal O}(\alpha^1)$)
it is also quite easy to upgrade the ISR and FSR radiator functions
in Eqs.~(\ref{eq:kkfoam2}) to ${\cal O}(\alpha^2)$:
\begin{equation}
\begin{split}
& 
  \rho(\gamma_I,v_I) \to 
  \rho_I^{(2)}(\gamma_I,v_I)= 
  \rho(\gamma_I,v_I) 
  \exp\Big[ \frac{1}{4}\gamma_I
           +\frac{\alpha}{\pi}\Big( -\frac{1}{2} +\frac{\pi^2}{3} \Big)  \Big]
   \Big[ 1+\frac{\gamma_I}{4} +\frac{\gamma_I^2}{8}
\\&~~~~~~~~~~~~~~~~~~~~
   +v_I\Big( -1 +\frac{v_I}{2} \Big) 
   +\gamma_I \Big( -\frac{v_I}{2} -\frac{1+3(1-v_I)^2}{4} \ln(1-v_I)  \Big)
  \Big],
\\&
  \rho(\gamma_F,v_F) \to
  \rho_F^{(2)}(\gamma_F,v_F)=
  \rho(\gamma_F,v_F) 
  \exp\Big[ \frac{1}{4}\gamma_I
           +\frac{\alpha}{\pi}\Big( -\frac{1}{2} +\frac{\pi^2}{3} \Big)
           -\frac{\gamma_F}{2}\ln(1-v_F)
      \Big]
\\&~~~~~~~~~
  \Big[  1+\frac{\gamma_F}{4} +\frac{\gamma_F^2}{8}
  +v_F\Big( -1 +\frac{v_F}{2} \Big) 
       + \gamma_F \Big( -\frac{v_F}{2} 
                        +\frac{v_F(2-v_F)}{8} \ln(1-v_I)  \Big)
  \Big],
\end{split}
\label{eq:rho2}
\end{equation}
see Tables I and II  in Ref.~\cite{Jadach:2000ir}.

The resulting ISR+FSR+IFI formula of Ref.~(\ref{eq:kkfoam0})
with all the above upgrades of ISR and FSR radiator functions
(with the resummation of $\ln(\Gamma_Z/M_Z)$) 
is now ready for the MC implementation.

We are also going to implement the following
formula in which IFI is completely neglected:
\begin{equation}
\begin{split}
&\sigma^{(0)}_{\rm no IFI}(s,v_{\max})=
\frac{3\sigma_0(s)}{8}
\int dv\; dv_I\;  dv_F\;
\delta_{1-v=(1-v_I)(1-v_F)}\;
\theta_{v_{\max}>v}\;
\\&\times
\int \frac{d \cos\theta d\phi}{2}\;
 \rho(\gamma_I(s),v_I)\; \rho(\gamma_F(s(1-v_I)(1-v_F)),v_F)\;
 e^{Y(p_i,q_i)}\;
 \frac{1}{4} \sum_{\veps\tau}
 \big|\Mmf_{\veps\tau}\big(v_I, c) \big|^2,
\label{eq:kkfoam2}
\end{split}
\end{equation}

In the final push towards inclusion of as many known fixed
order results as possible into the analytic exponentiation formula,
we include the complete ${\cal O}(\alpha^1)$ virtual IFI contributions.
This amounts to adding the non-IR parts of the $\gamma\gamma$ and $\gamma Z$
box diagrams explicitly provided in Eqs.~\ref{eq:BoxesGGandGZ} in
Appendix~\ref{sec:appendixC}
to the Born spin amplitudes:
\begin{equation}
\begin{split}
& 
 \Mmf_{\veps\tau}(s,t) \to \Mmf_{\veps\tau} (s,t)
   +\Mmf_{\veps\tau}^{\{\gamma\gamma\}}(s,t,m_\gamma)
   +\Mmf_{\veps\tau}^{\{\gamma Z\}} (s,t,m_\gamma)
\\&~~~~~~~~~~~~~~~~~~
   -2\alpha B_4(s,t,m_\gamma)\;  \Mmf_{\veps\tau}(s,t)
   -\alpha \Delta B_4^Z(s,t)\;  \Mmf^Z_{\veps\tau}(s,t).
\end{split}
\label{eq:boxing}
\end{equation}
This is done in the framework of the standard YFS-inspired reorganization
of the IR singularities, the same way as in the CEEX matrix element of \kkmc,
without any danger of double counting.
The additional subtraction of $ \alpha \Delta B_4^V$ prevents double counting
with the resummation/exponentiation of this term in the semisoft regime.

In addition electroweak and QCD corrections are also included
in coupling constants of Born amplitudes, the same way as in \kkmc.
Both \kkmc\ and \kkfoam\ use the \dizet\ library of 
${\cal O}(\alpha^1)$ EW corrections~\cite{Bardin:1989tq}
(including some of ${\cal O}(\alpha^2)$)
and the method in which EW corrections are inserted into Born-like
parts of the spin amplitudes in \kkmc\ is essentially the same as 
in {\tt ZFITTER}~\cite{Bardin:1999yd}.
It is described in Eqs.~(21-25) of Ref.~\cite{Jadach:1999vf}.

Finally, with all the above changes due to matching with  ${\cal O}(\alpha^1)$
and  ${\cal O}(\alpha^2)$ known fixed-order corrections,
we are now ready to implement the results of analytic exponentiation
of Eqs.~(\ref{eq:kkfoam0}) and (\ref{eq:kkfoam2})
with radiator functions of Eq.~(\ref{eq:rho0},\ref{eq:rho2})
and box insertions, using the Monte Carlo method.

Coming back to
the extension of the phase space in Eqs.~(\ref{eq:kkfoam0}) and (\ref{eq:kkfoam2})
we see that it has now a well defined meaning: for IFI switched off these formulas
coincide with the well known QED convolution formulas for the total cross section
\cite{Z-physics-at-lep-1:89,Bardin:1999gt,Jadach:2000ir}
including hard photons.
However,
for the angular distributions Eq.~(\ref{eq:kkfoam2}) is not able to reproduce {\em exactly}
the \order{\alpha^{1,2}} angular distribution beyond the soft limit.
On the other hand, from the analysis of Ref.~\cite{Was:1989ce} it is known 
that it reproduces numerically very well
\order{\alpha^{1,2}} MC results without IFI near the Z resonance
for $\afb$ calculated practically for any choices on the muon angle,
within realistic cutoffs including hard photon emissions.
Unfortunately,
the known nonsoft \order{\alpha^{1}} IFI contributions to angular distributions,
see Ref.~\cite{Jadach:1988zp} and Appendix~\ref{sec:appendixC}
cannot be reproduced exactly%
\footnote{It would cost adding two extra phase space integration variables
  in \kkfoam\ in order to complete \order{\alpha^1}. }
by Eqs.~(\ref{eq:kkfoam0},\ref{eq:kkfoam2}).
The main aim of Eqs.~(\ref{eq:kkfoam0},\ref{eq:kkfoam2}) 
with radiator functions of Eqs.~(\ref{eq:rho0},\ref{eq:rho2})
is to test soft limit of IFI implementation
in \kkmc\ in the presence of \order{\alpha^{2}} ISR, FSR and \order{\alpha^{1}} EW corrections.

How different are
the above analytic resummations of the IFI effect in the semisoft approximation
from the known similar calculations in
the literature \cite{Greco:1975rm, Greco:1975ke,Greco:1980mh}?
Although the starting point in terms of multiphoton amplitudes is the same,
the semisoft resummation in
Ref.\cite{Greco:1975rm} exploits techniques of coherent states and Mellin 
transform
for dealing with multiple sums over photons and phase space integration,
while our approach is based on the straightforward combinatorics 
and direct phase space integration.
References\ \cite{Greco:1975rm, Greco:1975ke,Greco:1980mh} attempt to do final phase space
integration analytically, while in our approach we perform them numerically,
gaining more flexibility in the matching with finite order results
and in numerical comparisons with \kkmc.

As a parenthetical remark, let us remark that
one may try to do some extra {\em ad hoc} simplifications,
strictly speaking not justified in the semisoft regime,
which may have some practical advantages in the parametrization of the MC results or data.
One example is the following variant of Eq.~(\ref{eq:kkfoam0})
which implements IFI in the approximate form:%
\footnote{ There are more variants of this formula,
  for instance setting $u=0$ in Born matrix element and form factor, {\em etc}.}
\begin{equation}
\begin{split}
&\sigma^{(0)}(s,v_{\max})=
\frac{3\sigma_0(s)}{8}
\int dv\; dv_I\;  dv_F\; du\;
\delta_{1-v=(1-v_I)(1-v_F)(1-u)}\;
\theta_{v_{\max}>v}\;
\\&\times
\int \frac{d \cos\theta d\phi}{2}\;
 \rho(\gamma_I(s),v_I)\; \rho(\gamma_F(s(1-v_I)(1-v_F)),v_F)\; 
 \rho(2\gamma_X(c),u)\;
 e^{Y(s,c)}\;
\\&\times
 \frac{1}{4} \sum_{\veps\tau}
\big|\sum_{V}
     e^{2\alpha \Delta B_4^V(  s(1-v_I)(1-u))} 
     \Mmf^V_{\veps\tau}\big(c,1-(1-v_I)(1-u), c) \big|^2.
\label{eq:kkfoam0a}
\end{split}
\end{equation}
The simplification is due to neglecting $r$ and $r'$ dependence
in the Born matrix element and keeping the integration over $u=r+r'$.
The quality of the above approximation can only be judged
using numerical tests.

\subsection{Numerical integration methodology}
\label{subsec:NumIntegr}

Our aim is to perform numerically the 5-dimensional and 3-dimensional 
integrals in Eqs.~(\ref{eq:kkfoam0}) and (\ref{eq:kkfoam2})
using the Monte Carlo integrator FOAM~\cite{Jadach:1999sf,Jadach:2002kn}.
This is not quite trivial because the integrands of 
Eqs.~(\ref{eq:kkfoam0}) and(\ref{eq:kkfoam2})
are singular and nonpositive.
Singularities due to $\rho_I$ of ISR and $\rho_F$ of FSR
can be easily eliminated with 
the following simple mapping of variables%
\footnote{FOAM can cope with these singularities even
  without such a mapping.}
\[
  v=x_{\max}\; y_1^{1/\gamma_I},\quad 
  u=x_{\max}\; y_2^{1/\gamma_I},\quad y_i\in(0,1).
\]
The variable $x_{\max}=0.999...$ is a technical cutoff
introduced to avoid numerical instabilities near $v=1$.
The main problem is the integration over the two
more strongly singular and nonpositive $\rho_X$ factors.
This occurs when
\begin{equation}
 \gamma_X(\theta)= 
    2Q_eQ_f\frac{\alpha}{\pi} 
    \ln\left( \frac{1-\cos\theta}{1+\cos\theta} \right)
\end{equation}
becomes negative: $\gamma_X(\theta) = -\beta <0$ in the forward
hemisphere, where $\cos\theta>0$.

In fact, one may think that in such a case the integral 
of Eq.~(\ref{eq:compact}) does not make sense at all, because the
singularity $ r^{\gamma_X-1}=r^{-\beta-1}$ from $\rho_X$ is even not integrable!
However, a closer examination of the multiphoton integral which has led to 
$\rho_X(r,\gamma_X(\theta))$ reveals that
the original distribution is in fact regularized with the familiar
plus-prescription $(...)_+$.

In order to understand the problem better, it is worth examining
the generic YFS multiphoton integral
\begin{equation}
\begin{split}
&
\int_{K^0<E} \frac{d^4 K\; d^4z }{ (2\pi)^4} \;
e^{ iz\cdot K +\int \frac{d^3 k}{ k^0}\;   S(k) [e^{-ik \cdot z} -\theta_{k^0<E}]   }
\\&
=\int_{K^0<E}  d^4 K\;
e^{ -\int \frac{d^3 k}{ k^0}\;  S(k)  \theta_{\veps<k^0<E} }
 \sum_{n=0}^\infty \frac{1}{n!}
\prod_{i=1}^n \int_{\veps<k^0<E} \frac{d^3 k_i}{ k_i^0}\;  S(k_i)\;
\delta^4\left(K-\sum_{i=1}^n k_i\right)
\\&
=\int_0^{E}  d K^0\;
e^{ -\int \gamma \frac{d k^0}{ k^0}\;   \theta_{\veps<k^0<K^0} }
 \sum_{n=0}^\infty \frac{1}{n!}
\prod_{i=1}^n \int_{\veps<k^0<K^0} \gamma\frac{d k_i^0}{ k_i^0}\;\;
\delta\left(K^0-\sum_{i=1}^n k_i^0\right)
\\&
=\int_0^{E}  dK^0\;
 \int \frac{dz }{2\pi} \;
 e^{ iz K^0 +\int \frac{d k^0}{ k^0}\;  \gamma [e^{-ik^0  z} -\theta_{k^0<K^0}]   }
\\&
= \int_0^{E} \frac{d K^0}{K^0}\;  \gamma   F(\gamma)
  \left( \frac{K^0}{E} \right)^{\gamma}
= \int_0^1 dv\; \gamma v^{\gamma-1}   F(\gamma)
= \int_0^1 dv\; \rho(\gamma,v)
=  F(\gamma).
\end{split}
\label{eq:generic}
\end{equation} 
It is easy to check that the above integral
is always finite and well-defined for any choice of $S=S_I,S_F,S_X$, 
even for negative $S$ and for negative $\gamma$!
Obviously, for $\gamma>0$, the singularity $v^{\gamma-1}$ is integrable and does
not require any regulation.  Closer inspection of Eq.~(\ref{eq:generic})
with an explicit IR regulator $\epsilon \ll 1$ reveals that for any $\gamma$,
including $\gamma=-\beta<0$, the following holds:
\begin{equation}
\begin{split}
&\rho(\gamma,v)=
e^{ -\int \gamma \frac{d k^0}{ k^0}\;   \theta_{\veps<k^0<K^0} }
 \sum_{n=0}^\infty \frac{1}{n!}
\prod_{i=1}^n \int_{\veps<k^0<K^0} \gamma\frac{d k_i^0}{ k_i^0}\;\;
\delta\left(K^0-\sum_{i=1}^n k_i^0\right)
\\&
= \delta(v) F(\gamma) \Big[1- \int_\veps^1 dv' \gamma {v'}^{\gamma-1} \Big]
 +\theta(v-\veps) F(\gamma) \gamma v^{\gamma-1}
= F(\gamma) \big[ \delta(v) + (\gamma v^{\gamma-1})_+ \big].
\end{split}
\label{eq:rho}
\end{equation} 
The standard plus prescription can be formulated
either in a regulator-independent way
\begin{equation}
\begin{split}
& \int_0^1 dv\; \phi(v) (\gamma v^{\gamma-1})_+ =
  \int_0^1 dv\; [\phi(v)-\phi(0)] \gamma v^{\gamma-1},
\end{split}
\label{eq:rhoplus}
\end{equation} 
or with an explicit regulator $\veps \ll 1$
\begin{equation}
\begin{split}
& (\gamma v^{\gamma-1})_+ 
  = \gamma v^{\gamma-1} \theta(v-\veps)
       - \delta(v) \int_\veps^1 dv' \gamma {v'}^{\gamma-1}
  = \gamma v^{\gamma-1} \theta(v-\veps) -\delta(v)[1-\veps^\gamma].
\end{split}
\label{eq:rhoplus2}
\end{equation} 
Of course, for $\gamma>0$, it becomes simpler,
because for $\veps \to 0$, we get
$(\gamma v^{\gamma-1})_+ \to \gamma v^{\gamma-1} -\delta(v)$
and $\rho(\gamma,v) \to  F(\gamma) \gamma v^{\gamma-1}$.
However, the explicit IR regulator remains mandatory for $\gamma<0$.

As a closing crosscheck, let us verify that for $\gamma<0$ and regularized
\begin{equation}
  \rho(-\beta,v) = F(-\beta)\; 
         [\delta(v) \veps^{-\beta} -\theta(v>\veps) \beta v^{-1-\beta}] ,
\end{equation}
the basic convolution rule of Eq.~(\ref{eq:rho_convol}) still holds:
\footnote{In the $\veps\to 0$ limit, of course.}
\begin{equation}
\int dv_1 dv_2 \delta(v-v_1-v_2)\; \rho(-\beta,v_1) \rho(\gamma,v_2) 
 = \rho(\gamma-\beta,v).
\label{eq:rho_convol2}
\end{equation}
In terms of the Markovian process, the function $\rho(\gamma,v)$ for $\gamma>0$
represents adding more (soft) photons.
The other  $\rho(-\beta,v)$ function is undoing that (backward evolution).

\begin{figure}[t]
  \centering
  \includegraphics[height=50mm,width=60mm]{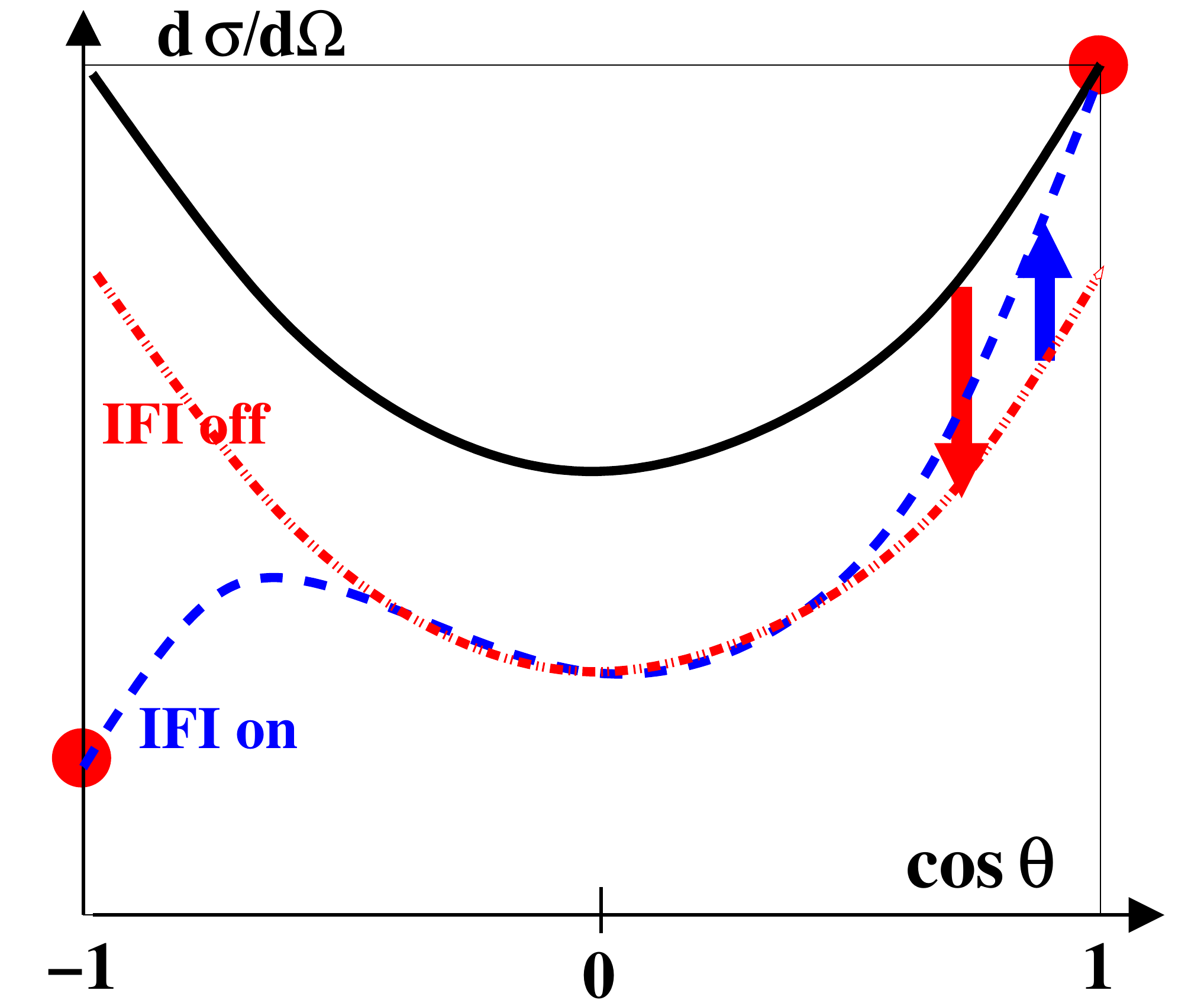} 
  \caption{\sf
  The role of IFI. In forward scattering,
  the upward arrow (IFI) counteracts partly the action of the
  downward arrow (ISR+FSR).
  }
  \label{fig:arrows}
\end{figure}

In the context of the explanation of the physics of IFI in 
Sec.~\ref{sec:physIFI},
the presence of $\rho(\gamma,v)$ with negative $\gamma$ in the
forward hemisphere in Eq.~(\ref{eq:compact})
is now perfectly understandable:
$\rho(\gamma_X,r) \rho(\gamma_X,r') $ is undoing
part of the ISR and FSR photon emission coming from
$\rho(\gamma_I,v)\rho(\gamma_F,u) $!
See the upward (blue) arrow in Fig.~\ref{fig:arrows} 
for the corresponding graphical illustration.

In the numerical MC integration, it is not difficult to introduce a small 
IR regulator $\veps$ into $\rho(\gamma_X,r)$ when $\gamma_X<0$.
In the integrand for FOAM, this is done as a part of the mapping of
the integration variables $r$ and $r'$ into internal variables of FOAM.

Another issue is that the integrand becomes negative for $\gamma_X<0$, 
for $r>\veps$, or for $r'>\veps$.
This is handled in a standard way using
using weighted MC events with a nonpositive weight.
In the actual integration by means of FOAM, the modulus of the integrand
is used during the exploration stage,
while in the following MC calculation of the integral,
the MC events are weighted with the true signed distribution.
The distribution of the MC weights in the second stage has two peaks:
\footnote{This entails a certain loss of integration precision, but it
 turns out to be affordable.}
the bigger one close to $+1$ and smaller one near $-1$;
see Fig.~\ref{fig:WTdist}.
More details on the mappings used in the construction
of the integrand for FOAM in \kkfoam\ program are given 
in Appendix~\ref{AppendixFoam}.

\begin{figure}[h]
  \centering
  \caption{\sf
  The right-hand plot is an example of the MC weight distribution for 
  calculating the total cross section using FOAM according to 
  Eq.~(\ref{eq:kkfoam0}).
  The left-hand plot presents the MC weight distribution without IFI, 
  see Eq.~(\ref{eq:kkfoam2}).
  }
  \label{fig:WTdist}
\end{figure}

\section{Numerical results from \kkfoam\ and \kkmc}
\label{sec:Numeric}

In this section, we present results from
the updated v4.22 of \kkmc, also referred to as {\tt KKMCee},
the non-MC integrator \kksem.%
\footnote{\kksem\ uses Gauss quadrature programs to integrate
 analytic formulas up to 3 dimensions.
 It was developed at the time of preparing Ref.~\cite{Jadach:2000ir}
}
and the newly developed  MC integrator (simulator)
program \kkfoam, 
based on the C++ version of {\tt FOAM} ~\cite{Jadach:2002kn}.
\kkfoam implements the 5-dimensional integral 
of Eq.~(\ref{eq:kkfoam0}) including IFI,
together with its 3-dimensional variant without IFI
of Eq.~(\ref{eq:kkfoam2}).
They will be often nicknamed in the following 
as \kkfoam5 and \kkfoam3 correspondingly.

Another subgenerator in \kkfoam\ taking care of 2-dimensional
integration over $v$ and $\cos\theta$ will be used for
reproducing and/or implementing old pure ${\cal O}(\alpha^1)$ 
results without resummation.

In \kkfoam5 and \kkfoam3, one may choose ISR and FSR structure functions
with soft photon exponentiation and QED corrections up to
${\cal O}(\alpha^0)$, ${\cal O}(\alpha^1)$ and ${\cal O}(\alpha^2)$,
as defined in Tables I and II in Ref.~\cite{Jadach:2000ir}.
Pure QED nonlogarithmic ${\cal O}(\alpha^2)$ corrections are $<10^{-5}$,
hence are neglected for ISR, FSR and IFI.
They should be included and evaluated more precisely in the future.

The Born cross section in both \kkfoam5 and \kkfoam3
is implemented using two types
of subprograms of \kkmc,
both of them using spin amplitudes: either calculated in the scheme%
\footnote{This is a variant of Kleiss-Stirling method of Ref.~\cite{Kleiss:1985yh}.}
of Ref.~\cite{Jadach:2000ir} and labeled with GPS or CEEX,
or using spin amplitudes of KORALZ \cite{Jadach:1999tr}
and labeled as EEX.
Note that it is not possible to use EEX Born for IFI implementation;
hence in \kkfoam5 only GPS/CEEX Born amplitudes are implemented.

Electroweak and QCD corrections are included in \kkfoam\ in the rescaled
coupling constants of Born amplitudes,
both for CEEX/GPS and EEX type, the same way as in \kkmc.
Contributions from nonfactorizable $\gamma\gamma$ and $\gamma Z$ boxes
are also included; see also Eq.~(\ref{eq:boxing}) for the details.
Both \kkmc\ and \kkfoam\ use {\tt DIZET} library of 
the ${\cal O}(\alpha^1)$ EW corrections~\cite{Bardin:1989tq}
(including some of ${\cal O}(\alpha^2)$)
and the method in which EW corrections are inserted into Born-like
parts of the spin amplitudes in \kkmc\ is essentially the same as 
in {\tt ZFITTER}~\cite{Bardin:1999yd}.
It is described in Eqs.~($21-25$) of Ref.~\cite{Jadach:1999vf}.
This method protects completeness of the 
${\cal O}(\alpha^1)$ content of the EW corrections.
If there is any bias introduced in this method, 
then it has to be of ${\cal O}(\alpha^2)$.

It should be kept in mind that in \kkfoam5, hard photon corrections
are included in integrated form in the structure functions up to 
${\cal O}(\alpha^2)$ for ISR and FSR, while for IFI they are not 
included--only the finite parts of the virtual ${\cal O}(\alpha^1)$ 
IFI corrections ($\gamma-Z$ boxes) are included there.
(In \kkfoam3, IFI is completely absent.)

The immediate short-term aim in this section is to prove 
that these programs correctly calculate
$\sigma(v<v_{\max})$ and $\afb(v_{\max})$
with physical and technical precision $\delta\afb \sim 10^{-4}$
and $\delta\sigma/\sigma \sim 3\cdot 10^{-4}$.
This is a factor of 10 better than at LEP, but still a factor of 10
short of what needed for FCCee near the $Z$ resonance.
An additional cutoff $|\cos\theta|<c_{\max}$ will sometimes be imposed.
An analysis for more realistic cuts will be presented
in a separate publication.  
The IFI effect in $\afb$ depends strongly on the
cutoff on the total photon energy $v_{\max}$,
which will typically be varied between $v_{\max}=0.002$ and $v_{\max}=0.200$.
As already pointed out in the Introduction, 
such a cutoff, stronger that in typical LEP data analysis, 
may be necessary at FCCee for the sake of better control of backgrounds 
and higher order QED effects.
Moreover, the expectation is that semisoft photon resummation employed in \kkfoam5
(taking into account the energy shift due to ISR in the $Z$ propagator)
will work fairly well in this cutoff range near the $Z$ pole.%
\footnote{On the other hand, strict YFS soft photon approximation
  neglecting the ISR energy shift in the $Z$ propagator is expected to be
  adequate for our precision requirements 
  only for $v_{\max}\leq 10^{-4}$.}

In the following analysis, 
event selection will be examined in terms of two variables only, 
$\cos\theta$ for the angle between $e^-$ and $\mu^-$
and $v=1-M^2_{\mu\mu}/s$.
The variable $v$ represents approximately the total energy
of all ISR and FSR photons, in units of the beam energy.
(More results for realistic selection cuts will be shown in the next paper.)
Of course, once harder photons are allowed,
the definition of  $\cos\theta$ is no longer unique.
For the \kkmc\ results, we will use the $\cos\theta$ 
definition of Ref.~\cite{Jadach:1988zp}  
unless otherwise stated--see Sec.~\ref{sec:choice_angle}
for more discussion of other choices of $\cos\theta$ and their precise definitions.

\begin{table}[ht!]
\centering
\begin{tabular}{|c|c|c|c|c|c|c|}
\hline\hline
 MC Prog. & M.E.  & Resum. & ISR & FSR & IFI & EW \\
\hline\hline
\kkmc  & CEEX2  & Semisoft  & \order{\alpha^2}  
                       & \order{\alpha^2} & \order{\alpha^1} &  Yes  \\
\hline
\kkmc  & CEEX1  & Semisoft  & \order{\alpha^1}  
                       & \order{\alpha^1} & \order{\alpha^1} &  Yes  \\
\hline
\kkmc  & CEEX0  & Semisoft  & \order{\alpha^0}  
                       & \order{\alpha^0} & \order{\alpha^0} &  Yes  \\
\hline\hline
\kkmc  &  EEX3  & Soft.+Col.   & \order{\alpha^3}  
                       & \order{\alpha^2} & None &  Yes  \\
\hline
\kkmc  &  EEX2  & Soft.+Col.   & \order{\alpha^2}  
                       & \order{\alpha^2} & None &  Yes  \\
\hline
\kkmc  &  EEX1  & Soft.+Col.   & \order{\alpha^1}  
                       & \order{\alpha^1} & None &  Yes  \\
\hline
\kkmc  &  EEX0  & Soft.+Col.  & \order{\alpha^0}  
                       & \order{\alpha^0} & None &  Yes  \\
\hline\hline
\kksem2 &  EEX Born & Soft.+Col.  & \order{\alpha^2}  
                           & \order{\alpha^2} &  None &  Yes  \\
\hline
\kksem0 &  EEX Born & Soft.+Col.  & \order{\alpha^0}  
                           & \order{\alpha^0} &  None &  Yes  \\
\hline\hline
\kkfoam5 &  GPS Born  & Soft.+Col.  & \order{\alpha^2}  
                           & \order{\alpha^2} & \order{\alpha^1} &  Yes  \\
\hline
\kkfoam3 &  EEX Born  & Semisoft & \order{\alpha^2}  
                           & \order{\alpha^2} &   None  &  Yes  \\

\hline\hline
\kkfoam2 &  EEX Born  & None & \order{\alpha^1}  
                           & \order{\alpha^1} &  \order{\alpha^1} &  Yes  \\
\hline\hline
\end{tabular}
\caption{\sf
 Table of various types of QED matrix elements, resummation methodology
and phase space integration methods in the following numerical studies.
``Semisoft''  indicates that exact multiphoton M.E. with narrow resonance effects included.
``Soft.+Col.'' indicates the use of collinear PDFs for ISR and FSR.
``GPS Born'' means the use of Born spin amplitudes as in CEEX, 
while ``EEX Born'' indicates the use of Born in the EEX scheme.
EW corrections are placed in the Born-like part of the spin amplitudes.
}
\label{tab:table1}
\end{table}

\subsection{Outline of the numerical investigations}

We have conducted numerical studies with three different 
programs, \kkmc, \kksem\ and \kkfoam\ 
featuring several variants of QED matrix elements
and different types of phase space integration.
For the convenience of the reader, we summarize in Table~\ref{tab:table1}
all types of programs and QED matrix elements (M.E.) used in them.
The CEEX matrix element of \kkmc\ for IFI component
is rated in the table as \order{\alpha^1} because
of missing nonsoft \order{\alpha^2} parts of the pentabox diagrams
specified in Fig.~5 in Ref.~\cite{Jadach:2000ir},
but the soft/infrared parts of these diagrams are included thanks
to the semisoft resummation technique.
It would be desirable to include these pentaboxes 
in a future version of \kkmc%
\footnote{Another urgent desirable update of the M.E. in \kkmc\ would
be inclusion of the $\alpha^3\ln^3(s/m_e^2)$ corrections.}.

Let us outline the plan of the following tests
which will lead to new estimates of the theoretical uncertainty
of the IFI calculation:
\begin{itemize}
\item
In Sec.~\ref{sec:choice_angle}, we shall find 
that the influence of the choice of the muon scattering angle on
the measurement of $\afb$ is negligible.
\item
Section~\ref{subsec:baseline} is devoted to a calibration exercise
in which the correctness of the MC integration is checked by
comparing the cutoff dependence of $\sigma(v_{\max})$ and $\afb(v_{\max})$
from three programs, \kkmc, \kksem\ and \kkfoam\ with IFI switched off.
It is done first for a maximally simple variant of the QED matrix element 
with resummation and then for the best one.
\item
In the next step, in Sec.~\ref{subsec:IFItoAFB}, the
IFI effect in $\afb$ is examined in the results for  $\afb(v_{\max})$
from \kkmc\ and \kkfoam\ for a maximally simple and the best QED matrix element
separately for three energies $s^{1/2}=10,87.9,94.3$GeV.
\item
Section~\ref{subsec:AFBdif} is devoted to the difference
$\Delta A_{\rm FB}^{\rm IFI}(v_{\max})=A_{\rm FB}^{\rm IFI}(v_{\max},s_+)-A_{\rm FB}^{\rm IFI}(v_{\max},s_-)$
in which IFI effect cancels. 
Results from \kkmc\ and \kkfoam\ for this difference will be compared.
$\Delta A_{\rm FB}$ is directly related to the measurement of $\alpha_{QED}(M_Z)$ at FCCee.
\item
Finally, in Sec.~\ref{subsec:AFBord} results for 
the energy difference $\Delta A_{\rm FB}^{\rm IFI}(v_{\max})$ from \kkmc\
will be analyzed for QED matrix elements with an increasing level of 
sophistication in order to estimate its theoretical uncertainty due to 
missing higher orders of QED.
\end {itemize}

\subsection{On the choice of the scattering angle $\theta$}
\label{sec:choice_angle}
In the limit when all photons are very soft, the
momenta of the final muons are back to back and the scattering angle $\theta$
between $e^-$ and $\mu^-$ is unique.
Once at least one photon becomes energetic, the final muons are 
not back to back and there are many possible definitions
of the effective $\theta$.
Using $\theta^{(1)}=\angle(e^-,\mu^-)$
or    $\theta^{(2)}=\angle(e^+,\mu^+)$
is not a favorable choice experimentally,
because it does not exploit fully the power of the tracker detector,
which detects both $\mu^\pm$ equally well. 

An example of a choice favorable for experiments,
taking full advantage of the very good
angular resolution of the muon detectors (trackers), 
which is much higher than the energy resolution,
is that of ref~\cite{Was:1989ce}
\begin{equation}
\begin{split}
& \cos\theta^\bullet = y_1\cos\theta_1 - y_2 \cos\theta_2,
\\&
  y_1 = \frac{\sin\theta_2}{\sin\theta_1+\sin\theta_2},\quad
  y_2 = \frac{\sin\theta_1}{\sin\theta_1+\sin\theta_2},\quad
  y_1+y_2 = 1.
\end{split}
\end{equation}

However, it was shown in Ref.~\cite{Jadach:1988zp} 
that analytic evaluation
of the IFI effect according to the ${\cal O}(\alpha^1)$ QED matrix element
can be easily done using
\begin{equation}
\cos\theta^\star = x_1 \cos\theta^{(1)} +x_2 \cos\theta^{(2)},\quad
x_i= q_i^0/(q_1^0+q_2^0),\quad x_1+x_2=1.
\end{equation}
We use this choice for
most of the numerical results presented in this work,
unless otherwise stated.
Moreover, in Ref.~\cite{Jadach:1988zp} compact analytic results 
were obtained for a charge asymmetry defined using the first moment 
\begin{equation}
\tilde{A}^*_{\rm FB}= \frac{3}{2} \int_{-1}^1 \cos\theta^* \frac{d\sigma}{\sigma}  
\end{equation}
instead of the conventional forward-backward asymmetry
$\afb =(\sigma_F-\sigma_B)/\sigma$.

For \kkfoam, the choice of $\cos\theta$ is irrelevant as long as
all photons are sufficiently%
\footnote{Sufficiently from the point of view of the FCCee precision}
soft.
Once at least one photon becomes
energetic, the ${\cal O}(\alpha^1)$ contribution calculated for
a well-defined choice of $\cos\theta$ should be included in \kkfoam.
So far, this is not yet done -- 
it should be done in the next version.
Most likely, the preferred choice for \kkfoam\ will be
$\cos\theta^\star$.

\begin{figure}[!t]
  \centering
  \includegraphics[width=100mm,height=70mm]{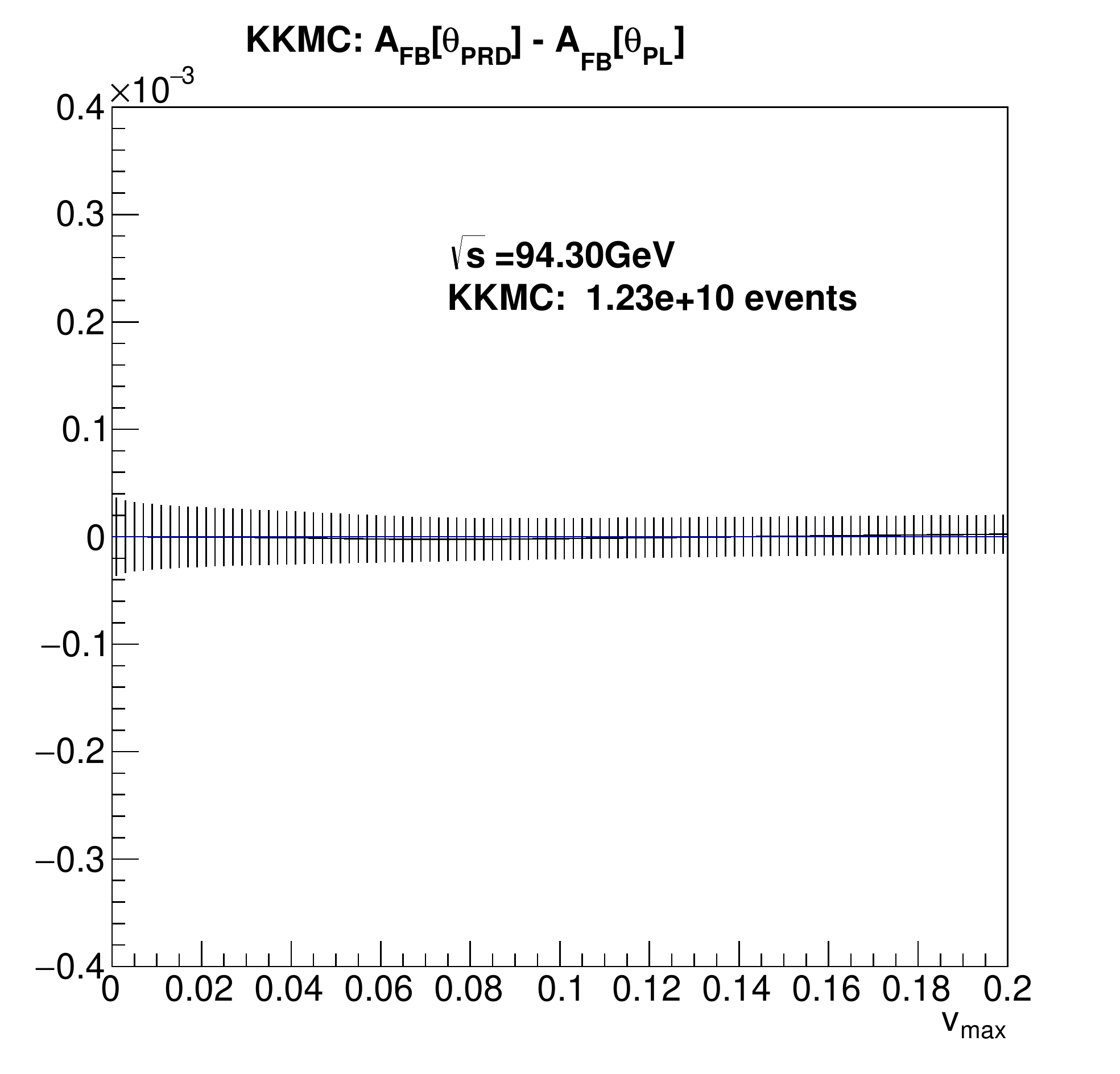}
  \caption{\sf
  The difference between $A_{\rm FB}^\star$ and $A_{\rm FB}^\bullet$.
  From \kkmc\  at 10 GeV with IFI on.
  }
  \label{fig:Afb3b_10_cPLGeV}
\end{figure}

On the other hand, \kkmc\ is a regular MC event generator providing four-momenta of
both muons (and all photons), 
hence it provides a prediction for $\afb$ with any definition of $\cos\theta$.
Let us examine, using \kkmc,
how different the QED predictions for $\afb$
are for the above two choices of $\theta$ when $v<0.2$. 
Figure~\ref{fig:Afb3b_10_cPLGeV} shows that the difference
between $A^\bullet_{\rm FB}$ and $A^*_{\rm FB}$ is below expected FCCee experimental 
precision of $\delta\afb \sim 3\cdot 10^{-5}$,
{\em i.e.} all of our analysis for $\cos\theta^\bullet$
is valid for $\cos\theta^\star$ and {\it vice versa}.

\begin{figure}[!t]
  \centering
  \includegraphics[width=100mm,height=70mm]{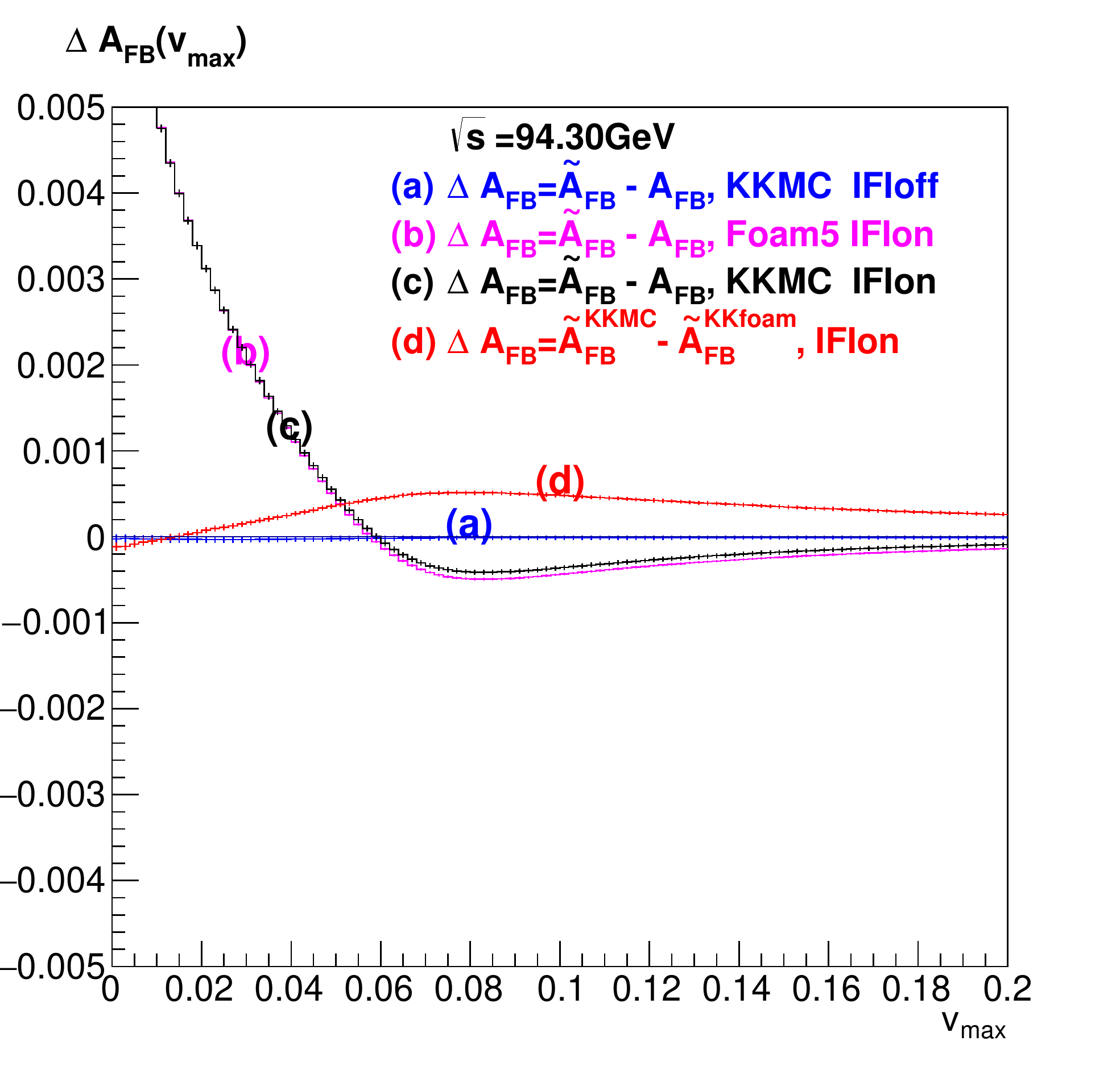}
  \caption{\sf
  The difference between $\afb$ and $\tilde{A}_{\rm FB}$  at 94.3 GeV with IFI on.
  }
  \label{fig:cFigAfb4}
\end{figure}

Using \kkmc, it is easy to examine the difference between
$\tilde{A}^*_{\rm FB}$ and $A^*_{\rm FB}$.
Figure~\ref{fig:cFigAfb4} shows that such a difference
might be sizable, up to $\sim 1\%$.
However, the difference in the IFI component could cancel between two 
calculations -- 
for instance, we have checked that it does cancel in the difference
between \kkmc\ and \kksem, for IFI switched on.

\subsection{Baseline calibration, ISR+FSR without IFI}
\label{subsec:baseline}
Let us start with the baseline calibration of the MC tools at 
the precision level $\sim 10^{-5}$
at $\sqrt{s_-}= 87.9$ GeV and $\sqrt{s_+}= 94.3$ GeV.
Although {\tt KKsem} does not include IFI, it is still useful
for checking the normalization of both \kkmc\ and \kkfoam.
Of course, normalization is irrelevant for our main observable, 
$\afb(v_{\max})$, but it is still profitable to test it,
simply because some technical problem 
that would be evident in $\sigma(v_{\max})$ could
produce a small annoying effect in $\afb$ as well.
Thus, it is better to keep an eye on both of these.

As already underlined, our main aim in the present study 
is a precise prediction for $\afb$ at two energies
$\sqrt{s_\pm}= 87.9$ near $Z$ resonance.
However,  in order to get better confidence in the implementation
of the QED matrix element,
we will also check $\afb$ at $\sqrt{s}= 10$ GeV,
where the $Z$ resonance is negligible, and at $\sqrt{s}=M_Z$,
where the suppression of IFI due to the long life time of the $Z$ is maximal.

\begin{figure}[!ht]
  \centering
  \includegraphics[width=165mm,height=66mm]{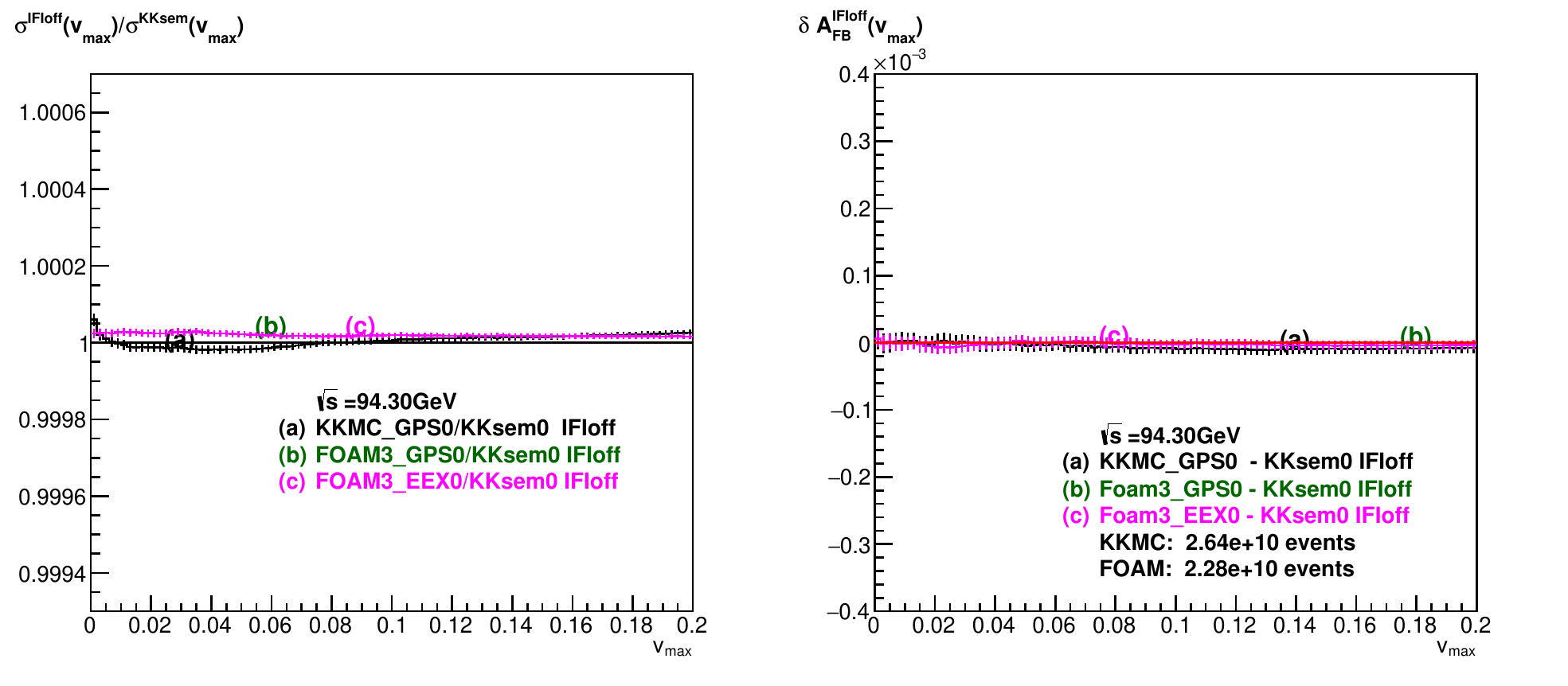}
  \includegraphics[width=165mm,height=66mm]{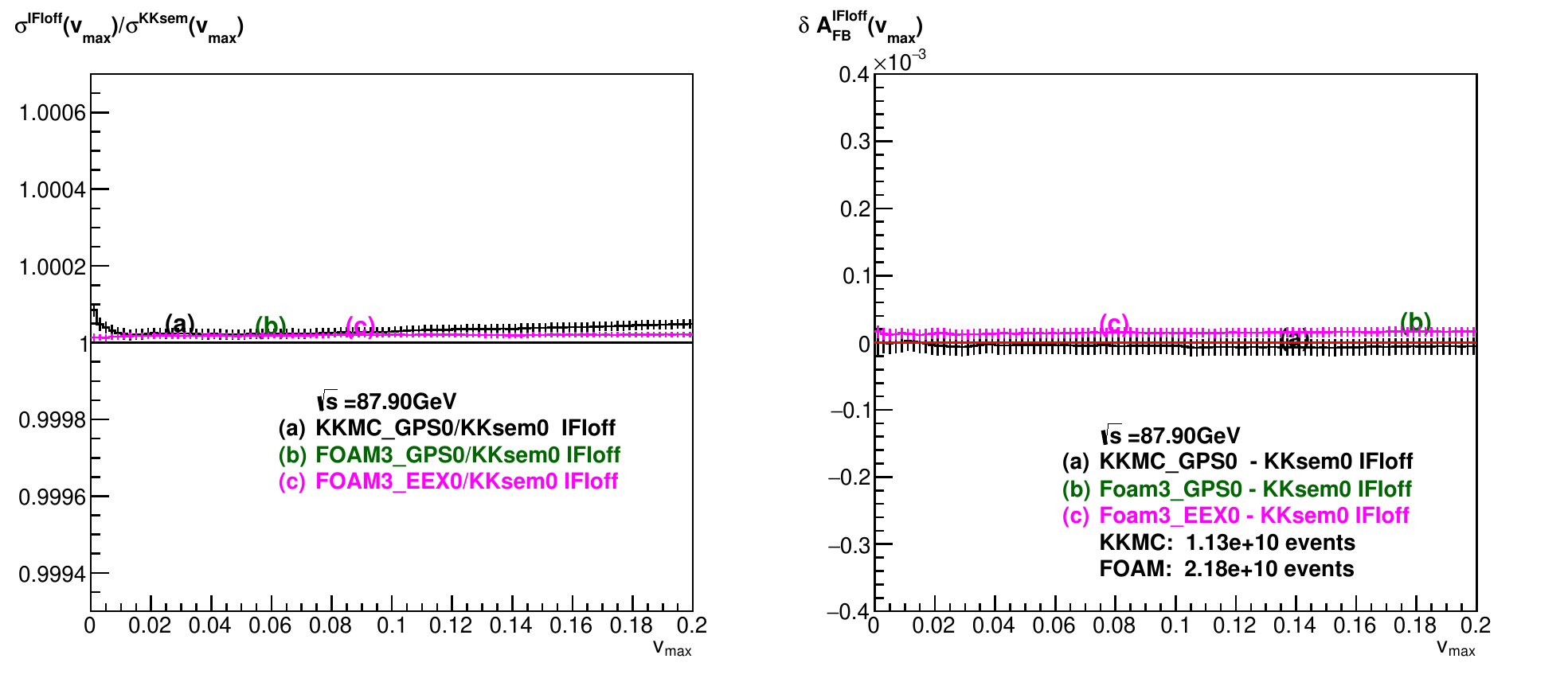}
  \includegraphics[width=165mm,height=66mm]{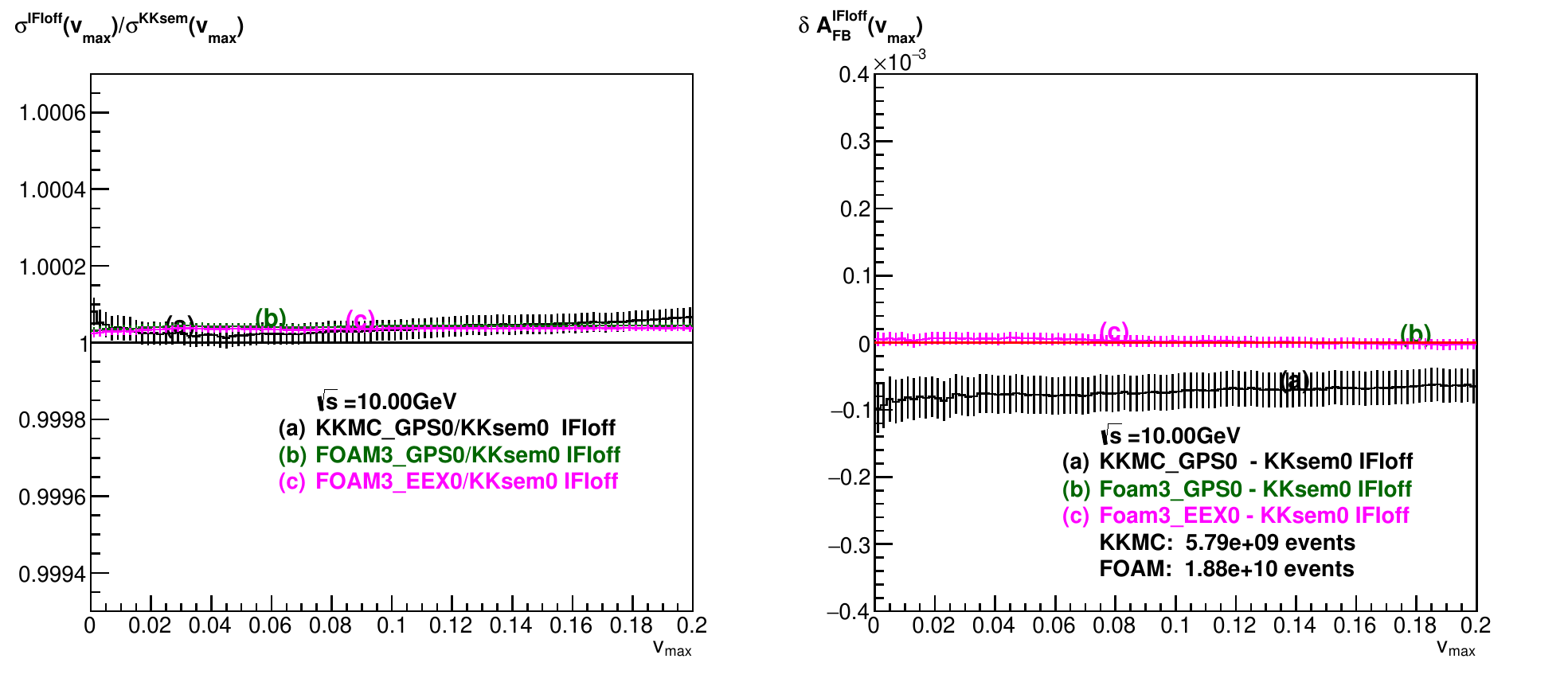}
  \caption{\sf
  Technical test,  ${\cal O}_{\rm exp}(\alpha^0)$ ISR+FSR
  without IFI at 94.3, 87.9 and 10 GeV.
  }
  \label{fig:cFigSigAfb0}
\end{figure}

Let us start with a purely technical test with IFI off 
at $\sqrt{s}=94.3,~87.9$ and 10 GeV,
presented in Fig.~\ref{fig:cFigSigAfb0}.
In the LHS plot, all cross sections $\sigma(v_{\max})$
are divided by the reference cross section from \kksem.
All calculations are at simplistic exponentiated ${\cal O}(\alpha^0)$ 
QED including ISR and FSR, but without IFI.
Different types of the Born matrix element, EEX or GPS, are used.
Very good agreement is seen, up to statistical error
$\delta\sigma/\sigma \sim 3\cdot 10^{-5}$.
The~agreement for $\afb$ is also very good,
essentially up to statistical error
$\delta\afb \sim 1\cdot 10^{-5}$ at $\sqrt{s_\pm}$.
The above equality of the \kkfoam\ and \kksem\ results
is very important, because it illustrates/proves the quality of the MC integrators
-- it should be kept in mind that
for IFI off they integrate exactly the same 3-dimensional integrand.
Even more significant is the agreement,
to within $\delta\sigma/\sigma \sim 1\cdot 10^{-5}$ near the $Z$ resonance,
 of \kkmc\ with the other two
programs for the simplified EEX0 matrix element.
This is because
for MC statistics of $2\cdot 10^{10}$ events one may expect problems with
rounding errors in the accumulation of the weights in the 
histograms.\footnote{Running in parallel on 100 nodes and combining the
  histograms afterwards helps to reduce this problem.}
The slightly bigger discrepancy 
beyond statistical error of $\delta\afb \sim 3\cdot 10^{-5}$
for 10 GeV is not yet statistically significant
and not so important for our aims.

We conclude that the technical precision of the MC numerical integration
in all three programs, \kksem, \kkfoam\ and \kksem, is satisfactory for our needs.
Moreover, the above test is also important due to the fact that the IFI effect
is added in \kkmc\ by reweighting MC events generated without IFI.
Hence, the technical precision established for the non-IFI mode persists 
when IFI is switched on.

\begin{figure}[!ht]
  \centering
  \includegraphics[width=165mm,height=66mm]{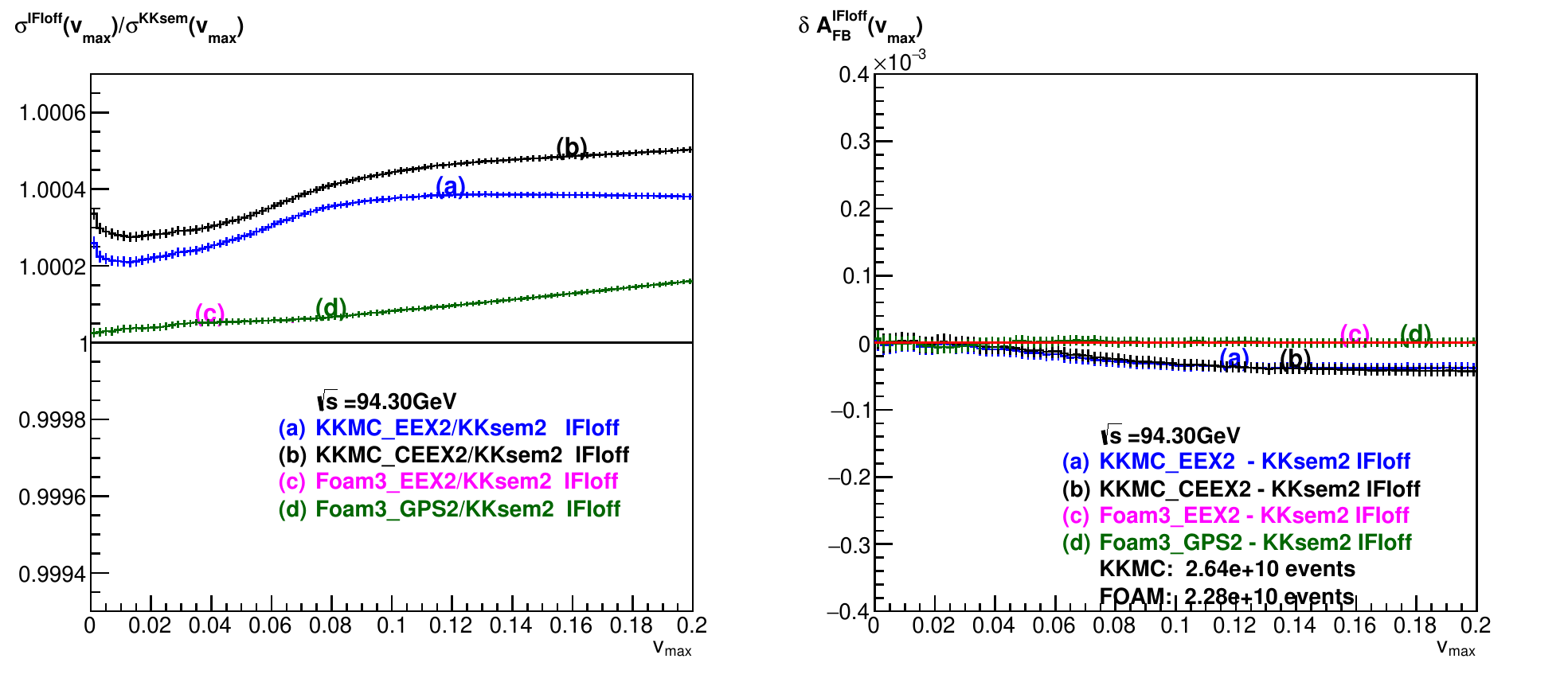}
  \includegraphics[width=165mm,height=66mm]{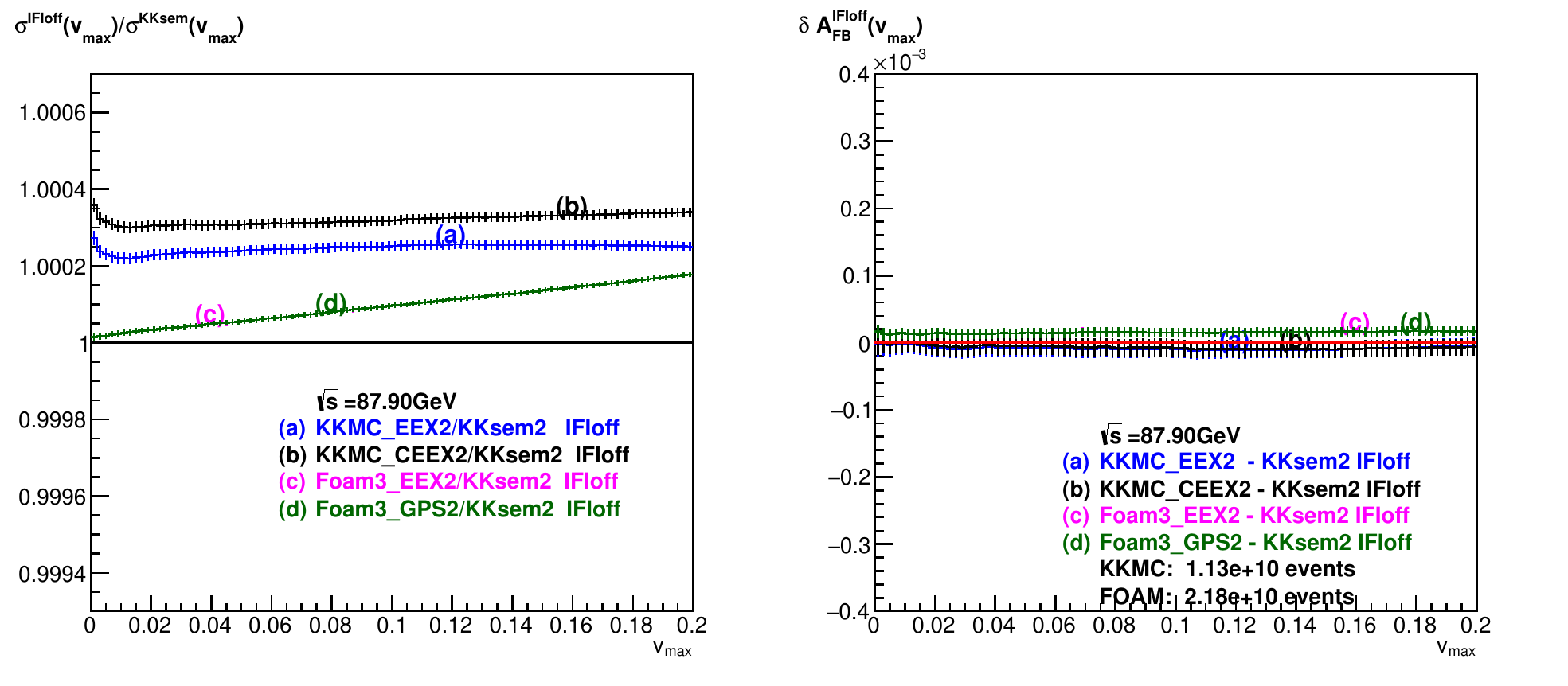}
  \includegraphics[width=165mm,height=66mm]{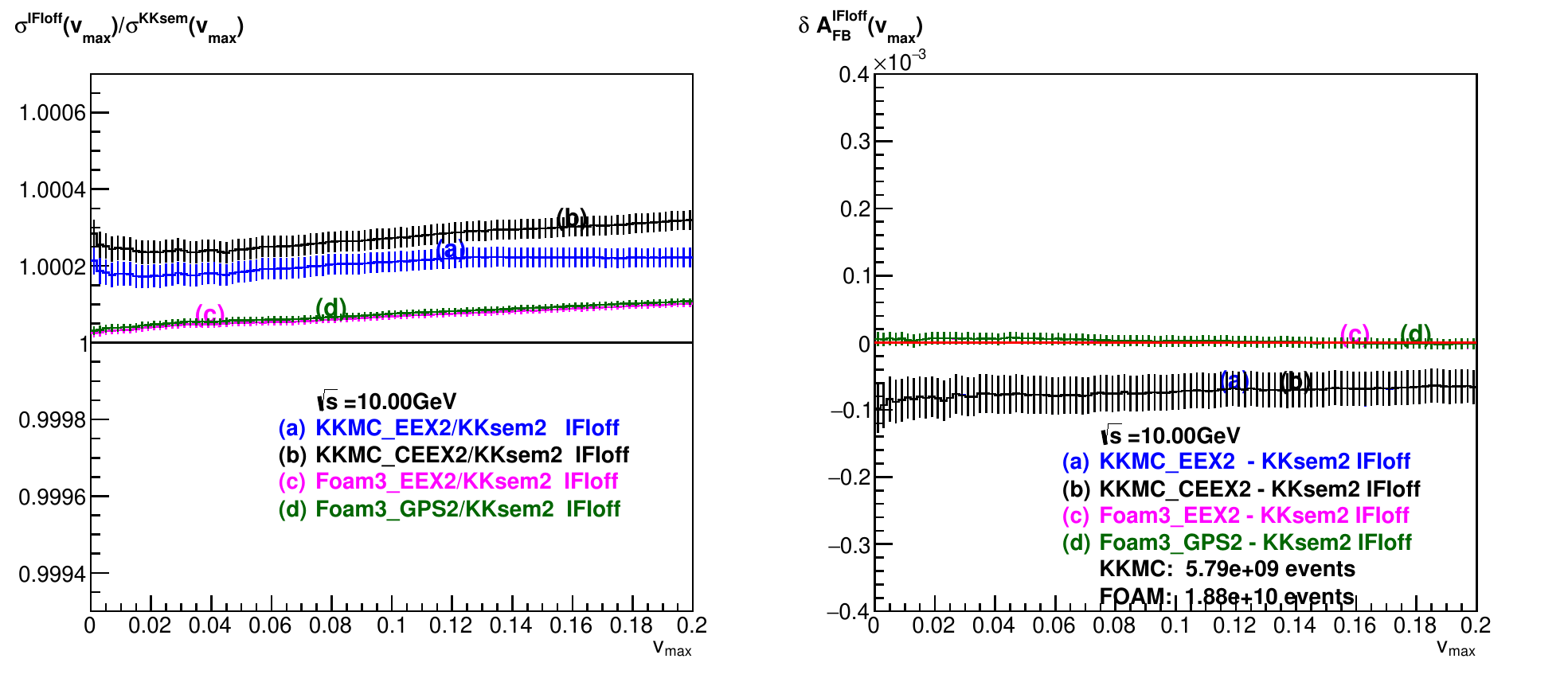}
 \caption{\sf
  Results with ${\cal O}_{\rm exp}(\alpha^2)$ ISR+FSR
  without IFI at 94.3, 87.9 and 10 GeV.
 }
  \label{fig:cFigSigAfb2}
\end{figure}

In Fig.~\ref{fig:cFigSigAfb2}, we continue baseline testing without IFI,
now with ${\cal O}(\alpha^2)$ exponentiated ISR and FSR.
The relative differences $\delta\sigma/\sigma$ between \kkmc\ and \kkfoam\
versus \kksem\ are examined.
It is done for the CEEX/GPS and EEX Born matrix element.
The relative difference $\delta\sigma/\sigma \sim 3\cdot 10^{-4}$ for \kkmc\
confirms all older tests in Ref.~\cite{Jadach:2000ir},
rated at the $\sim 1 \cdot 10^{-3}$ level.%
\footnote{
  This is not a high priority, but we shall try later 
  to find the source of these differences in the normalization.}
On the other hand, the differences in $\afb$ between \kkmc\
and \kkfoam\ or \kksem\ are again of the order of the statistical error,
which is $\sim 3\cdot 10^{-5}$, except $\sqrt{s}=10$ GeV, where it is slightly bigger.

The main result of the tests presented in
Figs.~\ref{fig:cFigSigAfb0} and ~\ref{fig:cFigSigAfb2} is that the basic technical
precision (in the MC integration) of \kkmc\ and \kkfoam3 
near the $Z$ resonance is generally better than $\delta\afb \sim 3\cdot 10^{-5}$.
The implementation of QED photonic corrections for ISR and FSR (no IFI)
up to ${\cal O}(\alpha^2)$ 
was also tested at this precision level.%
\footnote{ This is not true for \kkfoam, where subprograms
 with and without IFI are independent modules generating 
 their own different MC events.}

\subsection{IFI contribution to $\afb$ from \kkmc\ and \kkfoam}
\label{subsec:IFItoAFB}
Let us now increase the level of sophistication by one important step --
including IFI.  
This will be done first in the simpler case
(A) for ISR, FSR and IFI at the level ${\cal O}_{\rm exp}(\alpha^0)$ with exponentiation,
and next in the case 
(B) for exponentiated IFI at the level ${\cal O}_{\rm exp}(\alpha^1)$,
accompanied by ISR and FSR up to exponentiated ${\cal O}_{\rm exp}(\alpha^2)$.

\begin{figure}[!t]
  \centering
  \includegraphics[width=165mm,height=66mm]{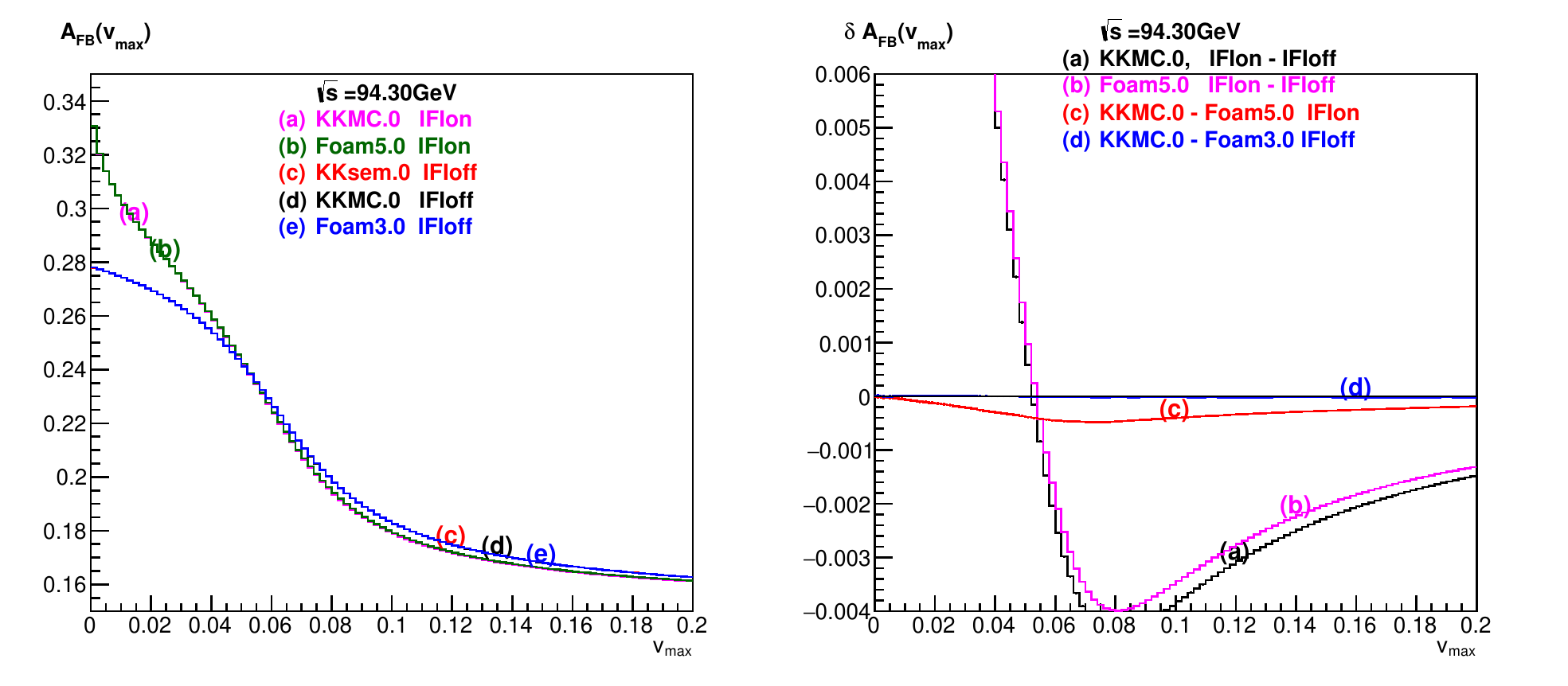}
  \includegraphics[width=165mm,height=66mm]{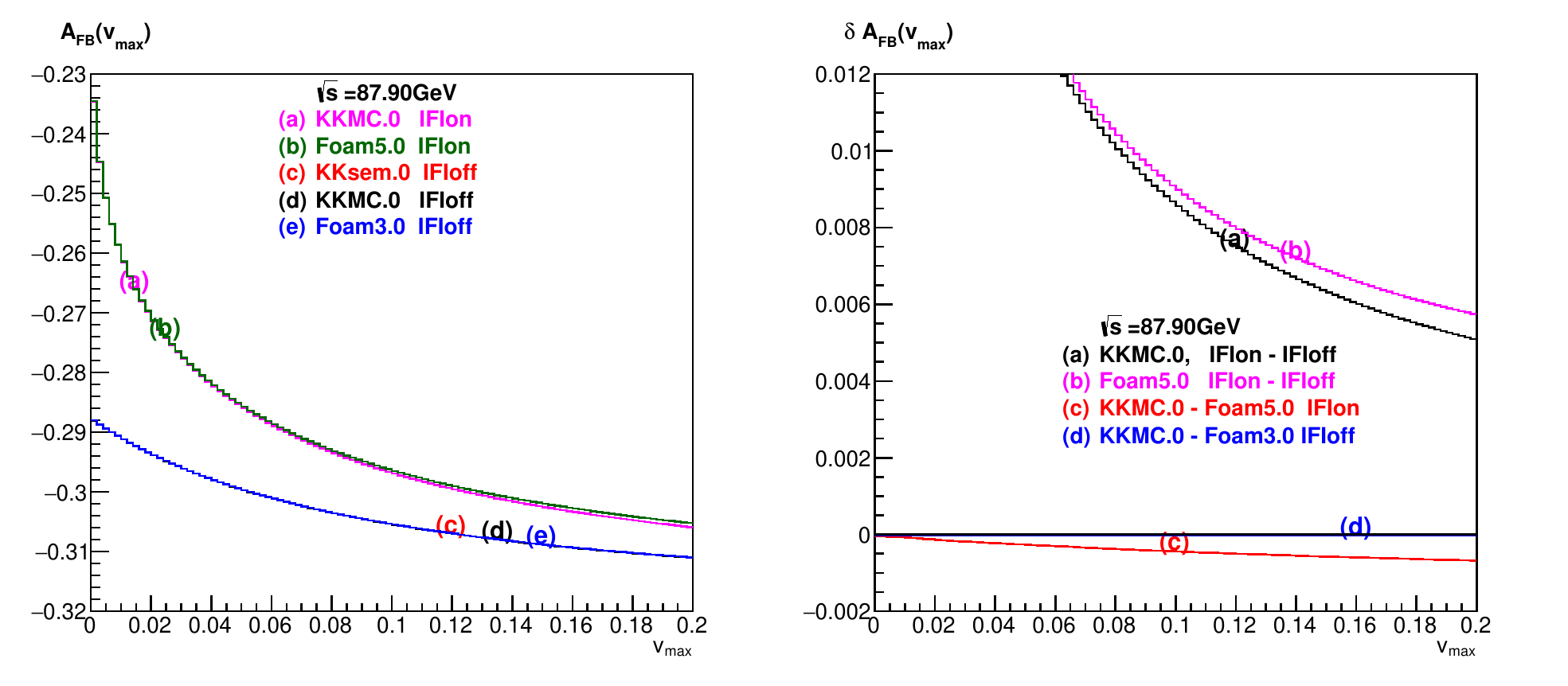}
  \includegraphics[width=165mm,height=66mm]{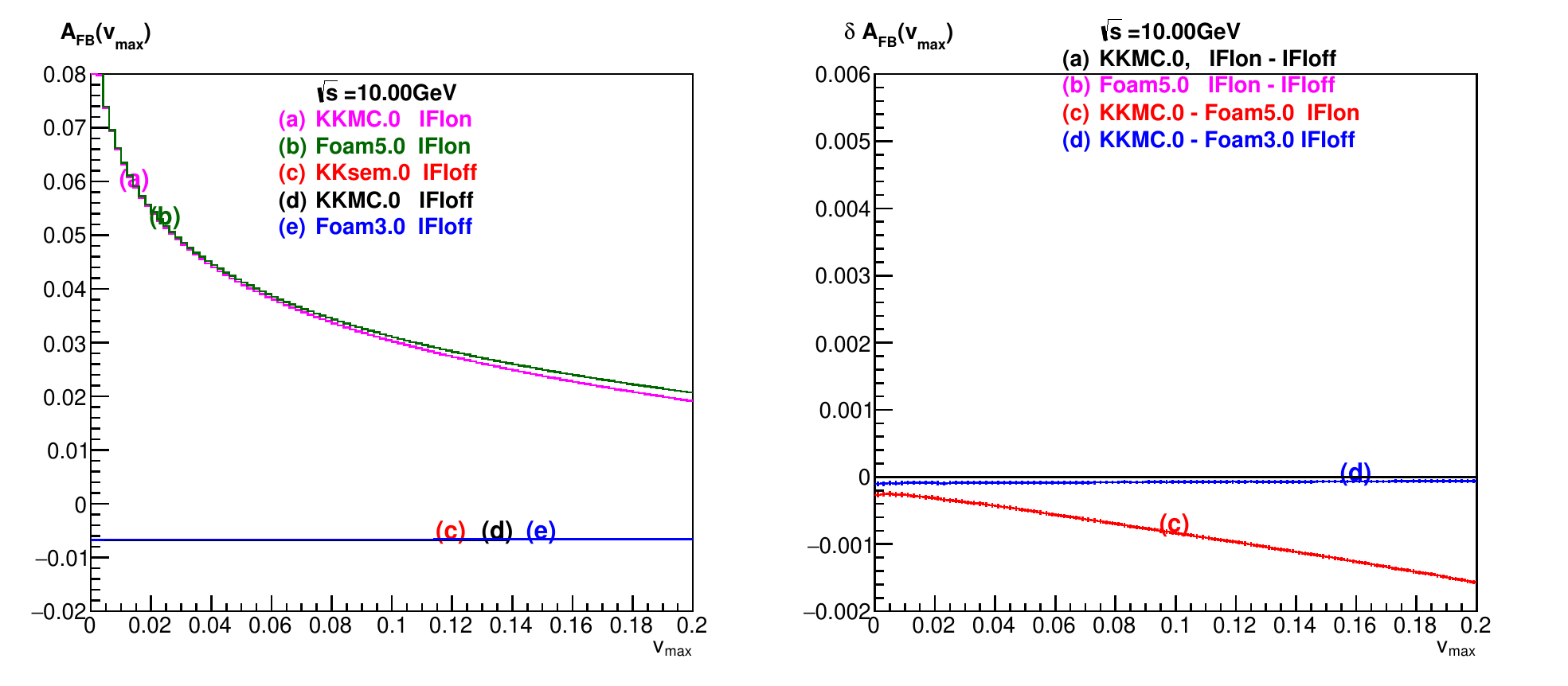}
 \caption{\sf
  Results with ${\cal O}_{\rm exp}(\alpha^0)$ ISR+FSR and IFI
  at 94.3 GeV, 87.9 GeV and 10 GeV.
 }
  \label{fig:FigAfb0}
\end{figure}

\begin{figure}[!t]
  \centering
  \includegraphics[width=165mm,height=66mm]{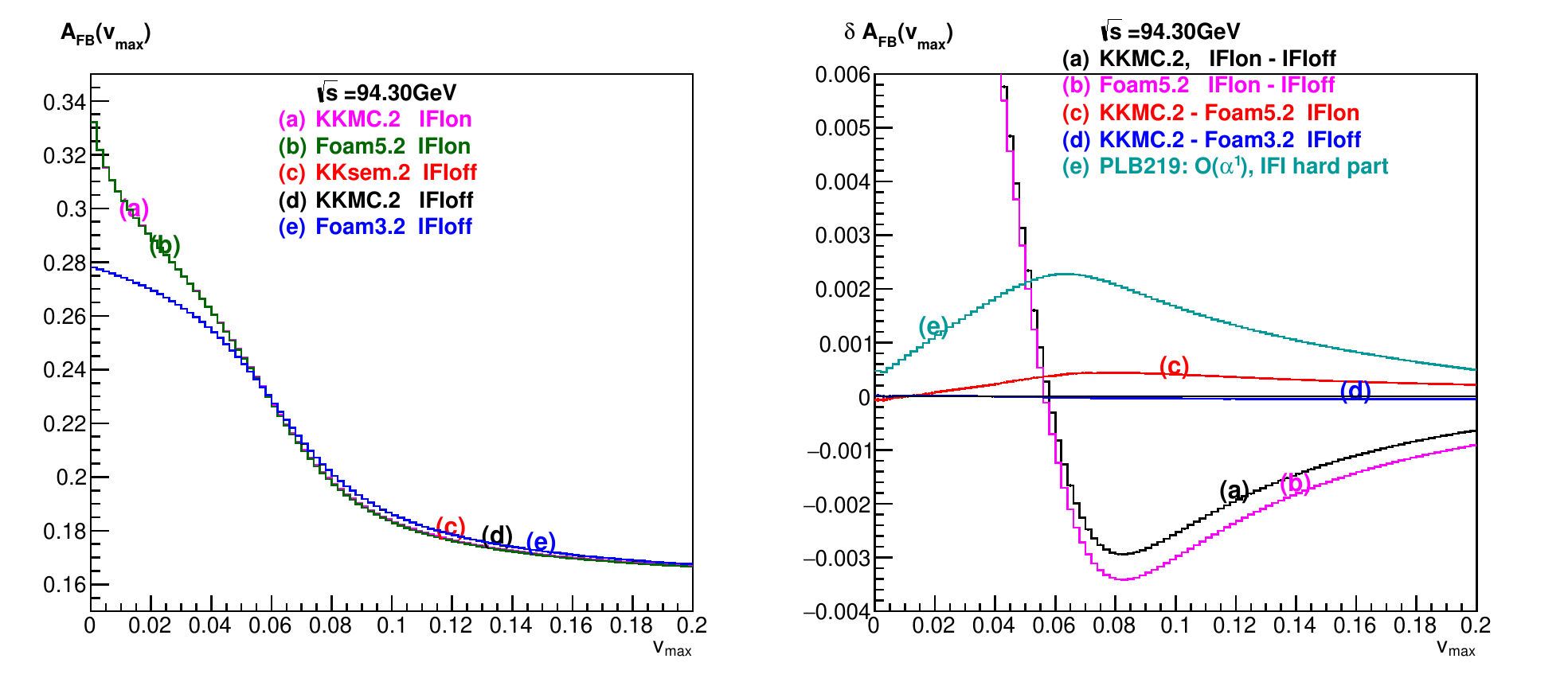}
  \includegraphics[width=165mm,height=66mm]{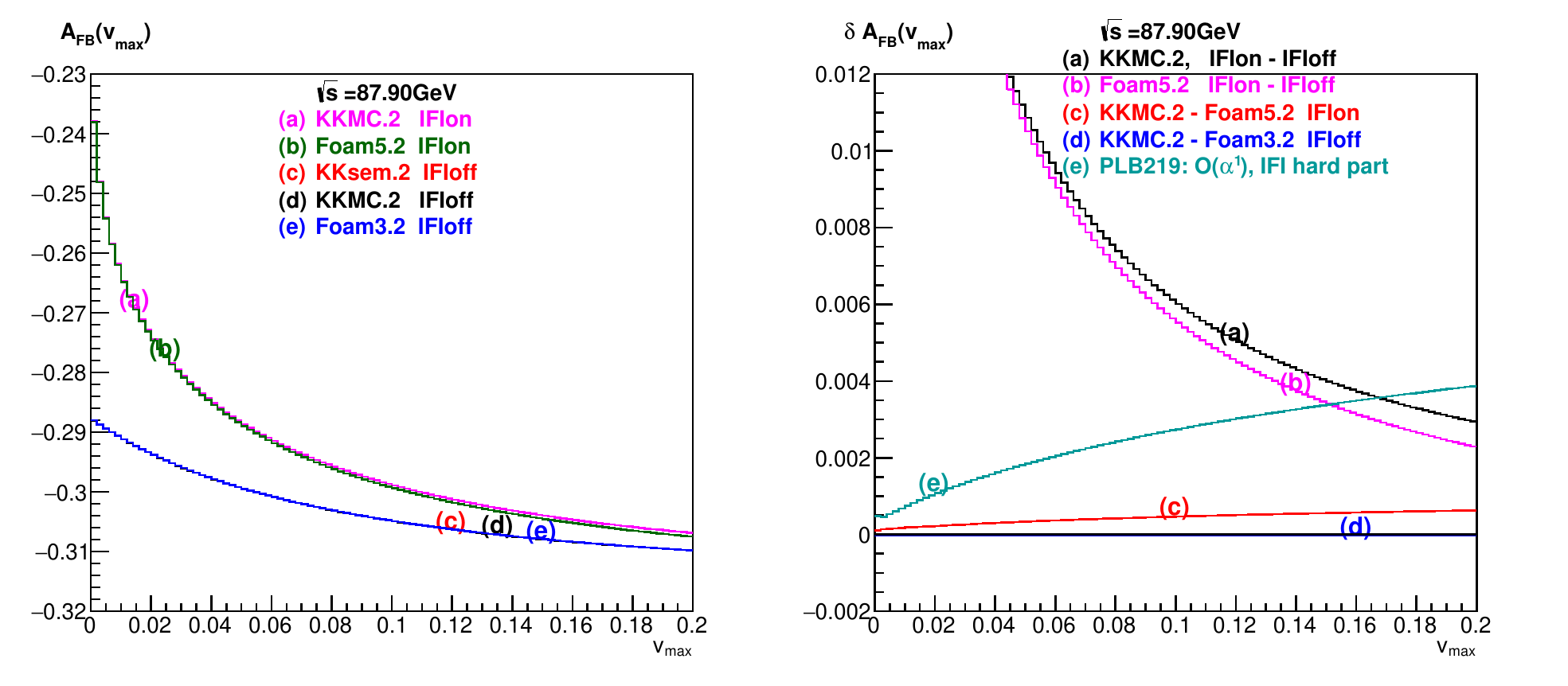}
  \includegraphics[width=165mm,height=66mm]{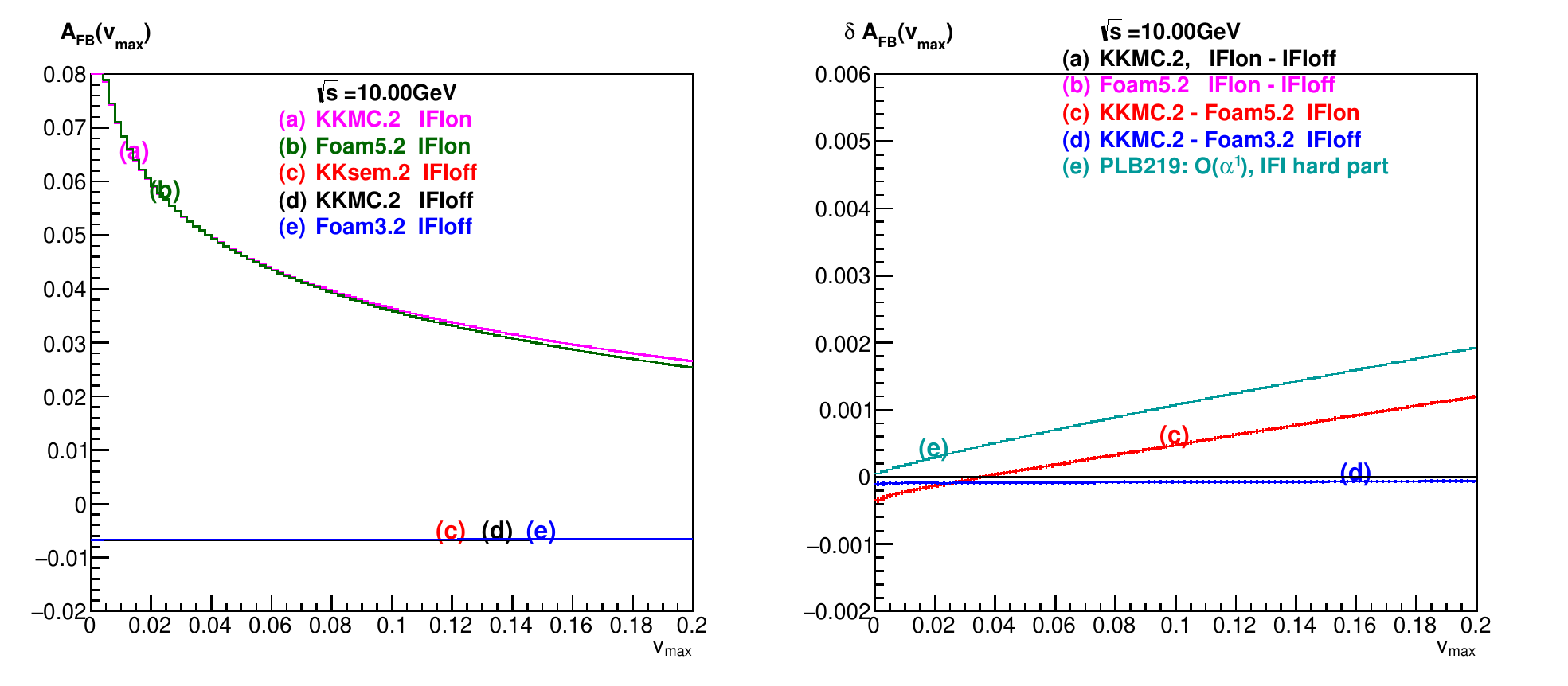}
 \caption{\sf
  Results with ${\cal O}_{\rm exp}(\alpha^2)$ ISR+FSR and  ${\cal O}_{\rm exp}(\alpha^1)$ IFI
  at 94.3, 87.9 and 10 GeV.
 }
  \label{fig:FigAfb2}
\end{figure}

In case (A), results for $\afb(v_{\max})$ from \kkmc\ and \kkfoam\ are
shown in Fig.~\ref{fig:FigAfb0},
while in case (B), the results are shown in Fig.~\ref{fig:FigAfb2},
for energies $\sqrt{s}= 87.9,\; 94.3,\; 10$ GeV in both cases.
The absolute predictions for $\afb$ from \kkmc\ and \kkfoam\ are
seen in the LHS plots of the these figures.
The differences in $\afb$ due to switching on the IFI contribution are
quite sizable and rising quickly for $\vmax \leq 0.06$,
up to 5\% for $\vmax \leq 0.002$.

The IFI contribution to $\afb$ is shown more clearly in the RHS plots of 
Figs.~\ref{fig:FigAfb0} and \ref{fig:FigAfb2}, 
where the absolute predictions for the IFI effect in $\afb$
from \kkmc\ and \kkfoam\ are presented.
The most important result is the one represented by the
red curve (c) in the RHS figures in Fig.~\ref{fig:FigAfb2}.
It represents the difference between \kkmc\ and \kkfoam\ for the IFI contribution.
This crucial difference is up to
$\delta\afb \sim 5\cdot 10^{-4}$.
(It will be analyzed carefully one more time in the next section.)
It is definitely above the technical precision level
$\delta\afb < 3\cdot 10^{-5}$,
established previously in case of IFI switched off.

How can we understand the above result?
In the case of Fig.~\ref{fig:FigAfb0} where both  \kkmc\ and \kkfoam\ 
are at the same ${\cal O}_{\rm exp}(\alpha^0)$ level for ISR, FSR and IFI,
with semisoft resummation of IFI, the source of the difference is a
different treatment of the matrix element 
far away from the infrared point $v_{\max}=0$.
Remembering that
the energy shift in the $Z$-resonance propagator is properly taken into account
in the semisoft approximation,
the difference between \kkmc\ and \kkfoam\ [curve (c)]
should be proportional to $v_{\max}$ and should vanish for $v_{\max}\to0$.
This is what we see in Fig.~\ref{fig:FigAfb0}.%
\footnote{The slight difference at $v_{\max}\to 0$ for $\sqrt{s}=10$ GeV
 can be traced to small spikes in the $0.99<|\cos\theta|<1$ range,
 to be examined separately. It goes away for realistic experimental cutoffs.}

In the case of Fig.~\ref{fig:FigAfb2}, the difference between \kkmc\ and 
\kkfoam\ should reflect the fact that in \kkmc\ the entire
${\cal O}_{\rm exp}(\alpha^1)$ real and virtual contributions are included,
while in \kkfoam5, the ${\cal O}_{\rm exp}(\alpha^1)$ real contribution is 
incomplete.  
This could increase the difference between \kkmc\ and \kkfoam.
In fact it changes sign and increases the difference by at most a factor of 2.  
This can be seen as unexpected.
In order to have an idea how big the ${\cal O}_{\rm exp}(\alpha^1)$ 
real photon IFI contribution can be, 
we have also included this contribution
(curve (e)) in  Fig.~\ref{fig:FigAfb2},
subtracting the soft component in an ad-hoc manner.
As we see, curve (e) typically has the same sign as 
the difference between \kkmc\ and \kkfoam\ shown in curve (c),
but is a factor of $3-4$ bigger.
Apparently, \kkfoam\ includes most of the ${\cal O}(\alpha^1)$ 
hard photon IFI contribution.%
\footnote{It would be interesting to include this missing 
  ${\cal O}_{\rm exp}(\alpha^1)$ real photon IFI contribution in \kkfoam.}

The inclusion of QED  ${\cal O}_{\rm exp}(\alpha^1)$ virtual corrections and
box diagrams was done in \kkfoam\
following the prescription of Eq.~(\ref{eq:boxing}).
The pure ${\cal O}(\alpha^1)$ numerical results
in  Fig.~\ref{fig:FigAfb2} were reproduced
using analytic formulas of Refs.~\cite{Jadach:1988zp,Was:1989ce},
which are collected and tested numerically
one more time in Appendix~\ref{sec:appendixC}.

In spite of the incompleteness 
of the ${\cal O}(\alpha^1)$ IFI in \kkfoam, the above result
makes us confident that we are quite close to reaching our first
intermediate goal of controlling the IFI effect in $\afb$
at the level of $\delta\afb \sim 10^{-4}$
in the semisoft resummation regime ($v_{\max} \leq 0.06$).

\begin{figure}[!t]
  \centering
  \includegraphics[width=165mm]{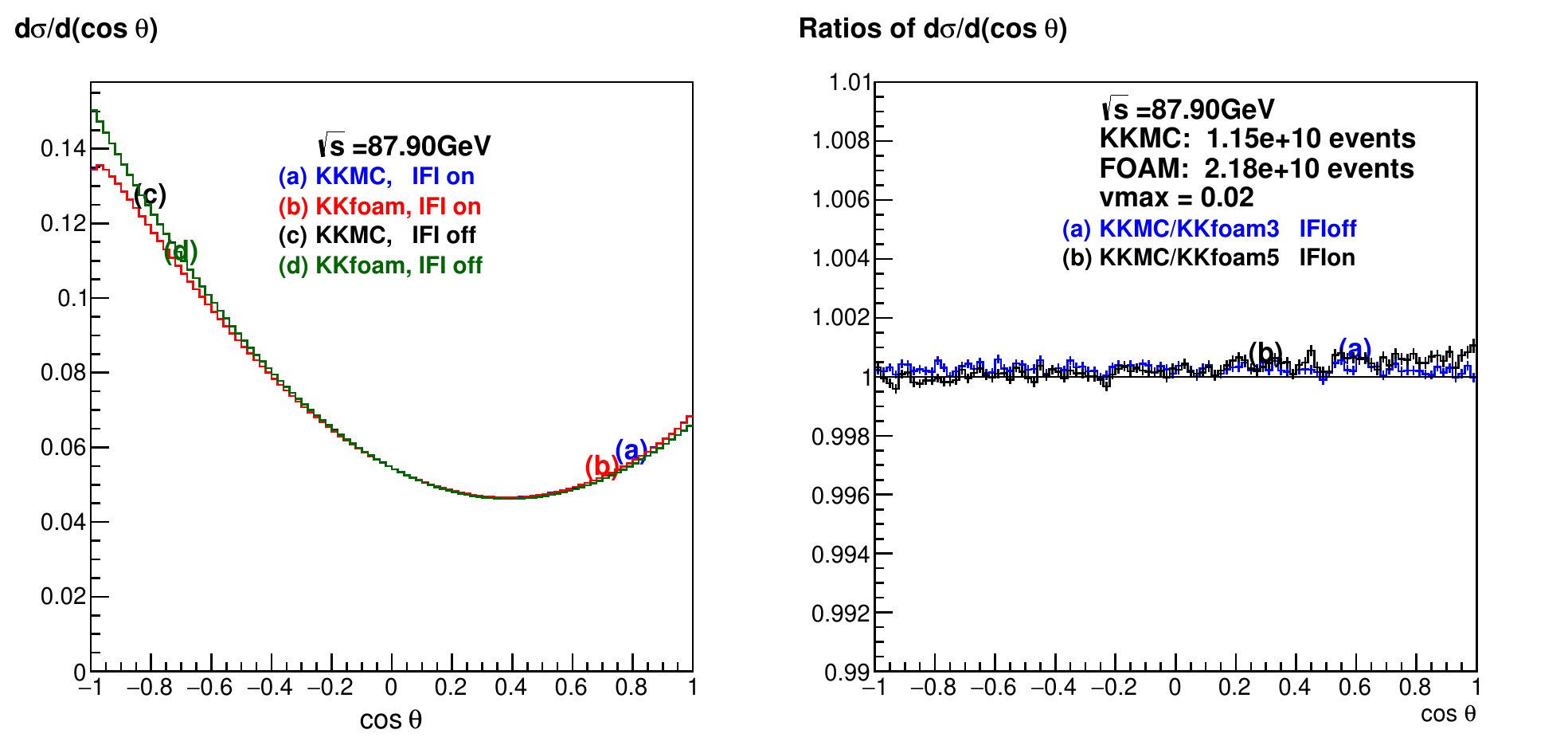}
\caption{\sf
  Comparison between \kkmc\ and \kkfoam\ for the angular distribution
  for the cutoff on total photon energy $v_{\max}=0.02$.
 }
  \label{fig:FigCosThe2}
\end{figure}

Let us also finally show just one example of the entire angular
distribution $d\sigma/d\cos\theta$ from \kkfoam\ and \kkmc,
simply because agreement in $\afb$ does not necessarily imply 
agreement in the angular distributions.
In Fig.~\ref{fig:FigCosThe2}, such a comparison is done for a
relatively mild cutoff $v_{\max}=0.02$ on the total photon energy.
The angular distributions agree to within 0.005\% as expected.

\subsection{$\afb(s_\pm)$ from \kkmc\ and \kkfoam\ in presence of IFI.}
\label{subsec:AFBdif}
As explained in Ref.~\cite{Janot:2015gjr}, the QED coupling constant
$\alpha_{\rm QED}(M_Z)$  is closely related to
$\afb(s_\pm)$, but the exact relation is not straightforward
and we are not trying to reproduce it.
We limit our interest to the propagation of errors from $\afb(s_\pm)$
to $\alpha_{\rm QED}(M_Z)$, which is simpler and can be read from Eq.~(4.9)
in Ref.~\cite{Janot:2015gjr}.
For our purpose, it will be enough to use a simplified version of this equation,
\begin{equation}
  \frac{\delta \alpha_{\rm QED}}{\alpha_{\rm QED}}\Bigg|_{M_Z}
  \simeq \frac{ \delta \afb(s_+) - \delta \afb(s_-)}{\afb(s_+) - \afb(s_-) },
\end{equation}
which is valid for small $\delta \afb(s_\pm)$ and/or when there are no
strong cancellations between them.%
\footnote{We thank P. Janot for pointing this out to us.}
This will be true in the following numerical examples,
 and we shall show typically 
the numerator $\delta \afb(s_+) - \delta \afb(s_-)$
along with the uncertainties  $\delta \afb(s_\pm)$.

Having the above in mind we reexamine the comparisons 
between \kkmc\ and \kkfoam\ of the previous section for this $\Delta\afb$.

\begin{figure}[!t]
  \centering
  \includegraphics[width=165mm]{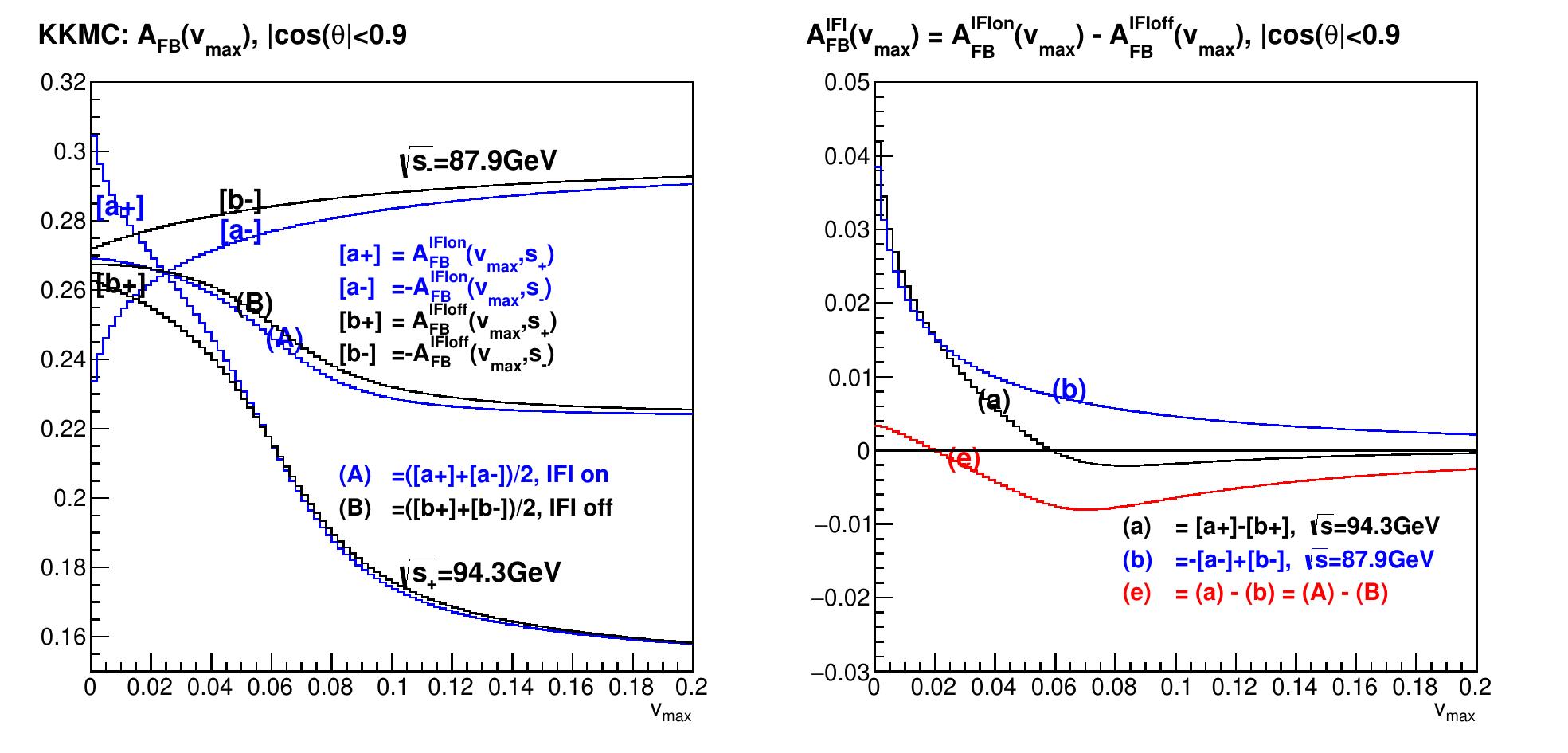}
\caption{\sf
  $\afb(v_{\max},s_\pm)$ from \kkmc\ with ${\cal O}_{\rm exp}(\alpha^2)$ ISR+FSR and  
  ${\cal O}_{\rm exp}(\alpha^1)$ IFI.
 }
  \label{fig:AfbIFIvT2}
\end{figure}

From now on, we impose a realistic cutoff $|\cos\theta|<0.9$ in the tests;
however the cutoff has little influence on the resulting $\afb$.
To start with, in Fig.~\ref{fig:AfbIFIvT2}
we show $\afb(v_{\max})$ from \kkmc\ at $\sqrt{s_\pm}$
with IFI switched on/off and with the best QED matrix element in \kkmc.
The $\afb$ changes sign between these two energies.
On the other hand, IFI keeps the same sign, hence we expect partial
cancellation of the IFI effect in the $\alpha_{\rm QED}(M_Z)$.
We do not pursue the reconstruction of  $\alpha_{\rm QED}(M_Z)$
and only plot the difference of IFI effect between two energies
in the LHS of Fig.~\ref{fig:AfbIFIvT2} as a guide.

\begin{figure}[!t]
  \centering
  \includegraphics[width=90mm]{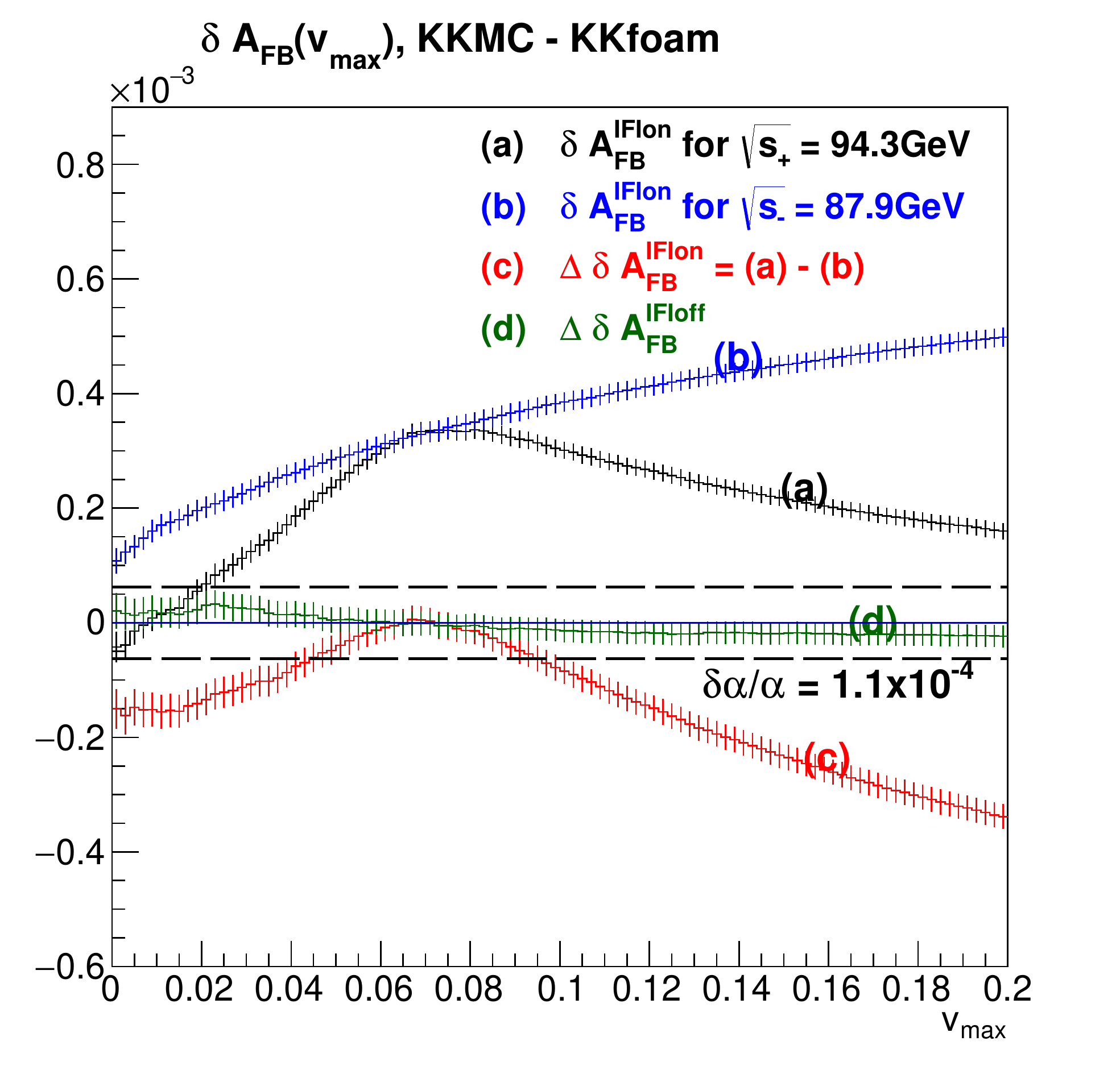}
\caption{\sf
  Difference $\delta\afb(v_{\max})$ between $\afb$ from \kkmc\ and \kkfoam\
  and their difference between two energies $\sqrt{s_\pm}$. IFI is switched
  on/off in both \kkmc\ and \kkfoam.
  The band marked with dashed line corresponds to the precision
  estimate of the $\alpha_{\rm QED}(M_Z)$ of Ref.~\cite{Davier:2010nc}.
 }
  \label{fig:AfbDifPat}
\end{figure}

We have produced the same figure for \kkfoam, but we do not show it
here, because it looks essentially the same as Fig.~\ref{fig:AfbIFIvT2}.
What is more interesting is to reexamine the difference between
\kkmc\ and \kkfoam\
\begin{equation}
 \delta A_{\rm FB}(s_\pm) = A_{\rm FB}(s_\pm)\big|_{\kkmc}-A_{\rm FB}(s_\pm)\big|_{\kkfoam},
\end{equation}
already shown in curve (c) of Fig.~\ref{fig:FigAfb2},
and its difference between two energies $\sqrt{s_\pm}$
\begin{equation}
 \Delta \delta A_{\rm FB} = \delta A_{\rm FB}(s_+) -  \delta A_{\rm FB}(s_-)
\end{equation}
relevant for the uncertainty in the measurement of $\alpha_{\rm QED}(M_Z)$.
We are interested in the above quantity primarily for IFI switched on.
This quantity is shown in Fig.~\ref{fig:AfbDifPat}, see curve (c) there.
It turns out that $\Delta \delta A_{\rm FB} \leq 2\cdot 10^{-4}$ 
within the interesting range of photon energy cutoff $v_{\max}\leq 0.1$.
In Fig.~\ref{fig:AfbDifPat}, we have also included two
dashed lines%
\footnote{The dashed lines of the band are at  
 $\pm 1.1 \cdot 10^{-4}\; |A_{\rm FB}(s_+)-A_{\rm FB}(s_+)|_{v_{\max}\to 0} = \pm0.57 \cdot 10^{-4}$.
}
marking the band of the present uncertainty 
$\delta \alpha_{\rm QED} /  \alpha_{\rm QED}(M_Z) = 1.1 \cdot 10^{-4}$
according to Ref.~\cite{Davier:2010nc}.
The aim of FCCee is of course to get substantially smaller error than that.

The main contribution to $\Delta \delta A_{\rm FB}$ in curve (c) comes
from the uncertainty
in the IFI implementation (most likely in \kkfoam),
as can be seen from curve (d) in Fig.~\ref{fig:AfbDifPat},
which represents $\Delta \delta A_{\rm FB}$ for IFI switched off.
The aim of future work will be to get
$\Delta \delta A_{\rm FB} \leq 3\cdot 10^{-5}$ for IFI on,
that is to the same level as for IFI off,
in the semisoft regime $v_{\max}\leq 0.06$.

The above 
$\Delta \delta A_{\rm FB} \leq 2\cdot 10^{-4}$
can be treated as an (over)conservative estimate of the
uncertainty of the IFI prediction for \kkmc\
in the semisoft regime, which is much better
than the LEP-era estimate but still not up to the needs of FCCee.
A less conservative estimate will be provided in the next section.

\subsection{On $\afb$ for ${\cal O}_{\rm exp}(\alpha^i)$, $i=0,1,2$ in \kkmc}
\label{subsec:AFBord}
The differences between \kkmc\ and \kkfoam\ provide much valuable information,
because the two programs differ quite a lot technically 
(MC soft photon phase space integration versus analytic integration),
while implementing the same physics of QED corrections.
However, \kkmc\ alone offers interesting insight into missing higher order
QED corrections related to IFI.%
\footnote{Provided we trust the smallness of the technical precision 
error of \kkmc.}

\begin{figure}[!t]
  \centering
  \includegraphics[width=165mm]{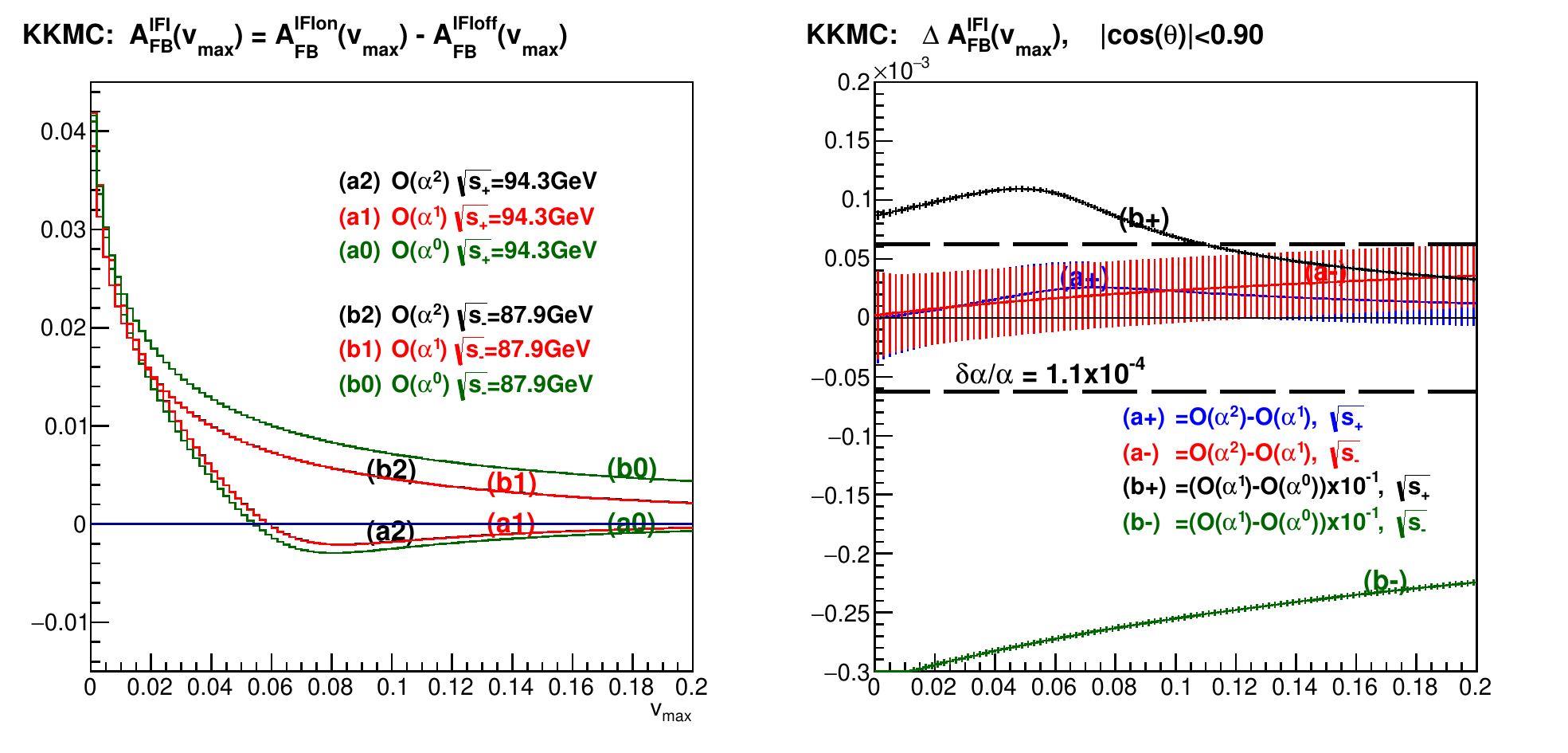}
\caption{\sf
  IFI component in $\afb(s_\pm)$ obtained using \kkmc\ program
  with three types of increasingly sophisticated QED matrix element,
  ${\cal O}_{\rm exp}(\alpha^i)$, $i=0,1,2$.
  The band between the dashed lines corresponds to the precision
  estimate of the $\alpha_{\rm QED}(M_Z)$ of Ref.~\cite{Davier:2010nc}.
 }
  \label{fig:AfbIFI_KKmc4}
\end{figure}

In \kkmc, one may choose three types of the QED multiphoton matrix element
with resummation at increasing sophistication levels, 
${\cal O}_{\rm exp}(\alpha^i)$, $i=0,1,2$.
In Fig.~\ref{fig:AfbIFI_KKmc4} we examine differences
in the IFI contribution $A_{\rm FB}^{IFI}(v_{\max})$ 
between ${\cal O}(\alpha^2)$ and ${\cal O}(\alpha^1)$
and also between ${\cal O}(\alpha^1)$ and ${\cal O}(\alpha^0)$.
In all of them, IFI may be switched on or off.
Complete non-IR  ${\cal O}(\alpha^1)$ corrections are included in the
${\cal O}(\alpha^i),\; i=1,2$ case while in the ${\cal O}(\alpha^0)$
case, only the IR part of exponentiated IFI is implemented.
In the most sophisticated case of the ${\cal O}_{\rm exp}(\alpha^2)$ QED 
matrix element in \kkmc,
only pure nonlogarithmic photonic corrections are missing.%
\footnote{In particular, the non-IR parts of QED pentaboxes are missing; 
  see Fig.~5 in Ref.~\cite{Jadach:2000ir}.}

In the LHS of Fig.~\ref{fig:AfbIFI_KKmc4}, we show plots of the IFI component
for all three cases ${\cal O}_{\rm exp}(\alpha^i),\; i=0,1,2$,
while in the RHS we see the differences, for the two energy 
points $\sqrt{s_\pm}$.%
\footnote{Differences in Fig.~\ref{fig:AfbIFI_KKmc4} are obtained using MC weights
 event per event, so statistical errors are grossly overestimated.
 This explains the lack of fluctuations among bins.}

The most important difference in Fig.~\ref{fig:AfbIFI_KKmc4}, 
between $\afb$ for
${\cal O}(\alpha^2)$ and ${\cal O}(\alpha^1)$, is below the
statistical error of $10^{-4}$.
This can be treated as a measure of the missing QED photonic higher order
corrections in the \kkmc\ predictions for $\afb$ for this particular type of
experimental cutoff, $v_{\max}<0.2$ and $|\cos\theta|<0.9$,
near $Z$ resonance, $|M_Z-\sqrt{s}| \leq 3.5$~GeV.

Finally, let us remark that the MC results for $\afb$ presented here
with a statistical precision of $10^{-4}$ were obtained using $\sim 10^{10}$
MC events generated in parallel runs on PC farms.
Reducing the statistical error to $10^{-5}$ will be feasible, but not trivial.
However, higher precision may be also feasible with less MC events
using the technique of recording differences of the MC weights, 
as it was done in some plots shown in the following.

\subsection{More on the uncertainty of the ISR effect in $\afb$.}
\label{subsec:ISRinAFB}
In this section, we will present a few results from \kkmc\ 
which, in particular, will
give us more insight on the ISR effects in $\afb$ when 
IFI is switched on and off.
 
\begin{figure}[!t]
  \centering
  \includegraphics[width=165mm]{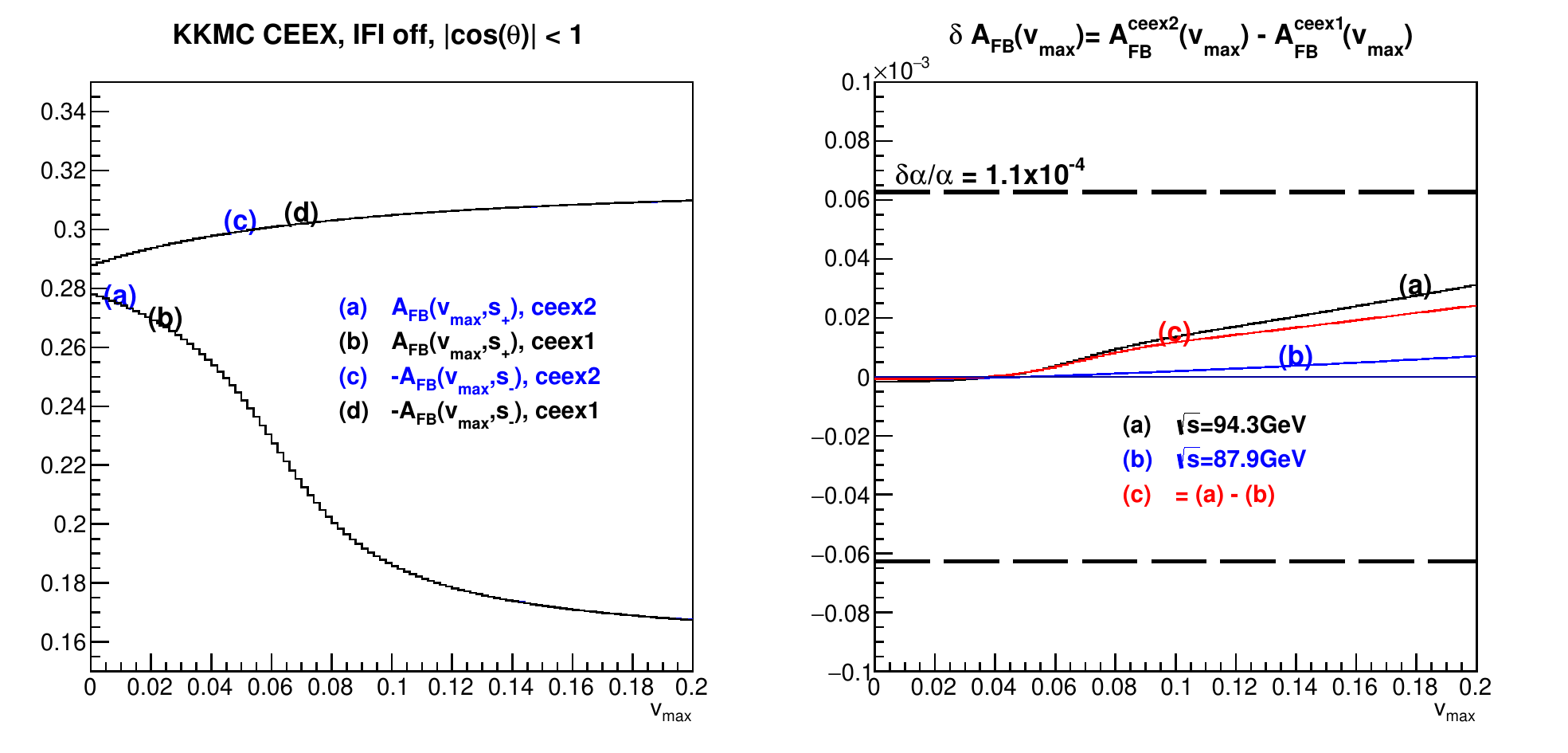}
\caption{\sf
  Differences between $\afb$ calculated using CEEX matrix element 
  ${\cal O}_{\rm exp}(\alpha^i)$, $i=1,2$ with IFI switched off.
  The band between the dashed lines represents the precision
  estimate of $\alpha_{\rm QED}(M_Z)$ of Ref.~\cite{Davier:2010nc}.
 }
  \label{fig:Afb_ceex21_wtd}
\end{figure}

In Fig.~\ref{fig:Afb_ceex21_wtd}, we show differences
between ${\cal O}_{\rm exp}(\alpha^i)$, $i=1,2$ results
from \kkmc\ with the CEEX matrix element in the case of IFI switched off --
that is pure ISR and FSR effects.
In fact, the ISR effect is dominant here.
The variation is smaller than $3\cdot 10^{-5}$ and cancels
between the two energies $\sqrt{s_\pm}$.%
\footnote{Such a cancellation of the ISR effect 
  was already noticed in Ref.~\cite{Janot:2015gjr}.}

\begin{figure}[!t]
  \centering
  \includegraphics[width=165mm]{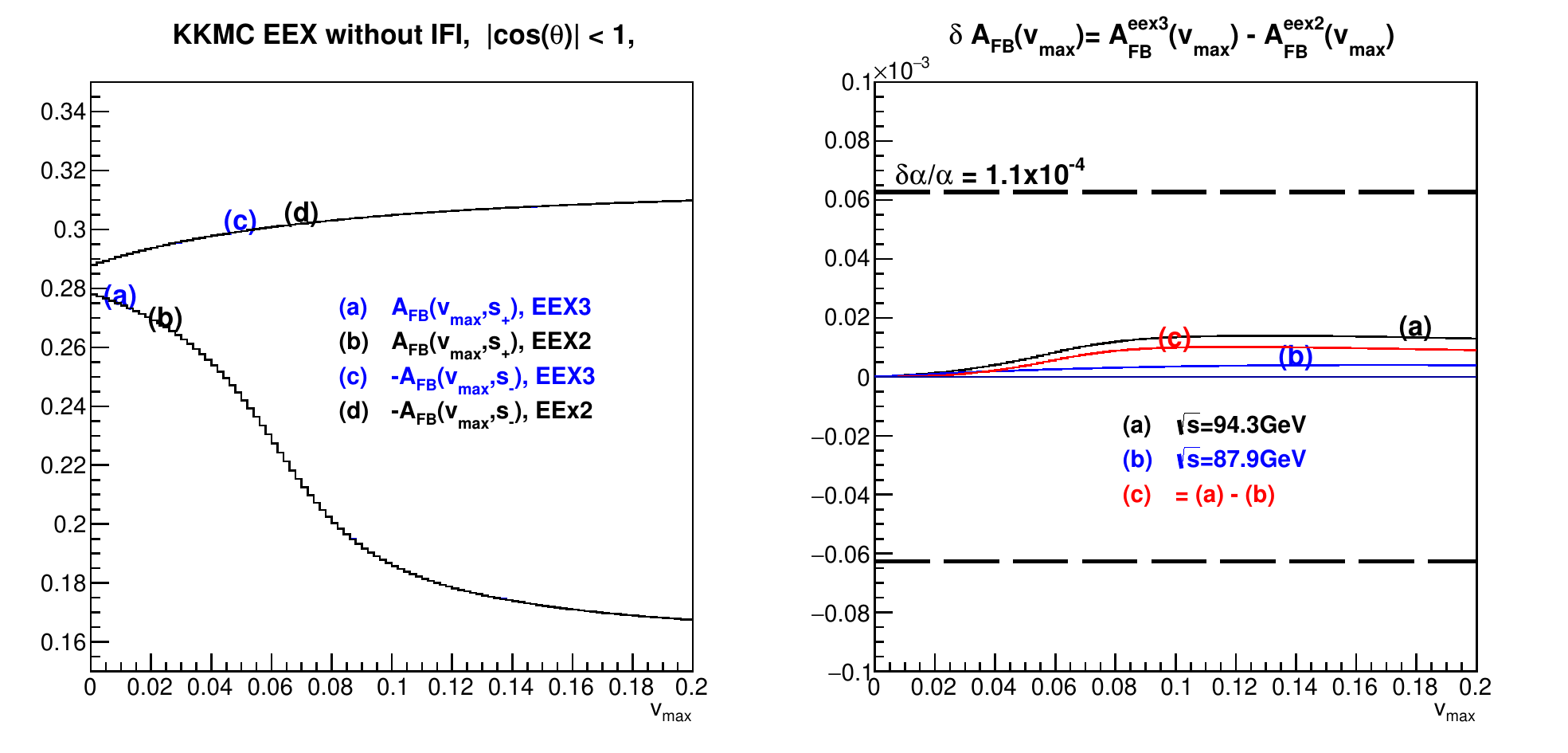}
\caption{\sf
  Differences between $\afb$ calculated using the EEX matrix element 
  ${\cal O}_{\rm exp}(\alpha^i)$, $i=2,3$, which is without IFI.
 }
  \label{fig:Afb_eex32_wtd}
\end{figure}

The same phenomenon is seen in Fig.~\ref{fig:Afb_eex32_wtd},
albeit the differences are smaller, as expected.
Note also that in both of the above cases, the effect of ISR
is completely negligible for $v_{\max}\leq 0.05$,
that is for cutoffs on photon energy interesting experimentally!

\begin{figure}[!t]
  \centering
  \includegraphics[width=165mm]{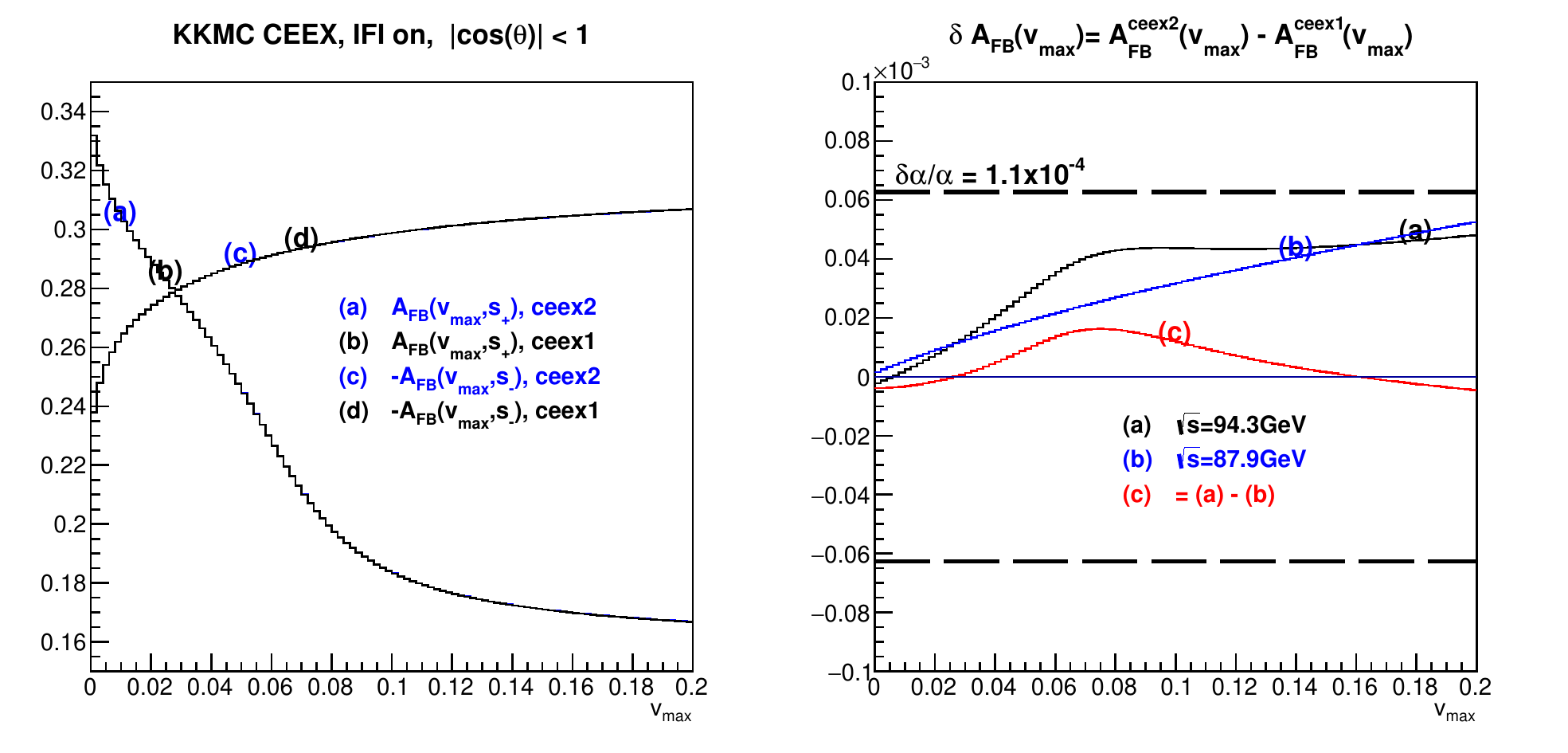}
\caption{\sf
  Differences between $\afb$ calculated using the CEEX matrix element 
  ${\cal O}_{\rm exp}(\alpha^i)$, $i=1,2$
  with IFI switched on.
 }
  \label{fig:Afb_ceex21_ifi_wtd}
\end{figure}

Finally, we switch on IFI and examine again the differences
between ${\cal O}_{\rm exp}(\alpha^i)$, $i=1,2$ results
from \kkmc\ with the CEEX matrix element in the case of IFI switched on.
The results are shown in Fig.~\ref{fig:Afb_ceex21_ifi_wtd}.
This is the most interesting result, because it shows
the indirect influence of ISR on the IFI contribution to $\afb$.
Curve (c) shows that for the difference in $\afb$ between the two 
energies $\sqrt{s_\pm}$, the first and second order results agree 
to within $\leq 2\cdot 10^{-5}$.
The disagreement is larger than was seen in the previous graph  with IFI off
in the semisoft region $v_{\max}\leq 0.06$.

One may conclude that the above result provides us strong indication that
the QED uncertainty in $\afb$ from \kkmc\ is 
of the order of the expected FCCee experimental error $\delta\afb \simeq 3\cdot 10^{-5}$.
In the above plots, statistical MC errors are negligible,
because all differences between the various QED matrix elements
are calculated using differences between the weights for the same sets of
weighted MC events.

\begin{table}[ht!]
\centering
\begin{tabular}{|c|c|c|c|c|c|c|}
\hline\hline
 No. & Uncert. & IFI & Source  & Ref. & $v_{\max} \leq 0.05$ & $v_{\max} \leq 0.20$ \\
\hline\hline
\hline
1. &  Tech. & OFF &$ |\afb(s_{\pm}) |_{\tt KKMC_{ceex0}-KKsem0} $  
            & Fig.~\ref{fig:cFigSigAfb0}  & $2\cdot 10^{-5}$ 
                                          & $2\cdot 10^{-5}$ \\
\hline
2. &  Tech. & OFF &$ |\afb(s_{\pm}) |_{\tt KKMC2_{ceex2}-KKsem2} $  
            & Fig.~\ref{fig:cFigSigAfb2}  & $2\cdot 10^{-5}$ 
                                          & $3\cdot 10^{-5}$ \\
\hline
3. &  Phys. & ON &$ |\Delta A^{\rm IFI}_{\rm FB} |_{\tt KKMC_{cce2}-KKfoam5} $  
            & Fig.~\ref{fig:AfbDifPat}  & $2\cdot 10^{-4}$ 
                                        & $3\cdot 10^{-4}$ \\
\hline
4. &  Phys. & ON &$ | A^{\rm IFI}_{\rm FB}(s_+)  |_{\rm CEEX2-CEEX1} $  
            & Fig.~\ref{fig:AfbIFI_KKmc4}  & $0.3\cdot 10^{-4}$ 
                                                 & $0.40\cdot 10^{-4}$ \\
\hline
5. &  Phys. & ON &$ | A^{\rm IFI}_{\rm FB}(s_-)  |_{\rm CEEX2-CEEX1} $  
            & Fig.~\ref{fig:AfbIFI_KKmc4}  & $0.3\cdot 10^{-4}$ 
                                                 & $0.30\cdot 10^{-4}$ \\
\hline
6. &  Phys. & OFF &$ |\Delta A_{\rm FB}^{\rm CEEX2}-\Delta A_{\rm FB}^{\rm CEEX1}|$  
            & Fig.~\ref{fig:Afb_ceex21_wtd}  & $ 0.01\cdot 10^{-5}$ 
                                             & $ 0.20\cdot 10^{-4}$ \\
\hline
7. &  Phys. & ON &$ |\Delta A_{\rm FB}^{\rm CEEX2}-\Delta A_{\rm FB}^{\rm CEEX1}|$  
            & Fig.~\ref{fig:Afb_ceex21_ifi_wtd}  & $0.15\cdot 10^{-4}$ 
                                           & $0.15\cdot 10^{-4}$ \\
\hline\hline
\end{tabular}
\caption{\sf
 Table of the most important estimates of technical and physical uncertainties 
of $\afb$ due to IFI.
 We define $\Delta \afb \equiv \afb(s_+) - \afb(s_-)$.
The quoted numbers represent maximum values for the $v_{\max}\in(0.0-0.05)$ 
and $v_{\max}\in(0.0-0.20)$ ranges in the corresponding figures.
}
\label{tab:table2}
\end{table}

\newpage

\section{Summary and outlook}
\label{sec:Summary}
The extensive numerical results
presented in this work and summarized in Table~\ref{tab:table2} allow us to
conclude that the technical and physical uncertainty of the prediction of \kkmc\
for the IFI component of $\afb$ near the $Z$ resonance is of order $\sim 10^{-4}$
(row 3 in Table~\ref{tab:table2}).
This is definitely better than the state of art in the LEP era 
$\sim 2\cdot 10^{-3} - 3\cdot 10^{-3}$.
For IFI switched off the technical precision $\sim 10^{-5}$ was obtained
(rows $1-2$ in Table~\ref{tab:table2}).
Some of the results presented here indicate that the physical precision (higher orders)
of  the IFI component in $\afb$ near the $Z$ resonance from \kkmc\ is in fact
at the level $\sim 1.5 \cdot 10^{-5}$ 
(row 7 in Table~\ref{tab:table2}).
{\em I.e.}, what is needed in the FCCee experiment proposed to
measure the QED coupling constant at the scale $M_Z$ with this precision.
This would allow $\alpha_{\rm QED}(M_Z)$ to be determined to a precision
significantly better than the present estimate of Ref.~\cite{Davier:2010nc},
which is $\frac{\delta\alpha_{\rm QED}}{\alpha_{\rm QED}} \simeq 1.1\cdot 10^{-4}$.
However, more work is needed for achieving better confidence
in the technical precision and higher order
photonic QED corrections in the \kkmc\ results.
More work is also needed to estimate other missing nonphotonic QED corrections
({\em e.g.} pair emission) and electroweak corrections.
Extension of the presented analysis to more realistic
experimental selections (cuts) is also desirable.

The newly developed auxiliary MC program \kkfoam\ was instrumental in
the above achievement.  For more precise tests of \kkmc,
it would be profitable to include in the phase space of \kkfoam\ 
the exact contribution from nonsoft real ${\cal O}(\alpha^1)$ emission
matched with semisoft analytic resummation.

\vspace{10mm}
\noindent {\bf\Large Acknowledgment }\\
The authors would like to thank Patrick Janot, Maciej Skrzypek
and Zbigniew W\c{a}s
for critical reading of the manuscript and useful comments.

\newpage
\appendix
\noindent {\bf\Large APPENDIX}

\section{Factorizing the exponentiated formula}
\label{appendixA}
Starting from Eq.~(\ref{eq:expon0}), let us introduce 
$\int d^4z\;  \delta^4(z-y-x+x')=1$,
$\int d^4u\;  \delta^4(u-y+x)  =1 $ and
$\int d^4u'\; \delta^4(u-y+x') =1 $
in the Fourier expression, obtaining
\begin{equation}
\begin{split}
&\sigma(s)=
\frac{1}{{\rm flux}(s)}
\sum_{V,V'}
\int \frac{d^3 q_1}{ q_1^0} \frac{d^3 q_2}{ q_2^0}\;
\frac{ d^4 Q  d^4 x }{(2\pi)^4}
\frac{ d^4 Q' d^4 x'}{(2\pi)^4}
e^{ix \cdot (P-Q) -ix' \cdot (P-Q')}
\frac{ d^4 y}{(2\pi)^4} e^{ iy \cdot (P-q_1-q_2) }
\\&~~\times
\int \frac{ d^4 K d^4 z}{ (2\pi)^4}\;
     \frac{ d^4 R d^4 u}{ (2\pi)^4}\;
     \frac{ d^4 R' d^4 u'}{ (2\pi)^4}\;
e^{ i(z-y-x+x')\cdot K + i(u-y-x) \cdot R +i(u'-y +x')\cdot R' }
\\&~\times 
\exp\Big\{ \int \frac{d^3 k}{ k^0}\;
\big[ e^{-ik \cdot (y+x-x')} S_I(k) 
     +e^{-ik \cdot (y+x)}    S_X(k)
     +e^{-ik \cdot (y-x')}   S_X(k)
     +e^{-ik \cdot y}       S_F(k)
 \big]
\Big\}
\\&~\times
\exp\big\{
  \alpha B_4^V(Q^2,t,m_\gamma) +\alpha( B_4^{V'}(Q'^2,t,m_\gamma))^*
\big\}\;
\Mcal_V(Q,t) \Mcal^*_{V'}(Q',t)\;
\end{split}
\label{eq:exponA1}
\end{equation}
The lowest-order spin amplitudes $\Mcal_V,\; V=\gamma,Z$ 
are, up to a normalization constant,
equal to the amplitudes $\Mmf_{\veps\tau}^V$ defined in 
Appendix \ref{sec:appendixC},
but fermion helicities are temporarily suppressed.

Thanks to the above reorganization, we may clearly factorize the result into
contributions due to the ISR, FSR, and IFI components of 
multiphoton emission:
\begin{equation}
\begin{split}
&\sigma(s)=
\frac{1}{{\rm flux}(s)}
\int \frac{d^3 q_1}{ q_1^0} \frac{d^3 q_2}{ q_2^0}\;
\frac{ d^4 Q d^4 x}{(2\pi)^4}
\frac{ d^4 Q' d^4 x'}{(2\pi)^4}
e^{ix \cdot (P-Q) -ix' \cdot (P-Q')}
\sum_{V,V'}
\\&\times
\int d^4 K\;
     d^4 R\;
     d^4 R'\;
e^{ i(-x+x')\cdot K + i(-x) \cdot R +i( +x')\cdot R' }
\\&\times
\int \frac{d^4z }{ (2\pi)^4} \;
e^{ iz\cdot K +\int \frac{d^3 k}{ k^0}\; e^{-ik \cdot z}  S_I(k) }
\int \frac{ d^4u}{ (2\pi)^4} \;
e^{ iu\cdot R +\int \frac{d^3 k}{ k^0}\; e^{-ik \cdot u}  S_X(k) }
\\&\times
\int \frac{d^4u'}{ (2\pi)^4}\;
e^{ iu'\cdot R' +\int \frac{d^3 k}{ k^0}\; e^{-ik \cdot u'} S_X(k) }
\int \frac{d^4y}{ (2\pi)^4} \;
e^{  iy \cdot (P-q_1-q_2 -K -R -R')  +\int \frac{d^3 k}{ k^0}\; e^{-ik \cdot y}  S_F(k) }
\\&\times
\exp\big\{
  \alpha B_4^V(Q^2,t,m_\gamma) +\alpha( B_4^{V'}(Q'^2,t,m_\gamma))^*
\big\}\;
\Mcal_V(Q) \Mcal^*_{V'}(Q')
\\&=
\frac{1}{{\rm flux}(s)}
\sum_{V,V'}
\int \frac{d^3 q_1}{ q_1^0} \frac{d^3 q_2}{ q_2^0}\;
d^4 Q\;  d^4 Q'\;
\\&\times
\int d^4 K\;
     d^4 R\;
     d^4 R'\;
\delta^4(P-K-R-Q) \delta^4(P-K-R'-Q')
\\&\times
\int \frac{d^4z }{ (2\pi)^4} \;
e^{ iz\cdot K +\int \frac{d^3 k}{ k^0}\; e^{-ik \cdot z}  S_I(k) }
\int \frac{ d^4u}{ (2\pi)^4} \;
e^{ iu\cdot R +\int \frac{d^3 k}{ k^0}\; e^{-ik \cdot u}  S_X(k) }
\\&\times
\int \frac{d^4u'}{ (2\pi)^4}\;
e^{ iu'\cdot R' +\int \frac{d^3 k}{ k^0}\; e^{-ik \cdot u'} S_X(k) }
\int \frac{d^4y}{ (2\pi)^4} \;
e^{  iy \cdot (P-q_1-q_2 -K -R -R')  +\int \frac{d^3 k}{ k^0}\; e^{-ik \cdot y}  S_F(k) }
\\&\times
\exp\big\{
  \alpha B_4^V(Q^2,t,m_\gamma) +\alpha( B_4^{V'}(Q'^2,t,m_\gamma))^*
\big\}\;
\Mcal_V(Q) \Mcal^*_{V'}(Q').
\end{split}
\label{eq:appA-final}
\end{equation}

\section{Mappings in the \foam\ integrand}
\label{AppendixFoam}

The regularized radiator distribution for the IFI component
in the semisoft photon analytical exponentiation
\begin{equation}
\rho(\gamma,v) = F(\gamma)\big( \delta(v)\; \veps^{\gamma} 
    + \theta(v-\veps)\; \gamma v^{\gamma-1}\big),\qquad
\int_0^1 dv\; \rho(\gamma,v) =F(\gamma) \equiv F_\gamma,
\end{equation}
is valid for both positive and negative $\gamma$.
The regulator $\veps$ should be smaller that any scale dependence
in the Born cross section times the target precision of the calculation.
In our case it should be below $\Gamma_Z/M_Z$ by factor of at least
$10^{-4}$, {\it i.e.}\ $\veps < 10^{-5}$ is recommended.%
\footnote{In the actual MC runs, we use $\veps = 10^{-6}$.
}
The distribution for \foam\ should be positive in the exploration phase; hence,
\begin{equation}
\tilde\rho(\gamma,v)=|\rho(\gamma,v)|
	= F_\gamma\; \big[ \delta(v)\; \veps^{\gamma} 
                   + \theta(v-\veps)\; |\gamma| v^{\gamma-1}\big]
\end{equation}
is used.
The mapping from $v$ to the internal variable $r\in (0,1)$ of \foam\
is chosen such that its Jacobian compensates exactly $\tilde\rho(v)$.
More precisely, $v(r)$ is the solution of the equation
\begin{equation}
r \int_0^{1} dv'\; \tilde\rho(\gamma,v') =
  \int_0^{v} dv'\; \tilde\rho(\gamma,v') = F_\gamma\; R(v).
\label{eq:B3}
\end{equation}
Note that for $\gamma>0$ we have $R(1)=1$,
while for  $\gamma<0$ we get $R(1)=2 e^{\gamma}-1 > 1$.
Differentiating Eq.~(\ref{eq:B3}) we get
$ F_\gamma R(1) dr = \tilde\rho(\gamma,v) dv ; $
hence the Jacobian is
\begin{equation}
 J(v) = |dv/dr| = F_\gamma R(1) (\tilde\rho(\gamma,v))^{-1}.
\end{equation}

For $\gamma>0$, the mapping 
[with $R(1)=1$ and $R(\veps)=\veps^{\gamma} $]
is simply
\begin{equation}
\begin{split}
&v(r)=0,\quad {\rm for}\quad r<R(\veps)=\veps^{\gamma},
\\&
v(r)= r^{1/\gamma},\quad {\rm for}\quad r>R(\veps).
\end{split}
\end{equation}
The corresponding Jacobian is
\begin{equation}
J(v) = 1/R(\veps)= \veps^{-\gamma}\; {\rm for}\quad v=0\quad
{\rm and}\quad
J(v) = F_\gamma\; (\tilde\rho(\gamma,v))^{-1}=\frac{v}{r\gamma}\;  {\rm for}\quad v>\veps.
\end{equation}

For $\gamma<0$ the mapping 
[with $R(1)=2\veps^{\gamma}-1 $ and $R(\veps)=\veps^{\gamma} $]
is more complicated:
\begin{equation}
\begin{split}
&v(r)=0,\quad {\rm for}\quad 
 r<\frac{R(\veps)}{R(1)}=\frac{\veps^{\gamma}}{2 \veps^{\gamma}-1},
\\&
v(r)= \big[ 2R(\veps) -r R(1) \big]^{1/\gamma}\!\!
    = \big[ 2\veps^\gamma -r (2\veps^\gamma-1) \big]^{1/\gamma} \quad 
{\rm for}\quad  r> \frac{R(\veps)}{R(1)}.
\end{split}
\end{equation}
The corresponding Jacobian reads
\begin{equation}
\begin{split}
J(v) = \frac{R(1)}{R(\veps)}= \frac{2\veps^{\gamma}-1}{\veps^{\gamma}}\;     
     {\rm for}\quad v=0\quad
{\rm and}\quad
J(v) = \frac{F_\gamma  R(1)}{\tilde\rho(\gamma,v)}\;\;  {\rm for}\quad v>\veps.
\end{split}
\end{equation}

In the second simulation stage, \foam\ generates weighted MC events
with the IFI component being $w=J(v) \rho(\gamma,v)$.
In the case of $\gamma>0$, the weight (component) in \foam\ will be $w=1$ 
for any $v$, while for $\gamma<0$ it will be $w=R(1)$
for $v=0$ and $w=-R(1)$ for $v>\veps$.

In addition, special care has to be taken in the case of $\gamma\to 0$,
that is for $\cos\theta \simeq 0$, because in this region
the above mappings can be numerically unstable due to the limited
range of the exponent in floating-point arithmetic.
Because of that, when $|\gamma\ln\veps|<\Delta \ll 1$,
many of the above formulas have to be expanded accordingly.%
\footnote{
 The value $\Delta=10^{-4}$ used now looks OK,
 as the error of $\sim\Delta^2=10^{-8}$ is more than acceptable.}

For $|\gamma\ln\veps|<\Delta \ll 1$ and $\gamma >0$,
the expanded distribution, mapping, and Jacobian read: 
\begin{equation}
\begin{split}
& \tilde\rho(v)=\rho(v)
  = F_\gamma \Big[
  \delta(v) (1+\gamma\ln\veps) +\theta(v>\veps) \frac {\gamma}{v} \Big],\quad
  R(\veps)= (1+\gamma\ln\veps),\; R(1)=1,
\\&
v(r)=0\quad {\rm for}\quad r<R(\veps),\qquad
v(r)= \exp\Big[-\frac{1}{\gamma}(1-r) \Big]\quad {\rm for}\quad r>R(\veps),
\\&
J(v) = 1/R(\veps)\; {\rm for}\quad v=0\quad {\rm and}\quad
J(v) = F_\gamma\; (\tilde\rho(\gamma,v))^{-1}\;  {\rm for}\quad v>\veps.
\end{split}
\end{equation}
For $\gamma<0$, the expanded expressions 
with $ R(\veps)= 1+\gamma\ln\veps,\; R(1)    = 1+2\gamma\ln\veps>1,$
read:
\begin{equation}
\begin{split}
& \tilde\rho(v)=|\rho(v)|
 =F_\gamma \Big[
    \delta(v) (1+\gamma\ln\veps) -\theta(v>\veps) \frac {\gamma}{v} \Big],\quad
\\&
v(r)=0,\quad {\rm for}\quad 
 r < \frac{R(\veps)}{R(1)}=\frac{1+\gamma\ln\veps}{1+2\gamma\ln\veps},
\\&
v(r)= \exp\Big[ \frac{1}{\gamma}(1-r) R(1) \Big] ,\quad 
{\rm for}\quad  r> \frac{R(\veps)}{R(1)},
\\&
J(v) = \frac{R(1)}{R(\veps)}=     
     {\rm for}\quad v=0\quad
{\rm and}\quad
J(v) = \frac{ F_\gamma R(1)}{\tilde\rho(\gamma,v)}\;  {\rm for}\quad v>\veps.
\end{split}
\end{equation}

\section{Zero and first order amplitudes without resummation}
\label{sec:appendixC}

For constructing the semisoft photon 
analytical resummation and matching
with the fixed-order ${\cal O}(\alpha^1) $ result,
we need the zeroth and first order amplitudes and distributions
in analytical form.
In particular, we will need the differential cross section
of the final muons, integrated over photon angles,
but keeping control over the photon energy.
The relevant results are scattered over several papers
\cite{Jadach:1988zp, Was:1989ce, Jadach:1987ws}.
See also Refs.~\cite{Bohm:1989pb, Bardin:1990de, Bardin:1990fu},
where they are sometimes incomplete,
or given in a form not suitable for our purposes;
hence it is worth collecting them once more in this appendix.

Following the notation of Ref.~\cite{Was:1989ce}, the Born cross section
and charge asymmetry read as follows:
\begin{equation}
\begin{split}
&\frac{d \sigma^{(0)}(s(1-v))}{dc} = 
  \frac{3\sigma_0(s) }{8}   \frac{1}{4} \sum_{\veps,\tau=\pm}
     \big| \Mmf_{\veps\tau}(v,c)  \big|^2 
 = \frac{3\sigma_0(s)}{8}  
   \big[ (1+c^2)\; \Dmf(v)  +2c \overline\Dmf(v) \big],
\\&
 \Mmf_{\veps\tau}(v,c)
 = \Mmf^\gamma_{\veps\tau}(v,c) +\Mmf^Z_{\veps\tau}(v,c)
 = (\veps\tau + c)  D_{\veps,\tau}(v),
\\&
  D_{\veps,\tau}(v) 
 = D^{\gamma}_{\veps,\tau}(v)+ D^{Z}_{\veps,\tau}(v)
 = \frac{q\tilde{q}}{1-v} + \frac{ g_\veps \tilde{g}_\tau }{\zeta -v},
\\&
 \zeta = \frac {s-M_Z^2+i \Gamma_Z M_Z} {s},\quad
 g_\tau = g_V +\tau g_A,\quad  \tilde{g}_\tau = \tilde{g}_V +\tau \tilde{g}_A,\quad 
 \sigma_0= \frac{4 \alpha \pi^2}{3s},
\end{split}
\end{equation}
where $c = \cos\theta$, $q=Q_e,\; \tilde{q}=Q_\mu$ are electric charges,
$\veps,\tau=\pm$ are twice the helicity of $e^-$ and $\mu^-$, and
\begin{equation}
\begin{split}
&
\Dmf(v)=
\frac{1}{4} \sum_{\veps\tau} \big| D_{\veps,\tau}(v) \big|^2 =
 \frac{ c_0}{ (1-v)^2}
+ \Re \frac{ 2c_1 }{(1-v)(\zeta-v) }
+ \frac{ c_2 }{ |\zeta-v |^2 },
\\&
\overline\Dmf(v)=
\frac{1}{4} \sum_{\veps\tau}  \veps\tau \big| D_{\veps,\tau}(v) \big|^2 =
  \Re \frac{ 2d_1 }{(1-v)(\zeta-v) }
+ \frac{ d_2 }{ |\zeta-v |^2 },
\\&
c_0 = (q\tilde{q})^2,\;\; 
c_1 = q\tilde{q} g_v \tilde{g}_v,\;\; 
c_2 = (g_v^2+g_a^2)( \tilde{g}_v^2+\tilde{g}_a^2) ),\;\;
\\&
d_1 = q\tilde{q} g_a \tilde{g}_a,\;\; 
d_2 = 4 g_v g_a  \tilde{g}_v \tilde{g}_a.
\end{split}
\end{equation}

The integration over $\cos\theta$ results in
\begin{equation}
\begin{split}
& \sigma^{(0)}=
  \sigma_0\; \frac{1}{4} \sum_{\veps\tau} | D_{\veps,\tau}(0)|^2,\quad
\sigma^{\cst (0)}= \int 2\cos\theta^\cst d\sigma^{(1)}
= \sigma_0\; \frac{1}{4} \sum_{\veps\tau} \veps\tau | D_{\veps,\tau}(0)|^2
\\&
 A_{\rm FB}^{(0)} = \frac{3}{4}  \langle 2\cos\theta^\cst \rangle^{(0)}
   = \frac{3}{4}  \frac{ \int 2\cos\theta^\cst\; d\sigma^{(0)}}{\sigma^{(0)}}
   = \frac{3}{4}\; \frac{\sum_{\veps\tau}  \veps\tau | D_{\veps,\tau}(0)|^2 }%
                      {\sum_{\veps\tau}  | D_{\veps,\tau}(0)|^2 }
   = \frac{3}{4}\; \frac{ \overline\Dmf(0)}{ \Dmf(0)}.
\end{split}
\end{equation}
Following the notation of Ref.~\cite{Was:1989ce},
the {\em noninterference} ${\cal O}(\alpha^1) $
results with implicit integration over photon angles
and explicit integration over photon energy up to $x=v_{\max}$ read:
\begin{equation}
\begin{split}
& \overline{A}_{\rm FB}^{(1)}(x)
 = \frac{3}{4} \frac{\overline\sigma^{\cst(1)}(x) }%
                    {\overline\sigma^{(1)}(x)},\qquad
\overline\sigma^{\cst (1)}(x)
  =  \int_{v<x} 2\cos\theta^\cst d \overline\sigma^{(1)},
\\& 
\frac{\overline\sigma^{(1)}(x)}{ \sigma_0}
= [1 +F(x) ]\; \Dmf(0) + W (x),\;
 \frac{\overline\sigma^{\cst(1)}(x)}{ \sigma_0}
 =[1 +F^\cst(x) ]\; \overline\Dmf(0) + W^\cst(x),
\\&
 W( x)= \int_0^x dv\; \left[\gamma_I(s) P(v) 
   + q^2 \frac{\alpha}{\pi} \Delta_s\delta(v) \right]\; (1-v)\; \Dmf(v),
\\&
W^\cst(x)=
\int_0^x dv\;\left[\gamma_I(s) P(v) 
   +q^2 \frac{\alpha}{\pi} \Delta_s\delta(v) 
   -q^2 \frac{\alpha}{\pi}\;v \right]\; (1-v)\; \overline\Dmf(v),
\end{split}
\end{equation}
\begin{equation}
\begin{split}
& F(x) = \int_0^x dv\; [\gamma_F(s(1-v)) P(v) 
  +\tilde{q}^2\frac{\alpha}{\pi}  \Delta_s\delta(v) ],\quad
 \quad \Delta_s =   -\frac{1}{2} +\frac{\pi^2}{3} ,
\\&
 F^\cst(x) =
 \int_0^x dv\; \big[\gamma_F(s(1-v)) P(v) 
  +\tilde{q}^2\frac{\alpha}{\pi}  \Delta_s\delta(v) 
  -\tilde{q}^2 \frac{\alpha}{\pi}\;v  \big],
\\&
 P(v)= \Big(\frac{1+(1-v)^2}{v} \Big)_+=
  -\delta(v) \frac{3}{4} \ln\frac{1}{\veps}
  +\theta(v-\veps) \frac{1+(1-v)^2}{v},
\end{split}
\end{equation}
In Ref.~\cite{Was:1989ce}, analytical integrations over $v$ were done,
but for the purpose of the present resummation, we are more 
interested in the above unintegrated version.

In Ref.~\cite{Jadach:1988zp},
the contribution of IFI was added to the above charge asymmetry, 
 but in a version that was integrated over $v$.
The unintegrated version%
\footnote{The unintegrated version of $U^\cst_{\veps,\tau} (x)$
   was obviously used in Ref.~\cite{Jadach:1988zp}, but was not explicitly 
   shown there. Also, $U_{\veps,\tau} (x)$ was not provided there.}
including ISR+FSR+IFI with complete ${\cal O}(\alpha^1)$
for $v\in (0,1)$,
needed for resummation is as follows:
\begin{equation}
\begin{split}
&A_{\rm FB}^{(1)} (x) 
= \frac{3}{4} \frac{\sigma^{\cst (1)}(x)}{\sigma^{(1)}(x) }
= \frac{3}{4} \frac{ \int_{v< x } \! \cos\theta^\cst\; d\sigma^{(1)}}%
       {\sigma^{(1)}(x) },
\\& 
\sigma^{(1)}(x)   = \overline\sigma^{(1)}(x)     +\sigma_0 U(x),\quad
\sigma^{\cst(1)}(x)= \overline\sigma^{\cst (1)}(x) +\sigma_0 U^\cst(x),
\\&
 U(x)= \int_0^x dv\; \rho_X^{(1)}(v)\; (1-v)\overline\Dmf(v,0)
  + 3q\tilde{q} \frac{\alpha}{\pi} \Re\Big\{
           A_\gamma\; \overline\Bmf^{\gamma}(0)
          +A_Z\;     \overline\Bmf^{Z}(0) \Big\},
\\& 
 U^\cst (x)= \int_0^x dv\;\rho_X^{\cst (1)}(v)\; (1-v)\Dmf(v,0)
  +  2 q\tilde{q}\frac{\alpha}{\pi} \Re\Big\{
           A^\cst_\gamma\; \Bmf^{\gamma}(0)
         + A^\cst_Z\;     \Bmf^{Z}(0) \Big\},
\\&
\rho_X^{(1)}(v)= 2q\tilde{q}\frac{\alpha}{\pi}
\bigg\{
 \delta(v) \Big[ 3\ln\frac{1}{\delta} \Big]
+\theta(v-\delta) (-3) \frac{2-v}{2v}
\bigg\}
\\&
\rho_X^{\cst (1)}(v)= 2 q\tilde{q} \frac{\alpha}{\pi}
\bigg\{
 \delta(v) \Big[ 5 \ln\frac{1}{\delta} \Big]
+\theta(v-\delta) \frac{(-1)}{(2-v) v} \big[ 10(1-v) + 3 v^2 \big]
\bigg\},
\end{split}
\end{equation}

The combined contributions to the total cross section 
from real soft emission (interference part)
and virtual $\gamma\gamma$ and $\gamma Z$ boxes 
can be deduced from the $k_{\max}\to 0$ limit of
formulas in Ref.~\cite{Jadach:1987ws}:
\begin{equation}
\begin{split}
& A_\gamma= -\frac{1}{2} ,\quad
  A_Z= -\ln|1-\zeta| -\zeta +(1-\zeta)(2-\zeta) \ln\frac{-\zeta}{1-\zeta}.
\\&
 \Bmf^{\gamma}(0)= c_0 +\frac{ c_1 }{\zeta^* },\quad
 \Bmf^{Z}(0)    = \frac{ c_1 }{\zeta^* }+\frac{ c_2 }{\zeta\zeta^*} ,
\end{split}
\end{equation}
The analogous contributions to $\sigma^{\cst (1)}$ can be obtained
from formulas in Ref.~\cite{Jadach:1988zp}:
\begin{equation}
\begin{split}
& A^\cst_\gamma= \frac{65}{36} -i\frac{2}{3}\pi,
\\&
A^\cst_Z=
   \frac{31}{9}\zeta -9\zeta^2+4\zeta^3
   -\ln(1-\zeta)\Big( \frac{15}{2} -13\zeta+12\zeta^2-4\zeta^3 \Big)
\\&~~~~~
   +\ln(-\zeta) \Big( 5 -\frac{17}{3}\zeta +2\zeta^2 \Big)
   +4\zeta (1-\zeta)^3  \Big( {\rm Li}_2 \Big( \frac{-\zeta}{1-\zeta} \Big)
           -\frac{\pi^2}{6} \Big).
\\&
\overline\Bmf^{\gamma}(0)= \frac{ d_1 }{\zeta^* },\quad
\overline\Bmf^{Z}(0)= \frac{ d_1 }{\zeta^* }+\frac{ d_2 }{\zeta\zeta^*}.
\end{split}
\end{equation}
The following combinations of the Born amplitudes are involved:
\begin{equation}
\begin{split}
&
\Dmf(v,u)=
 \Re \frac{1}{4} \sum_{\veps\tau} \big( D_{\veps,\tau}(v)^* D_{\veps,\tau}(u) \big)=
\\&~~~
=\Re \Big\{
 \frac{ c_0}{ (1-v)(1-u)}
+\frac{ c_1 }{(1-v)(\zeta^*-u) }
+\frac{ c_1 }{(\zeta-v)(1-u) }
+\frac{ c_2 }{(\zeta-v)(\zeta^*-u) }
\Big\}
\\&
\overline\Dmf(v,u)=
 \Re \frac{1}{4} \sum_{\veps\tau}  \veps\tau\; 
   \big( D_{\veps,\tau}(v)^* D_{\veps,\tau}(u) \big) =
\\&~~~
=\Re \Big\{
 \frac{ d_1 }{ (1-v)(\zeta^*-u) }
+\frac{ d_1 }{ (\zeta-v)(1-u) }
+\frac{ d_2  }{ (\zeta-v)(\zeta^*-u) }
\Big\},
\\&
\Bmf^V(0)=\frac{1}{4} \sum_{\veps\tau}   D^V_{\veps,\tau}(0)^* D_{\veps,\tau}(0),\quad
\overline\Bmf^V(0)=
 \frac{1}{4} \sum_{\veps\tau} \veps\tau  D^V_{\veps,\tau}(0)^* D_{\veps,\tau}(0),\;
V=\gamma,Z,
\end{split}
\end{equation}
Let us remark that the following relations hold:
\begin{equation}
\Dmf(v)= \Dmf(v,v),\;\; \overline\Dmf(v)= \overline\Dmf(v,v),
\end{equation}
\begin{equation}
\Re\Bmf^{\gamma}(0)+ \Re\Bmf^{Z}(0) = \Dmf(0),\quad
\Re\overline\Bmf^{\gamma}(0)+ \Re\overline\Bmf^{Z}(0) = \overline\Dmf(0),\quad
\end{equation}

We also need the virtual box and real soft contributions
before integration over $c=\cos\theta$.
Spin amplitudes for two $\gamma\gamma$ box diagram and two $\gamma Z$ 
box diagram contributions, 
normalized the same way as the Born spin amplitudes, read as follows:
\begin{equation}
\begin{split}
&\Mmf^{\{\gamma\gamma\}}_{\veps\tau} 
= (q\tilde{q})^2 (\veps\tau X_1(c) +X_2(c)),
\\&
\Mmf^{\{\gamma Z\}}_{\veps\tau} 
= q\tilde{q}\; g_\veps q_\tau  (\veps\tau Z_1(c) +Z_2(c)).
\end{split}
\label{eq:BoxesGGandGZ}
\end{equation}
Their interference with Born amplitudes leads to the following contributions%
\footnote{
We use
 $ (1/4)\sum_{\veps\tau} \veps\tau\; q\tilde{q}\; g_\veps \tilde{g}_\tau = d_1
  = q\tilde{q}\; g_a \tilde{g}_a $ and
$ (1/4)\sum_{\veps\tau} g_\veps \tilde{g}_\tau = c_1 = q\tilde{q}\; g_v \tilde{g}_v$
}
\begin{equation}
\begin{split}
&\frac{d\sigma^{\gamma\gamma}}{dc}
=\frac{3\sigma_0}{8} \frac{1}{4} \sum_{\veps\tau}
 2\Re \big[ \Mmf^{\{\gamma\gamma\}}_{\veps\tau}\; \Mmf^*_{\veps\tau}(0,c) \big]
\\&
=\frac{3\sigma_0}{8} \frac{1}{4} \sum_{\veps\tau}
2\Re \big\{
(q\tilde{q})^2 
    \big[ X_1(c)+cX_2(c) +\veps\tau (cX_1 +X_2(c)) \big]\;
    D^*_{\veps\tau}(0)
   \big\}
\\&
=\frac{3\sigma_0}{8} q\tilde{q}\;
2\Re \big\{ (c_0 +\frac{c_1}{\zeta^*} )[X_1(c)+cX_2(c)] 
           +\frac{d_1}{\zeta^*} [cX_1(c) +X_2(c)]  \big\}
\\&
=\frac{3\sigma_0}{8}
q\tilde{q}\;
2\Re \big\{
   (c_0 +\frac{c_1}{\zeta^*} ) [F^{\gamma\gamma}(c) -F^{\gamma\gamma}(-c)]
        +\frac{d_1}{\zeta^*}   [F^{\gamma\gamma}(c) +F^{\gamma\gamma}(-c)]
   \big\},
\end{split}
\end{equation}
where
\begin{equation}
\begin{split}
& X_1(c)+cX_2(c)= F^{\gamma\gamma}(c) -F^{\gamma\gamma}(-c),
\\&
  cX_1(c) +X_2(c)=F^{\gamma\gamma}(c) +F^{\gamma\gamma}(-c),\quad
  c_\pm = \frac{1 \pm c}{2},
\\&
F^{\gamma\gamma}(c)=
2\frac{\alpha}{\pi}  \Big\{
 2\Big( \ln\frac{m_\gamma^2}{s} +i\pi \Big) c_+^2 \ln\frac{c_-}{c_+}
  -\frac{1}{2} c   \Big( \ln^2 c_- +2i\pi \Big)
             + c_+ \Big( \ln c_- +i\pi \Big)
\Big\}.
\end{split}
\end{equation}
Similarly, for the $\gamma Z$ box, we have:
\begin{equation}
\begin{split}
&\frac{d\sigma^{\gamma Z}}{dc}
=\frac{3\sigma_0}{8} \frac{1}{4} \sum_{\veps\tau}
 2\Re \big[ \Mmf^{\{\gamma Z\}}_{\veps\tau}\; \Mmf^*_{\veps\tau}(0,c) \big]
\\&
=\frac{3\sigma_0}{8} \frac{1}{4} \sum_{\veps\tau}
2\Re \big\{
  q\tilde{q}
     \big[ Z_1(c)+cZ_2(c) +\veps\tau (cZ_1 +Z_2(c)) \big]\;
  D^*_{\veps\tau}(0)
\big\}
\\&
=\frac{3\sigma_0}{8} q\tilde{q}\;
2\Re \big\{ \Big( c_1 +\frac{c_2}{\zeta^*} \Big)[Z_1(c) +cZ_2(c)] 
          +\Big(  d_1 +\frac{d_2}{\zeta^*} \Big)[cZ_1(c) +Z_2(c)] \big\}
\\&
=\frac{3\sigma_0}{8} q\tilde{q}\;
2\Re \big\{ \Big( c_1 +\frac{c_2}{\zeta^*} \Big)
                       [F^{\gamma Z}(c) -F^{\gamma Z}(-c)]
          +\Big( d_1 +\frac{d_2}{\zeta^*} \Big)
                       [F^{\gamma Z}(c) +F^{\gamma Z}(-c)] \big\}
\end{split}
\end{equation}
where $ F^{\gamma Z}(c) $ is related  in a simple way to $f(s,t,u)$
of Ref.~\cite{Brown:1983jv}:
\begin{equation}
   F^{\gamma Z}(c)  = 2\frac{\alpha}{\pi}  c_+^2 s f(s,t,u),\quad
   F^{\gamma Z}(-c) = 2\frac{\alpha}{\pi}  c_-^2 s f(s,u,t).
\end{equation}
In the \kkmc\ code, the $\gamma Z$ box of Ref.~\cite{Brown:1983jv}
is programmed as follows:
\begin{equation}
\begin{split}
&F^{\gamma Z}(c)=
 \ln\frac{t}{u} \ln\frac{m_\gamma^2}{(tu)^{1/2}}
 -2\ln\frac{t}{u} \ln\frac{\Mb^2-s}{\Mb^2 }
 +{\rm Li}_2\Big( \frac{\Mb^2+u}{\Mb^2} \Big)
 -{\rm Li}_2\Big( \frac{\Mb^2+t}{\Mb^2} \Big)
\\&
+\frac{(\overline{M}^2-s)(u-t-\overline{M}^2)}{u^2}
 \bigg(
   \ln\frac{-t}{s} \ln\frac{ \overline{M}^2-s }{\overline{M}^2}
  +{\rm Li}_2\Big( \frac{\Mb^2+t}{\Mb^2} \Big)
  -{\rm Li}_2\Big( \frac{\Mb^2-s}{\Mb^2} \Big)
 \bigg)
\\&
+\frac{(\Mb^2-s)^2}{us} \ln\frac{\Mb^2-s}{\Mb^2 }
+\frac{(\Mb^2-s)}{u}    \ln\frac{-t}{\Mb^2 },
\end{split}
\end{equation}
where $\Mb^2= M_Z^2-M_Z\Gamma_Z$,  $t=-(1-c)s$ and $u=(1+c)s$.

Finally, the above box contributions have to be combined with (interference)
the corresponding soft real emission contribution
\begin{equation}
\begin{split}
&\frac{d\sigma^{\rm soft}_X}{dc}=
\frac{d\sigma^{(0)}}{dc}\; q \tilde{q}\frac{\alpha}{\pi} \; \delta^{\rm soft}_X(c),
\\&
\delta^{\rm soft}_X(c) =
  4\ln\frac{c_-}{c_+} \ln\frac{ s^{1/2} \epsilon}{m_\gamma}
 +\ln^2 c_- -\ln^2 c_+
 +2 {\rm Li}_2(c_+)-2 {\rm Li}_2(c_-)
\end{split}
\end{equation}
such that the usual cancellation of the IR regulator $m_\gamma$ occurs,
leaving out the IR cutoff on photon energy $v\leq \epsilon \ll 1$.

Let us finally define explicit relations between integrated
and unintegrated virtual+soft contributions:
\begin{equation}
\begin{split}
&
3A_\gamma=
\int dc\; \big( F^{\gamma\gamma}(c) - F^{\gamma\gamma}(-c)
                + \delta^{\rm soft}_X(c) \big),
\\&
2A^\cst_\gamma=
\int 2c dc\; \big( F^{\gamma\gamma}(c) + F^{\gamma\gamma}(-c)
                + \delta^{\rm soft}_X(c) \big),
\\&
3 A_Z=
\int dc\; \big( F^{\gamma Z}(c) - F^{\gamma\gamma}(-c)
                + \delta^{\rm soft}_X(c) \big),
\\&
2 A^\cst_Z=
\int 2c dc\; \big( F^{\gamma Z}(c) + F^{\gamma\gamma}(-c)
                + \delta^{\rm soft}_X(c) \big),
\end{split}
\end{equation}

\begin{figure}[hb!]
  \centering
  \includegraphics[width=165mm,height=75mm]{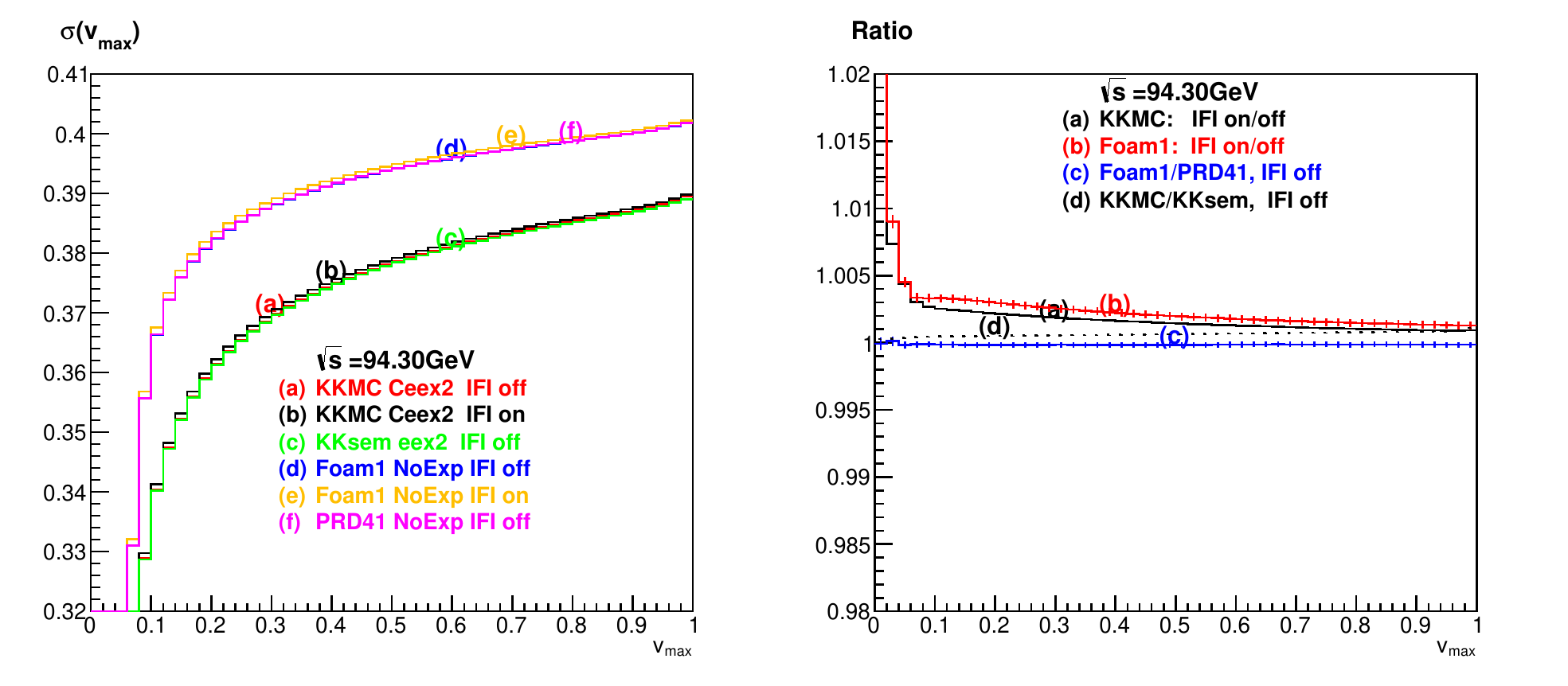}
  \caption{\sf
  ${\cal O}(\alpha^1)$ from old papers and \kkfoam.
  }
  \label{fig:cFigVcum}
\end{figure}

\begin{figure}[b!]
  \centering
  \includegraphics[width=165mm,height=75mm]{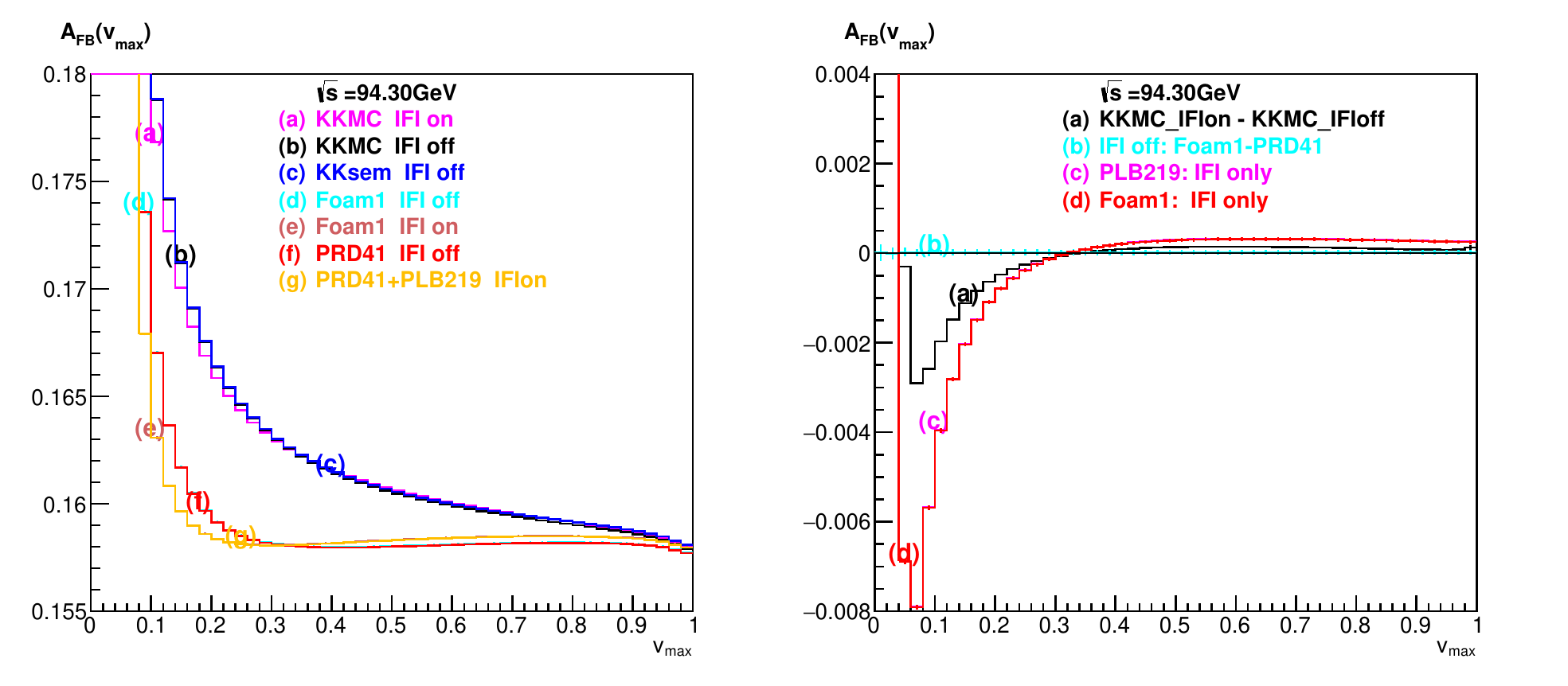}
  \caption{\sf
  ${\cal O}(\alpha^1)$ from old papers and \kkfoam.
  }
  \label{fig:cFigAfb10}
\end{figure}

Finally, in Figs.~\ref{fig:cFigVcum} and \ref{fig:cFigAfb10}
we crosscheck the old analytical results with \kkfoam,
in which the integration over $\cos\theta$ (virtual)
and over photon energy $v$ (real photon) is done numerically.
As we see, there is perfect agreement between old analytical
formulas and new results using \kkfoam2.

\bibliographystyle{h-physrev3}
\bibliography{bibafb}

\begin{thebibliography}{10}

\bibitem{Gomez-Ceballos:2013zzn}
TLEP Design Study Working Group, M.~Bicer {\em et~al.},
\newblock J. High Energy Phys. {\bf 01}, 164 (2014), [arXiv:1308.6176].

\bibitem{Mangano:2018mur}
FCC Collaboration, A.~Abada {\em et~al.},
\newblock Eur. Phys. J. C {\bf 79}, 474 (2019).

\bibitem{Benedikt:2018qee}
FCC Collaboration, A.~Abada {\em et~al.},
\newblock Eur. Phys. J. ST {\bf 228}, 261 (2019).

\bibitem{Blondel:2019qlh}
A.~Blondel {\em et~al.},
\newblock arXiv:1901.02648.

\bibitem{Janot:2015gjr}
P.~Janot,
\newblock J. High Energy Phys. {\bf 02}, 053 (2016), [arXiv:1512.05544].

\bibitem{ALEPH:2005ab}
ALEPH, DELPHI, L3, OPAL, SLD, LEP Electroweak Working Group, SLD Electroweak
  Group, SLD Heavy Flavour Group, S.~Schael {\em et~al.},
\newblock Phys. Rep. {\bf 427}, 257 (2006), [arXiv:hep-ex/0509008].

\bibitem{Bohm:1989pb}
M.~B{\"{ o}}hm {\em et~al.},
\newblock {Forward-backward asymmetries},
\newblock in {\em {LEP Physics Workshop Geneva, Switzerland, 1989}}, pp.
  203--234, 1989.

\bibitem{Jadach:1988zp}
S.~Jadach and Z.~W\c{a}s,
\newblock Phys. Lett. B {\bf 219}, 103 (1989).

\bibitem{Bardin:1990fu}
D.~{\relax Yu}. Bardin {\em et~al.},
\newblock Nucl. Phys. B {\bf 351}, 1 (1991), [arXiv:hep-ph/9801208].

\bibitem{Bardin:1999gt}
D.~{\relax Yu}. Bardin, M.~Grunewald and G.~Passarino,
\newblock arXiv:hep-ph/9902452.

\bibitem{Kobel:2000aw}
Two Fermion Working Group, M.~Kobel {\em et~al.},
\newblock {Two-Fermion Production in Electron-Positron Collisions},
\newblock in {\em {Proceedings, Monte Carlo Workshop: Report of the working
  groups on precision calculation for LEP-2 physics: CERN, Geneva, Switzerland,
  March 12-13, June 25-26, October 12-13 Oct 1999}}, 2000,
  [arXiv:hep-ph/0007180].

\bibitem{Jadach:2000ir}
S.~Jadach, B.~F.~L. Ward and Z.~Was,
\newblock Phys. Rev. D {\bf 63}, 113009 (2001), [arXiv:hep-ph/0006359].

\bibitem{Bardin:1999yd}
D.~{\relax Yu}. Bardin {\em et~al.},
\newblock Comput. Phys. Commun. {\bf 133}, 229 (2001), [arXiv:hep-ph/9908433].

\bibitem{Montagna:1995ja}
G.~Montagna, O.~Nicrosini, G.~Passarino and F.~Piccinini,
\newblock Comput. Phys. Commun. {\bf 93}, 120 (1996), [arXiv:hep-ph/9506329].

\bibitem{Montagna:1993py}
G.~Montagna, F.~Piccinini, O.~Nicrosini, G.~Passarino and R.~Pittau,
\newblock Nucl. Phys. B {\bf 401}, 3 (1993).

\bibitem{Jadach:1993yv}
S.~Jadach, B.~F.~L. Ward and Z.~Was,
\newblock Comput. Phys. Commun. {\bf 79}, 503 (1994).

\bibitem{Jadach:1999tr}
S.~Jadach, B.~F.~L. Ward and Z.~Was,
\newblock Comput. Phys. Commun. {\bf 124}, 233 (2000), [arXiv:hep-ph/9905205].

\bibitem{Greco:1975rm}
M.~Greco, G.~Pancheri-Srivastava and Y.~Srivastava,
\newblock Nucl. Phys. B {\bf 101}, 234 (1975).

\bibitem{Greco:1975ke}
M.~Greco, G.~Pancheri-Srivastava and Y.~Srivastava,
\newblock Phys. Lett. B {\bf 56}, 367 (1975).

\bibitem{Greco:1980mh}
M.~Greco, G.~Pancheri-Srivastava and Y.~Srivastava,
\newblock Nucl. Phys. B {\bf 171}, 118 (1980),
\newblock [Erratum: Nucl. Phys. B 197, 543 (1982)].

\bibitem{Jadach:1998jb}
S.~Jadach, B.~F.~L. Ward and Z.~Was,
\newblock Phys. Lett. B {\bf 449}, 97 (1999), [arXiv:hep-ph/9905453].

\bibitem{Jadach:1999vf}
S.~Jadach, B.~F.~L. Ward and Z.~Was,
\newblock Comput. Phys. Commun. {\bf 130}, 260 (2000), [arXiv:hep-ph/9912214].

\bibitem{Jadach:1999pp}
S.~Jadach, M.~Skrzypek and B.~Pietrzyk,
\newblock Phys. Lett. B {\bf 456}, 77 (1999).

\bibitem{Blondel:2018mad}
A.~Blondel {\em et~al.},
\newblock {Standard Model Theory for the FCC-ee: The Tera-Z},
\newblock in {\em {Mini Workshop on Precision EW and QCD Calculations for the
  FCC Studies : Methods and Techniques CERN, Geneva, Switzerland, January
  12-13, 2018}}, 2018, [arXiv:1809.01830].

\bibitem{Yennie:1961ad}
D.~R. Yennie, S.~C. Frautschi and H.~Suura,
\newblock Annals Phys. {\bf 13}, 379 (1961).

\bibitem{Jadach:2018jjo}
S.~Jadach, W.~Płaczek, M.~Skrzypek, B.~F.~L. Ward and S.~A. Yost,
\newblock Phys. Lett. B {\bf 790}, 314 (2019), [arXiv:1812.01004].

\bibitem{Jadach:2019bye}
S.~Jadach and M.~Skrzypek,
\newblock arXiv:1903.09895.

\bibitem{Bardin:1989tq}
D.~Y. Bardin {\em et~al.},
\newblock Comput. Phys. Commun. {\bf 59}, 303 (1990).

\bibitem{Jadach:1999sf}
S.~Jadach,
\newblock Comput. Phys. Commun. {\bf 130}, 244 (2000), [arXiv:physics/9910004].

\bibitem{Jadach:2002kn}
S.~Jadach,
\newblock Comput. Phys. Commun. {\bf 152}, 55 (2003), [arXiv:physics/0203033].

\bibitem{Jadach:1991ws}
S.~Jadach, B.~F.~L. Ward and Z.~Was,
\newblock Comput. Phys. Commun. {\bf 66}, 276 (1991).

\bibitem{Z-physics-at-lep-1:89}
R.~K. G.~Altarelli and C.~Verzegnassi, editors,
\newblock {\em Z-Physics at LEP1} (CERN, Geneva, 1989),
\newblock Vol. 1 - 3.

\bibitem{Was:1989ce}
Z.~W\c{a}s and S.~Jadach,
\newblock Phys. Rev. D {\bf 41}, 1425 (1990).

\bibitem{Kleiss:1985yh}
R.~Kleiss and W.~J. Stirling,
\newblock Nucl. Phys. B {\bf 262}, 235 (1985).

\bibitem{Davier:2010nc}
M.~Davier, A.~Hoecker, B.~Malaescu and Z.~Zhang,
\newblock Eur. Phys. J. C {\bf 71}, 1515 (2011), [arXiv:1010.4180],
\newblock [Erratum: Eur. Phys. J. C 72, 1874 (2012)].

\bibitem{Jadach:1987ws}
S.~Jadach, J.~H. Kuhn, R.~G. Stuart and Z.~W\c{a}s,
\newblock Z. Phys. C {\bf 38}, 609 (1988),
\newblock [Erratum: Z. Phys. C 45, 528 (1990)].

\bibitem{Bardin:1990de}
D.~{\relax Yu}. Bardin {\em et~al.},
\newblock Phys. Lett. B {\bf 255}, 290 (1991), [arXiv:hep-ph/9801209].

\bibitem{Brown:1983jv}
R.~W. Brown, R.~Decker and E.~A. Paschos,
\newblock Phys. Rev. Lett. {\bf 52}, 1192 (1984).

\end{thebibliography}
\end{document}